\documentclass[epsfig,12pt]{article}
\usepackage{amssymb}
\usepackage{amsfonts}
\usepackage{graphicx}
\usepackage{amsmath}

\input epsf.sty
\newtheorem{theorem}{Theorem}
\newtheorem{acknowledgement}[theorem]{Acknowledgement}

\input{tcilatex}
\makeatletter \@addtoreset{equation}{section}
\renewcommand{\theequation}{\thesection.\arabic{equation}}
\def\be{\begin{equation}}
\def\ee{\end{equation}}
\def\bea{\begin{eqnarray}}
\def\eea{\end{eqnarray}}

\newcommand{\nc}{\newcommand}
\nc{\al}{\alpha} \nc{\bib}{\bibitem} \nc{\la}{\lambda}
\nc{\C}{\mbox{\hspace{1.24mm}\rule{0.2mm}{2.5mm}\hspace{-2.7mm}
C}} \nc{\R}{\mbox{\hspace{.04mm}\rule{0.2mm}{2.8mm}\hspace{-1.5mm}
R}}

\begin{document}

\title{%
\rightline{\mbox{\small
{Lab/UFR-HEP0702-rev/GNPHE/0702-rev/VACBT/0702-rev}}} \textbf{\ }$\mathcal{N}%
\mathbf{=2}$\textbf{\ Supersymmetric Black Attractors in }\\
\textbf{Six and Seven Dimensions }}
\author{ A. Belhaj$^{1,2,3}${\small \thanks{%
belhaj@unizar.es}}, L.B. Drissi$^{1,3}${\small \thanks{%
drissilb@gmail.com}}, E.H. Saidi$^{1,3,4}${\small \thanks{%
h-saidi@fsr.ac.ma}}, A. Segui$^{2}${\small \thanks{%
segui@unizar.es} } \\
{\small 1. Lab/UFR- Physique des Hautes Energies, Facult\'{e} des Sciences,
Rabat, Morocco,}\\
{\small 2. Departamento de Fisica Teorica, Universidad de Zaragoza,
50009-Zaragoza, Spain }\\
{\small 3. GNPHE, Groupement National de Physique des Hautes Energies, Si%
\`{e}ge focal: FS, Rabat}\\
{\small 4. Coll\`{e}ge SPC, Acad\'{e}mie Hassan II des Sciences et
Techniques, Rabat, Morocco} }
\maketitle

\begin{abstract}
Using a quaternionic formulation of the moduli space $\boldsymbol{M}\left( 
{\small IIA/K3}\right) $ of 10D type IIA superstring on a generic K3 complex
surface with volume $\boldsymbol{V}_{0}$, we study extremal $\mathcal{N}=2$
black attractors in 6D space-time and their uplifting to 7D. For the 6D
theory, we exhibit the role played by $6D$ $\mathcal{N}=1$ hypermultiplets
and the $Z^{m}$ central charges isotriplet of the $6D$ $\mathcal{N}=2$
superalgebra. We construct explicitly the special hyperKahler geometry of $%
\boldsymbol{M}\left( {\small IIA/K3}\right) $ and show that the $SO\left(
4\right) \times SO\left( 20\right) $ invariant hyperKahler potential is
given by $\mathcal{H}=\mathcal{H}_{0}+\mathrm{Tr}\left[ \ln \left( 1-%
\boldsymbol{V}_{0}^{-1}\boldsymbol{S}\right) \right] $ with Kahler leading
term $\mathcal{H}_{0}=\mathrm{Tr}\left[ \ln \boldsymbol{V}_{0}\right] $ plus
an extra term which can be expanded as a power series in $\boldsymbol{V}%
_{0}^{-1}$ and the traceless and symmetric 3$\times $3 matrix $\boldsymbol{S}
$. We also derive the holomorphic matrix prepotential $\mathcal{G}$ and the
flux potential $\mathcal{G}_{BH}$ of the 6D black objects induced by the
topology of the RR field strengths $\mathcal{F}_{2}=d\mathcal{A}_{1}$ and $%
\mathcal{F}_{4}=d\mathcal{A}_{3}$ on the K3 surface and show that $\mathcal{G%
}_{BH}$ reads as $Q_{0}+\sum_{m=1}^{3}q^{m}Z^{m}$. Moreover, we reveal that $%
Z^{m}=\sum_{I=1}^{20}Q_{I}\left( \int_{C_{2}^{I}}J^{m}\right) $ where the
isotriplet $J^{m}$ is the hyperKahler 2- form on the K3 surface. It is found
as well that the uplifting to seven dimensions is quite similar to 4D/5D
correspondence for back hole potential considered in arXiv 0707.0964
[hep-th]. Then we study the $\mathcal{N}=2$ black object attractors in 6D
and 7D obtained respectively from type IIA string and M-theory on K3.

\ \ \newline
\textbf{Key words}: Type IIA superstring on the K3 surface, Special
hyperKahler geometry, 6D/7D$\mathcal{\ N}=2$ black objects, Attractor
mechanism.
\end{abstract}

\tableofcontents

\vspace{-1cm}

\section{Introduction}

\qquad Large distance of compactified ten dimensional ($10D$) type II
superstrings and $11D$ M- theory down to lower space time dimensions have
recently known a great revival of interest in connection with topological
string \ \cite{GV1,GV2,GV3,OSV,2,AOSV,CCGPSS1,CCGPSS2,3,4,5}, black objects
and attractor mechanism \cite{FK1,FK2,SV,OSV,MS,Wi,ABDS,AOO,GM,AO}. In the
context of $4D$ black holes, Calabi-Yau compactification of $10D$ type IIA
superstring ($11D$ M- theory) down to $4D$ $(5D)$ can be approximated at
large distances by $4D$ ($5D$) $\mathcal{N}=2$ supergravity whose low energy
theory include, in addition to the $4D$ ($5D$) $\mathcal{N}=2$ supergravity
multiplet, matter organized into a number $n_{V}$ of $4D$ ($5D$) $\mathcal{N}%
=2$ vector multiplets and $n_{H}$ hypermultiplets. In the 4D $\mathcal{N}=2$
black hole attractor mechanism, the hypermultiplets decouple and one is left
with gravity\ coupled to vector multiplets whose vacuum configuration is
nicely described by special Kahler (real) geometry \cite%
{CFM,L,KSM,AGS,Se,D,TT}. There, special complex (real) geometry plays a
crucial role in the study of $4D$ ($5D$) $\mathcal{N}=2$ extremal black
holes and in the understanding of their attractor mechanisms.

Motivated by the exploration of the attractor mechanism for black objects in 
$6D$ and $7D$ space-time as well as the basic role that have to play $6D$ $%
\mathcal{N}=1$ hypermultiplets in the attractor issue, we first study
extremal $\mathcal{N}=2$ black objects in six dimensional space time and
analyse the special \textit{d-geometry} of underlying $6D$ $\mathcal{N}=2$
supergravity describing the low energy limit of type superstring IIA on the
K3 surface. Then, we consider the uplifting of the $6D$ model to $7D$ which
can be understood as the compactifcation of $11D$ M-theory on the K3
surface. For the $6D$ case, we take advantage of the hyperkahler structure
of the moduli space $\boldsymbol{M}\left( \text{{\small IIA/K3}}\right) $ of 
$10D$ type IIA superstring on a generic K3 surface, with fixed string
coupling constant, to introduce a convenient basis of local quaternionic
coordinates $\left\{ w^{I},\text{ }0\leq I\leq 20\right\} $ built as, 
\begin{equation*}
w^{I}=\int_{C_{2}^{I}}\left( B^{NS}+i\sum_{s=0,\pm 1}\sigma ^{s}\Omega
^{\left( 1+s,1-s\right) }\right) ,\qquad I=1,\ldots ,h_{{\tiny K3}}^{\left(
1,1\right) },
\end{equation*}%
where the basis set $\left\{ C_{2}^{I}\right\} $ refer to the twenty real 2-
cycles of K3 dual to the Hodge $\left( 1,1\right) $- forms. In this
relation, the $2\times 2$ matrices $\sigma ^{0,\pm }$ are the usual Pauli
matrices in the Cartan basis, the $\left( p,q\right) $- form $\Omega
^{\left( p,q\right) }$, with $p+q=2$, are the Hodge 2-forms on the K3
surface and 
\begin{equation*}
\mathcal{J}=B^{NS}I_{2\times 2}+i\sum_{s=0,\pm 1}\sigma ^{s}\Omega ^{\left(
1+s,1-s\right) },
\end{equation*}%
is the \textit{"quaternionified"} Kahler 2-form on the K3 surface. This 2-
form is valued in the $2\times 2$ matrix algebra generated, in addition to
the identity $I_{2\times 2}$, by the three Pauli matrices $\sigma ^{m}$ and
should be compared with the complexified Kahler form%
\begin{equation*}
B^{NS}+i\Omega ^{\left( 1,1\right) },
\end{equation*}%
we encounter in the study of 10D type IIA superstring on Calabi-Yau
threefolds. The local coordinate basis $\left\{ w^{I}\right\} $ allows to
approach the underlying \ special hyperkahler \textit{d-geometry }of $%
\boldsymbol{M}\left( \text{{\small IIA/K3}}\right) $ in quite similar manner
as do the complexified Kahler moduli%
\begin{equation*}
z^{I}=\int_{C_{2}^{I}}\left( B^{NS}+i\Omega ^{\left( 1,1\right) }\right) ,
\end{equation*}%
in the study of $10D$ type IIA superstring on Calabi-Yau threefolds. It is a
basis set of non abelian 2$\times $2 matrices which permit to use the power
of the algebra of matrices to study the quaternionic geometry of $%
\boldsymbol{M}\left( \text{{\small IIA/K3}}\right) $. This matrix
formulation has several special properties mainly governed by the
quaternionic structure of $\boldsymbol{M}\left( \text{{\small IIA/K3}}%
\right) $ and captured in practice by the spin $\frac{1}{2}$ representation
of the $SU\left( 2\right) $ R- symmetry of the 6D $\mathcal{N}=2$
superalgebra and by the standard Clifford algebra identities of the $2\times
2$ Pauli matrices. Matrix formulation of the special quaternionic \emph{%
d-geometry} of $\boldsymbol{M}\left( \text{{\small IIA/K3}}\right) $ allows
in particular to: \newline
\textbf{(1)} exhibit manifestly the $SO\left( 4\right) \times SO\left(
20\right) $ gauge symmetry of the moduli space $\boldsymbol{M}\left( \text{%
{\small IIA/K3}}\right) $ with fixed string coupling constant.\newline
\textbf{(2)} permit to treat explicitly the computation of the hyperKahler
potential $\mathcal{H}$, the holomorphic matrix prepotenial $\mathcal{G}$
and the flux potential $\mathcal{G}_{BH}$ of the $6D$ $\mathcal{N}=2$ black
object attractors. \newline
\textbf{(3)} get the explicit moduli realization of the central charges of
BPS states of the 6D $\mathcal{N}=2$ superalgebra. \newline
\textbf{(4)} provide a natural way to deal with $7D$ supersymmetric black
objects by uplifting in the same spirit as recently done in \textrm{\cite%
{CFM}} for studying extremal black hole attractor mechanism in $4D/5D$
extended supergravities.

The organization of this paper is as follows. In section 2, we review the
compactification of 10D type IIA superstring on the K3 surface. Then we give
comments on the 6D $\mathcal{N}=2$ supersymmetric low energy limit; in
particular the aspect regarding the structure of central charges and 6D
supersymmetric BPS states. In section 3, we develop a matrix formulation to
analyse the corresponding moduli space $\boldsymbol{M}_{\text{{\small IIA/K3}%
}}$. In section 4, we introduce the quaternionified 2- form $\mathcal{J}$
and develop the special quaternionic \emph{d-geometry} in the matrix
formalism. In section 5, we compute the hyperKahler potential $\mathcal{H}$
as a power series in the inverse of the volume of the K3 surface and derive
the matrix holomorphic prepotential $\mathcal{G}$ using the 2-cycles
intersection matrix $d_{IJ}$ of generic K3. In section 6, we study the $6D$ $%
\mathcal{N}=2\ $ black object attractors and their 7D uplifting by using
type IIA D-branes wrapping cycles of the K3 surface. Extending the idea on
the flux compactification of 10D type IIA superstring on Calabi-Yau 4-folds
given in \cite{GVW} to the case of the K3 surface, we derive, amongst
others, the flux potential for $6D$ N=2 supergravity theory. In section 7,
we study the effective scalar potential and the attractor mechanism for the
6D and 7D black objects. In section 8, we give our conclusion and make a
discussion. In section 9, we give an appendix on $\mathcal{N}=2$
supersymmetry in six dimensions.

\section{Type II Superstrings on K3}

\qquad Low energy dynamics of $10D$ type II superstrings on the K3 complex
surface is described by $6D$ $\mathcal{N}=2$ supergravity coupled to
superYang-Mills \cite{W,V,A,A1,HT}.\ One distinguishes two six dimensional
models A and B depending on whether one started from 10D type IIA or 10D
type IIB superstrings. These models are respectively given by the usual 6D
non chiral $\mathcal{N}=\left( 1,1\right) $ and 6D chiral $\mathcal{N}%
=\left( 2,0\right) $ models and have different moduli spaces $\boldsymbol{M}%
_{\text{\textsc{IIA/K3}}}$ and $\boldsymbol{M}_{\text{\textsc{IIB/K3}}}$. In
present paper, we mainly consider the large distance limit described by non
chiral 6D $\mathcal{N}=\left( 1,1\right) $ ($\mathcal{N}=2$ for short)
supergravity and focus on the hyperKahler structure of $\boldsymbol{M}_{%
\text{\textsc{IIA/K3}}}$. The moduli space $\boldsymbol{M}_{\text{\textsc{%
IIB/K3}}}$ namely%
\begin{equation}
\boldsymbol{M}_{\text{\textsc{IIB/K3}}}\boldsymbol{=}\frac{S0\left(
5,21\right) }{S0\left( 5\right) \times S0\left( 21\right) },  \label{2b}
\end{equation}%
hasn't however such hyperKahler structure. The local coordinates of $%
\boldsymbol{M}_{\text{\textsc{IIB/K3}}}$ are in the bi- fundamental of $%
S0\left( 5\right) \times S0\left( 21\right) $ and so $\boldsymbol{M}_{\text{%
\textsc{IIB/K3}}}$ has a real dimension multiple of \emph{5} rather than a
multiple of \emph{4 }which is a necessary condition for having\emph{\ }%
quaternionic geometry.

\subsection{Type IIA superstring on K3}

\subsubsection{Moduli space $\boldsymbol{M}_{\text{\textsc{IIA/K3}}}$}

\qquad In 10D type superstring IIA on a generic K3 surface, which is known
to be dual to the 10D heterotic superstring on T$^{4}$, the moduli space $%
\boldsymbol{M}_{\text{\textsc{IIA/K3}}}$ of metric deformations and stringy
vacuum field configurations is given by the non compact real space 
\begin{equation}
\boldsymbol{M}_{\text{\textsc{IIA/K3}}}=\boldsymbol{M}\times SO\left(
1,1\right) ,
\end{equation}%
with $\boldsymbol{M}$ given by the homogeneous space 
\begin{equation}
\boldsymbol{M=}\frac{SO\left( 4,20\right) }{SO\left( 4\right) \times
SO\left( 20\right) },  \label{ms}
\end{equation}%
and where the extra $SO\left( 1,1\right) $ stands for the dilaton (i.e the
string coupling $g_{s}$). For fixed $g_{s}$, $\boldsymbol{M}_{\text{\textsc{%
IIA/K3}}}$ reduces to $\boldsymbol{M}$ and so the restricted moduli space $%
\boldsymbol{M}$ has a real eighty dimension capturing a quaternionic
structure. In addition to the NS-NS B-field 2-form B$^{NS}$ whose role will
be discussed later on, the hyper- structure of $\boldsymbol{M}$ should be
described by a real $SU\left( 2\right) $ isotriplet of 2-forms 
\begin{equation}
\mathrm{J}^{i}=\left( J^{1},\quad J^{2},\quad J^{3}\right) ,\qquad \mathrm{J}%
^{i}=\left( \mathrm{J}^{i}\right) ^{\dagger }\quad ,\quad d\mathrm{J}%
^{i}=0\quad ,  \label{ho}
\end{equation}%
rotated under the adjoint representation of $SU\left( 2\right) $ isometry
group of the K3 surface as shown below 
\begin{eqnarray}
\left[ D^{i},\mathrm{J}^{j}\right] &=&i\epsilon ^{ijk}\mathrm{J}^{k},\qquad
i,j,k=1,2,3,  \notag \\
\left[ D^{i},D^{j}\right] &=&i\epsilon ^{ijk}D^{k}.  \label{al}
\end{eqnarray}%
In the above relations, the 2-form $J^{3}$, to be denoted often as $\Omega
^{\left( 1,1\right) }$ or sometimes as $J^{0}$, is the usual hermitian
Kahler 2-form one encounters in generic complex $n$- dimensional Kahler
manifold. The two others are given\footnote{%
Given a generic Hodge $\left( p,q\right) $-form $\Omega ^{\left( p,q\right)
} $ on a Kahler manifold, one can associate to it two integers: the highest
weight $h=\frac{p+q}{2}$, of an underlying $SU\left( 2\right) $ group
representation that classify the forms, and the isospin $s=\frac{p-q}{2}$. A
Hodge multiplet consists of those $\left( p,q\right) $-form $\Omega ^{\left(
p,q\right) }$ with isospins as $-h\leq s\leq h$. For $h=1$, the
corresponding Hodge multiplet is an isotriplet with isospins $s=0,\pm 1$.}
by 
\begin{equation}
J^{1}=\func{Re}\Omega ^{2,0},\qquad J^{2}=\func{Im}\Omega ^{2,0},
\end{equation}%
where $\Omega ^{\left( 2,0\right) }$ is the complex holomorphic 2- form on
the K3 surface with complex conjugate $\overline{\Omega ^{\left( 2,0\right) }%
}=\Omega ^{\left( 0,2\right) }$. The $D^{i}$'s denote the generators of $%
SU\left( 2\right) $ symmetry group rotating the $\mathrm{J}^{i}$'s and $%
\epsilon ^{ijk}$ is the associated completely antisymmetric structure
constants.

For later use, we recall that in the K3 compactification of 10D type II
superstrings, the local Lorentz group $SO\left( 1,9\right) $ of the 10D
space time breaks down to $SO\left( 1,5\right) \times SU\left( 2\right) $.
The latter, which is required by supersymmetry and K3 holonomy, is contained
in $SO\left( 1,5\right) \times SO\left( 4\right) $. We also recall that the
homomorphism $SO\left( 4\right) \sim SU\left( 2\right) \times SU^{\prime
}\left( 2\right) $ allowing to put the 4-vector representation of $SO\left(
4\right) $ as a $2\times 2$ hermitian matrix.

\subsubsection{Central charges in 6D supersymmetric theories}

\qquad We first introduce the central charges as they appear in the standard
Haag-Lopuszanski-Sohnius (HLS) superalgebra. Then, we consider its extension
by implementing p- branes, $p>1$ \textrm{\cite{6}}.

\paragraph{\textbf{A}. Geometric central charges: \newline
}

There are different ways to introduce central charges $\mathcal{Z}_{m}$ of
extended supersymmetric algebras. A tricky way is to use their geometric
realization as translations along the transverse directions of 10D
superstrings compactified down to lower dimensions. Below, we describe this
realization for the case of supersymmetric field theory in six dimension
space time.\newline
$\mathcal{Z}_{m}$ \textbf{as translation operators}:\qquad\ \ \newline
First, recall that a generic ten dimensional real vector V$_{M}$ splits
generally into a real 6D space time vector V$_{\mu }$ and four real space
time scalars $\mathrm{f}_{m}$. In particular, this is valid for the 10D
space time coordinates $x^{M}\rightarrow $ $\left( x^{\mu },y^{m}\right) $;
but also for generic 10D Maxwell gauge fields $\mathcal{A}_{M}$ which then
split as 
\begin{equation}
\mathcal{A}_{M}\qquad \rightarrow \qquad \left( \mathcal{A}_{\mu },\phi
_{m}\right) .  \label{sg}
\end{equation}%
The same reduction can be done for the 10D energy momentum vector $P_{M}$
which decomposes into 6D energy momentum $P_{\mu }$ vector and four central
charges $P_{m}$ as shown below: 
\begin{equation}
P_{M}\qquad \rightarrow \qquad \left( P_{\mu },P_{m}\right) ,\qquad
m=6,7,8,9.  \label{im}
\end{equation}%
By central charge, we mean that, viewed as operators, the $P_{m}$'s commute
with all generators of the 6D supersymmetric Poincar\'{e} algebra $\mathcal{%
SP}_{6D}^{N=2}$ given in appendix and which will be considered later on. The 
$P_{m}$'s commute in particular with the bosonic $P_{\mu }$ and $M_{\left[
\mu \nu \right] }=x_{[\mu }P_{\nu ]}$ generators of the 6D Poincar\'{e}
algebra $\mathcal{P}_{6D}$;%
\begin{equation}
\left[ P_{\mu },P_{m}\right] =0,\qquad \left[ M_{\left[ \mu \nu \right]
},P_{m}\right] =0,\qquad \left[ P_{m},P_{n}\right] =0,  \label{sca}
\end{equation}%
In quantum physics where 10D energy momentum operator $P_{M}$ is realized as 
$\frac{\hbar }{i}\partial _{M}$, the central charge operators $P_{m}$ are
given by $\frac{\hbar }{i}\partial _{m}$. From this representation, we learn
that central charges $P_{m}$ may geometrically be interpreted as
translations along the $y^{m}$ coordinates of the 4D transverse space. 
\newline
$\mathcal{Z}_{m}$ \textbf{and the sgauginos} $\phi _{m}$:\qquad\ \ \newline
The above description can be pushed further by establishing a close link
between the central charges $\mathcal{Z}_{m}$ and the scalar gauge fields%
\footnote{%
In 6D $\mathcal{N}=2$ supersymmetric gauge theory, vector multiplets have,
in addition to gauginos, the usual 6D vector field A$_{\mu }$; but also four
scalars $\phi _{m}$.} (sgauginos) $\phi _{m}$ eq(\ref{sg}). The idea is to
use the same trick for the abelian gauge field strength $\mathcal{F}%
_{MN}=\partial _{M}\mathcal{A}_{N}-\partial _{N}\mathcal{A}_{M}$. We have
the decomposition 
\begin{equation}
\mathcal{F}_{MN}\qquad \rightarrow \qquad \left( \mathcal{F}_{\mu \nu },%
\text{ }\mathcal{F}_{\mu m}=\partial _{\mu }\phi _{m}\right) ,
\end{equation}%
where we have used the anzats $\partial _{m}\mathcal{A}_{\mu }=0=\partial
_{m}\phi _{n}$; that is the fields $\mathcal{A}_{\mu }$ and $\phi _{n}$\
have no dependence on the internal coordinates $y^{m}$:%
\begin{equation}
\mathcal{A}_{\mu }=\mathcal{A}_{\mu }\left( x\right) ,\qquad \phi _{n}=\phi
_{n}\left( x\right) .
\end{equation}%
Now, using the fact that $\mathcal{F}_{MN}$ is also the curvature of the
gauge covariant derivative $\mathcal{D}_{M}=\partial _{M}+\mathcal{A}_{M}$,
i.e 
\begin{equation}
\mathcal{F}_{MN}=\left[ \mathcal{D}_{M},\mathcal{D}_{N}\right]
\end{equation}%
we learn the two following: \newline
(i) the real four sgauginos $\phi _{m}$ are just "gauge fields" along the
transverse directions which appear in the covariantization of the central
charges 
\begin{equation}
\partial _{m}\rightarrow \mathcal{D}_{m}=\partial _{m}+\phi _{m}.
\label{213}
\end{equation}%
(ii) By introduction of the gauge fields, the commutator between the
covariantized space time and transverse translations get a non zero
curvature given by the gradient $\partial _{\mu }\phi _{m}$ of the
sgauginos, $\mathcal{F}_{\mu m}=\left[ \mathcal{D}_{\mu },\mathcal{D}_{m}%
\right] $. The square scalar $\sum_{\mu ,m}\left( \mathcal{F}_{\mu m}%
\mathcal{F}^{\mu m}\right) $ gives just the kinetic energy of the four real
sgauginos.\newline
\textbf{Supersymmetric algebra}:\qquad\ \newline
Under the compactification on the K3 surface, the 32 supersymmetries of 10D $%
\mathcal{N}=2$ superalgebra reduce to 16 supersymmetries forming a 6D $%
\mathcal{N}=2$ superalgebra with central charges. To get the defining
relations of this algebra, it is interesting to express space time and
transverse space translations $P_{\mu }$ and $P_{m}$ in terms of spinor
representations of the $SO\left( 1,5\right) \times SO\left( 4\right) $
group. Using $SO\left( 1,5\right) $ spinor representations, the 6D vector $%
P_{\mu }$ can be written as $P_{\left[ \alpha \beta \right] }$ ( roughly
speaking as a $SU\left( 4\right) $ antisymmetric representation). The
central charges are scalars under $SO\left( 1,5\right) $. Similarly, the
four real central charges $P_{m}$ can be also put as $Z^{ia}$; that is as a
spin $\left( \frac{1}{2},\frac{1}{2}\right) $ representation of $SO\left(
4\right) \sim SU\left( 2\right) \times SU^{\prime }\left( 2\right) $. As
spin $\frac{1}{2}$\ representations are complex, we need to impose the
reality condition 
\begin{equation}
\overline{Z^{ia}}=\epsilon _{ab}\epsilon _{ij}Z^{jb},\qquad i,j=1,2,\qquad
a,b=1,2,
\end{equation}%
where $\epsilon _{ij}$ ($\epsilon _{ab}$) is the metric tensor of $SU\left(
2\right) $ ($SU^{\prime }\left( 2\right) $). By identifying the two $%
SU\left( 2\right) $ factors of $SO\left( 4\right) $, the 2$\times $2
component matrix $Z^{ia}$ becomes $Z^{ij}$ and can be decomposed as a real
isosinglet $Z_{0}$ and a real isotriplet $Z^{\left( ij\right) }$ as given
below 
\begin{equation}
Z^{ij}=Z_{0}\epsilon ^{ij}+Z^{\left( ij\right) },\qquad i,j=1,2.  \label{cc}
\end{equation}%
We will show later, when we go into the details of 10D type IIA superstring
compactification on the K3 surface, that $Z_{0}$ can be related to the NS-NS
B- field and $Z^{\left( ij\right) }$ to the hyperKahler structure on K3.

In the appendix eq(\ref{n2}), see also eqs(\ref{mw}), we show that four real
central charges are allowed by 6D $\mathcal{N}=\left( 1,1\right) $
supersymmetric algebra. This non chiral superalgebra is generated by two
kinds of fermionic generators $Q_{\alpha }^{i}$ and $S_{i}^{\bar{\alpha}}$
whose basic graded commutation relations are given by,%
\begin{eqnarray}
\left\{ Q_{\alpha }^{i},Q_{\beta }^{j}\right\} &=&\epsilon ^{ij}P_{\left[
\alpha \beta \right] },\qquad \alpha ,\text{ }\beta =1,...4  \notag \\
\left\{ S_{i}^{\bar{\alpha}},S_{j}^{\bar{\beta}}\right\} &=&\epsilon _{ij}P^{%
\left[ \bar{\alpha}\bar{\beta}\right] },\qquad i,\text{ }j=1,2  \notag \\
\left\{ Q_{\alpha }^{i},S_{j}^{\bar{\alpha}}\right\} &=&\delta _{\alpha }^{%
\bar{\alpha}}Z_{j}^{i}  \label{n11} \\
\left[ P_{\left[ \alpha \beta \right] },Q_{\gamma }^{j}\right] &=&\left[ P_{%
\left[ \alpha \beta \right] },S_{j}^{\bar{\alpha}}\right] =0  \notag \\
\left[ P_{\left[ \alpha \beta \right] },Z_{j}^{i}\right] &=&\left[
Z_{a}^{i},Q_{\alpha }^{j}\right] =\left[ Z_{a}^{i},S_{b}^{\bar{\alpha}}%
\right] =0.  \notag
\end{eqnarray}%
The four central charge components $Z^{ij}$ appearing in above eqs can be
also given an interpretation from both the view of $SO\left( 1,5\right)
\times SU\left( 2\right) $ group theoretical representation and the view of
extended $\mathcal{N}=4$ and $\mathcal{N}=2$ supersymmetry in 4D.\newline
From the $SO\left( 1,5\right) \times SU\left( 2\right) $ view, the general
form of the anticommutation relation of the two $SO\left( 1,5\right) $ space
time spinors $Q_{\alpha }^{i}$ and $S_{j}^{\bar{\alpha}}$ should be as,%
\begin{equation}
\left\{ Q_{\alpha }^{i},S_{j}^{\bar{\alpha}}\right\} =Z_{j\alpha }^{i\bar{%
\alpha}},
\end{equation}%
where $Z_{j\alpha }^{i\bar{\alpha}}$ is constrained to commute with all
other generators of the 6D $\mathcal{N}=2$ superalgebra. However space time
interpretation which demand that the central charges to have no space time
index; that is $SO\left( 1,5\right) $ invariant eqs(\ref{im}-\ref{sca}),
requires then 
\begin{equation}
Z_{j\alpha }^{i\bar{\alpha}}=\delta _{\alpha }^{\bar{\alpha}}Z_{j}^{i}.
\end{equation}%
This restriction should be understood as associated with the space time
singlet $\left( 1,4\right) $ in the following tensor product 
\begin{equation}
\left( 4,2\right) \times \left( \overline{4},2\right) =\left( 1,4\right)
\oplus \left( 15,4\right) ,
\end{equation}%
where $\left( 4,2\right) $ stands for $Q_{\alpha }^{i}$ and $\left( 
\overline{4},2\right) $ for $S_{j}^{\bar{\alpha}}$. The $\left( 15,4\right) $
extra term in the above decomposition transforms non trivially under space
time symmetry. By roughly thinking about $SO\left( 1,5\right) $ as the
Euclidean $SO\left( 6\right) \simeq SU\left( 4\right) $, we see that the
term $\left( 15,4\right) $ transforms in the adjoint representation of $%
SU\left( 4\right) $ space time symmetry. It also corresponds to the
antisymmetric component in the reduction of the tensor product of two $%
SO\left( 1,5\right) $ vectors namely: $6\otimes 6=1+15+20$. \newline
From the view of 4D$\ \mathcal{N}=4$ superalgebra, it is interesting to
recall first that there, one has six central charges, 
\begin{equation}
Z^{\left[ ab\right] },\qquad a,\text{ }b=1,...,4,
\end{equation}%
transforming in the 6- dimensional representation of $SU\left( 4\right) \sim
SO\left( 6\right) $ R- symmetry\footnote{%
Notice that we have been using two kinds of $SO\left( 6\right) $ symmetry
groups which should not be confused. We have: (1) the usual $SO\left(
6\right) $ R- symmetry group of the compactification of 10D space time down
to 4D. (2) the Euclidean version of the 6D\ space time group $SO\left(
1,5\right) $.}. Under the breaking of the $SO\left( 6\right) $ R-symmetry as 
$SO\left( 2\right) \times SO\left( 4\right) $, the 6 central charges $Z^{%
\left[ ab\right] }$ of the 4D$\ \mathcal{N}=4$ superalgebra split as $6=4+2$
where the four central charges are precisely given by $Z^{ij}$, the same as
in eqs(\ref{n11}), and the extra two, which can be denoted as $\mathrm{Z}%
_{0}+i\mathrm{Z}_{e}$, are generated by the compactification from 6D down to
4D. To fix the ideas on the parallel between 6D $\mathcal{N}=2$ and 4D $%
\mathcal{N}=2$ superalgebras, we recall herebelow the 4D $\mathcal{N}=2$
supersymmetric algebra and the way in which the two central charges $\mathrm{%
Z}_{4D}^{\mathcal{N}=2}=\mathrm{Z}_{0}+i\mathrm{Z}_{e}$ enter in the game:%
\begin{eqnarray}
\mathrm{Q}_{a}^{i}\mathrm{Q}_{j}^{\dot{a}}+\mathrm{Q}_{j}^{\dot{a}}\mathrm{Q}%
_{a}^{i} &=&\delta _{j}^{i}\mathrm{P}_{a}^{\dot{a}},\qquad \qquad a,\text{ }%
\dot{a}=1,2,  \notag \\
\mathrm{Q}_{a}^{i}\mathrm{Q}_{b}^{j}+\mathrm{Q}_{b}^{j}\mathrm{Q}_{a}^{i}
&=&\epsilon ^{ij}\epsilon _{ab}\mathrm{Z}_{4D}^{\mathcal{N}=2},\qquad i,%
\text{ }j=1,2,  \notag \\
\mathrm{Q}_{i}^{\dot{a}}\mathrm{Q}_{j}^{\dot{b}}+\mathrm{Q}_{j}^{\dot{b}}%
\mathrm{Q}_{i}^{\dot{a}} &=&\epsilon ^{\dot{a}\dot{b}}\epsilon _{ij}\mathrm{Z%
}_{4D}^{\mathcal{N}=2}, \\
\left[ \mathrm{Z}_{4D}^{\mathcal{N}=2},\mathrm{Q}_{a}^{i}\right] &=&\left[ 
\mathrm{Z}_{4D}^{\mathcal{N}=2},\mathrm{Q}_{j}^{\dot{a}}\right] =\left[ 
\mathrm{Z}_{4D}^{\mathcal{N}=2},\mathrm{P}_{a}^{\dot{a}}\right] =0.  \notag
\end{eqnarray}%
We also recall that the positivity of the norm of the 10D energy momentum
vector ($\sum_{0}^{9}P_{M}P^{M}=\left( E^{2}-\mathbf{P}^{2}\right)
-\sum_{m}\left( P_{m}\right) ^{2}\geq 0$) puts a strong constraint on the $%
Z^{ij}$ central charges of 6D $\mathcal{N}=2$ supersymmetric
representations. We have 
\begin{equation}
Z_{0}^{2}+Z^{\left( ij\right) }Z_{\left( ij\right) }\leq \mathrm{M}_{6}^{2}
\label{bps}
\end{equation}%
where $\mathrm{M}_{6}$ is the mass of the 6D $\mathcal{N}=2$ supermultiplet;
that is energy $E=P_{0}$ in the rest frame of particles of the
supermultiplet. The 6D $\mathcal{N}=2$ BPS corresponds to supersymmetric
states with sutured bound; i.e $Z_{0}^{2}+Z^{\left( ij\right) }Z_{\left(
ij\right) }=\mathrm{M}^{2}$. We will turn later to this relation when we
consider black attractors in six dimensions.

For completeness of the study of the 6D$\ \mathcal{N}=2$ superalgebra, it is
useful to recall as well that a generic 10D Majorana-Weyl spinor with 16
components $Q_{A}$ generally decomposes, in six dimensions, into four $%
SO\left( 1,5\right) $ Weyl spinors $Q_{\alpha }^{i}$ and $Q_{\alpha }^{a}$,
constrained by a reality condition, according to the reduction rule $%
16=\left( 4,2\right) +\left( 4^{\prime },2^{\prime }\right) $. In the
particular case of the K3 compactification where only half of
supersymmetries survive, a 10D Majorana-Weyl spinor $Q_{A}$ reduces down to $%
8=\left( 4,2\right) $ real object $Q_{\alpha }^{i}$, $\alpha =1,...,4,$ $%
i=1,2$, and satisfies the following reality condition \textrm{\cite{HST}} 
\begin{equation}
\overline{\left( Q_{\alpha }^{i}\right) }=\epsilon _{ij}\mathrm{B}_{\alpha
}^{\beta }Q_{\beta }^{j},\qquad \text{with \ \ }\mathrm{B}^{+}\mathrm{B}=-1.
\label{mw}
\end{equation}%
The two real $Q_{\alpha }^{i}$'s can be complexified as $Q_{\alpha
}^{+}=Q_{\alpha }^{1}+iQ_{\alpha }^{2}$ to give one complex 4- Weyl spinor.
We will turn to eqs(\ref{cc}) and (\ref{mw}) later on when we consider the
supersymmetric algebra in six dimensions (see section 4 and appendix). For
completeness, it is also convenient to introduce the following basis for the
Kahler 2-forms 
\begin{eqnarray}
J^{+} &=&J^{1}+iJ^{2},\qquad J^{-}=J^{1}-iJ^{2},\qquad J^{0}=J^{3}  \notag \\
\left( J^{\pm }\right) ^{\dagger } &=&J^{\mp }\quad ,\qquad dJ^{\pm }=0\quad
,  \label{h1}
\end{eqnarray}%
where the charges $0$ and $\pm $ stand for the usual $U_{C}\left( 1\right) $
Cartan charge of the $SU\left( 2\right) $. Obviously $J^{+}=\Omega ^{\left(
2,0\right) }$ and $J^{-}=\Omega ^{\left( 0,2\right) }$; see also footnote 1.

\paragraph{\textbf{B}. Implementing p- branes \newline
}

\qquad In the study of black objects of supergravity theories, in particular
in 6D $\mathcal{N}=2$ non chiral supergravity, one has to distinguish
between two kinds of central charges: \newline
(\textbf{1}) the four usual central charges $Z^{a}\sim \sigma
_{ij}^{a}Z^{ij} $ considered above and transforming as scalars under space
time rotations. These central charges are strongly related with the gauge
fields $\left\{ \mathcal{A}_{\mu }^{ij},i,j=1,2\right\} $ belonging to the $%
\mathcal{N}=2$ non chiral supergravity multiplet%
\begin{equation}
\left( g_{\mu \nu },\mathcal{B}_{\mu \nu }^{\pm },\mathcal{A}_{\mu
}^{ij},\sigma ;\psi _{\pm \mu \alpha }^{i},\chi _{\pm \alpha }^{i}\right) .
\end{equation}%
These charges (dressed by scalars fields) appear in the supersymmetric
transformations of the gravitinos $\psi _{\pm \mu \alpha }^{i}$ and
gravi-gauginos $\chi _{\pm \alpha }^{i}$ (gravi-photinos/gravi-dilatinos)). 
\newline
\textbf{(2)} central charges $\mathcal{Z}_{\Lambda }$ that are associated
with (dressed) electric and/or magnetic charges of the $\left( p+2\right) $-
form gauge field strengths (and their duals) corresponding to the p- branes
and ($D-p-4$- branes) within the supergravity theory. These central charges,
which have space time indices, do not appear in the standard HLS
superalgebra; but rather in its extended version \textrm{\cite{6,7,03,8,9}}. 
\newline
In 6D $\mathcal{N}=2$ non chiral supergravity we are interested in here, we
have in addition to the four 1- form gauge fields $\mathcal{A}_{\mu }^{ij}$
of the supergravity multiplet, other gauge fields that contribute as well to
the full spectrum of the central charges of the theory and so to its
effective scalar potential that will be considered in section 7. These
charges are\ given by:\newline
(i) Two central charges $\mathcal{Z}_{+}$ and $\mathcal{Z}_{-}$ associated
with the 3- form field strengths $\mathcal{H}_{3}^{\pm }\sim d\mathcal{B}%
_{2}^{\pm }$ 
\begin{equation}
\mathcal{Z}_{\pm }\sim \int_{S^{3}}\mathcal{H}_{3}^{\pm },  \label{z+}
\end{equation}%
where $\mathcal{B}_{2}^{+}$ and $\mathcal{B}_{2}^{-}$ are respectively the
self dual and anti-self dual NS-NS B-field of the $\mathcal{N}=2$ non chiral
supergravity in six dimensional space time. \newline
(ii) Twenty central charges $\mathcal{Z}^{I}$, associated with the gauge
fields $\mathcal{A}_{\mu }^{I}$ of the twenty 6D $\mathcal{N}=2$ Maxwell
multiplets, that follow from the compactification of the 10D type IIA
superstring on K3.%
\begin{equation}
\mathcal{Z}^{I}\sim \int_{S^{2}}\mathcal{F}_{2}^{I},\qquad \mathcal{F}%
_{2}^{I}=d\mathcal{A}^{I}.  \label{zi}
\end{equation}%
Notice that the point-like states associated with fields are obtained by
wrapping D2- branes on the h$^{1,1}$ 2-cycles of K3, $h^{1,1}\left(
K3\right) =20$. More details on the relations (\ref{z+}-\ref{zi}) as well as
others will be given in section 7.

\subsection{$SO\left( 4\right) \times SO\left( 20\right) $ invariance}

\qquad A natural way to study the geometry of the moduli space $\boldsymbol{%
M=}S0\left( 4,20\right) /S0\left( 4\right) \times S0\left( 20\right) $ eq(%
\ref{ms}) is to use the real local coordinate system 
\begin{equation}
\left\{ x^{aI}\right\} ,\qquad a=1,...,4,\qquad I=1,...,20.
\end{equation}%
These coordinates transform in the bi-fundamental%
\begin{equation}
x^{aI}\sim \left( 4,20\right)
\end{equation}%
of the $SO\left( 4\right) \times SO\left( 20\right) $ gauge symmetry as
follows 
\begin{equation}
x^{aI}\qquad \rightarrow \qquad \widetilde{x}^{aI}=\left(
\sum_{b=1}^{4}\Lambda _{b}^{a}\right) \left( \sum_{J=1}^{20}\Gamma
_{J}^{I}\right) x^{bJ}.
\end{equation}%
where $\Lambda _{b}^{a}$ and $\Gamma _{J}^{I}$ are rotation matrices. The $%
SO\left( 20\right) $ symmetry corresponds to the arbitrariness in the choice
of the basis of Kahler deformations while $SO\left( 4\right) \sim
SU^{2}\left( 2\right) $ corresponds to the rotation symmetry of the
hyperKahler 2-form isotriplet.

The $SO\left( 4\right) $ symmetry of the moduli space (\ref{ms}) plays an
important role in the study of 6D supersymmetric field theory limit of 10D
type IIA superstring on the K3 surface. An immediate goal is to implemented
this symmetry in the formalism as a manifest covariance. A priori, one can
imagine different, but equivalent, ways to do it. Two methods seem
particularly interesting especially for the study of black objects in 6D
dimensions and the corresponding attractor mechanism. These methods are
given by: \newline
\textbf{(1)} Matrix formulation which has been motivated by the use of
quaternions to deal with eq(\ref{ms}). In this method, the $SU(2)$ symmetry
is captured by the Pauli matrices 
\begin{equation}
\sigma ^{1},\text{ \ }\sigma ^{2},\text{ \ }\sigma ^{3},
\end{equation}%
which, as it is well known, obey as well as a 2D Clifford algebra that is
used to realize the three complex structures $\mathbf{i}$, $\mathbf{j}$ and $%
\mathbf{k=i\wedge j}$ of the quaternions. This algebraic method will be
developed in the present paper. \newline
\textbf{(2)} Geometric method based on the geometrization of $SU\left(
2\right) $ symmetry. Instead of the Pauli matrices, the generators of the SU$%
\left( 2\right) $ algebra (\ref{al}) are realized in this method as follows%
\begin{eqnarray}
D^{++} &=&\sum_{i=1}^{2}u^{+i}\frac{\partial }{\partial u^{-i}},  \notag \\
D^{--} &=&\sum_{i=1}^{2}u^{-i}\frac{\partial }{\partial u^{+i}}, \\
D^{0} &=&\sum_{i=1}^{2}\left( u^{+i}\frac{\partial }{\partial u^{+i}}-u^{-i}%
\frac{\partial }{\partial u^{-i}}\right) .  \notag
\end{eqnarray}%
This approach has been motivated by the harmonic superspace method used for
the study of 4D $\mathcal{N}=2$ supersymmetric field theories \textrm{\cite%
{HS, SS,SB,BS}}. We suspect that this method to be the natural framework to
deal with the study of special hyperKahler \emph{d- geometry} and the
corresponding hyperKahler metrics building \cite{GT,MH}. This method is
powerful; but it requires introducing more technicalities. This will be
considered elsewhere \cite{BS2}.

\section{Matrix Formulation}

\qquad The key idea of the algebraic method we will develop in what follows
to deal with special hyperkahler geometry may be summarized as follows: 
\newline
First identify the $SU\left( 2\right) $ and $SU\left( 2\right) ^{\prime }$
subgroup factors of $SO\left( 4\right) $ so that the real $4$- vector of $%
SO\left( 4\right) $ splits under the $\frac{1}{2}\times \frac{1}{2}$ spin
representation of $SU\left( 2\right) $ as a isosinglet and isotriplet; i.e 
\begin{equation}
4=1\oplus 3.
\end{equation}%
Then think about $SU\left( 2\right) $ as an algebraic structure that allows
to put the isotriplet $\mathbf{J=}\left( J^{1},J^{2},J^{3}\right) $ into a
traceless hermitian $2\times 2$ matrix 
\begin{equation}
\mathbf{J}=\mathcal{J}-\left( \frac{1}{2}\left( \mathrm{Tr}\mathcal{J}%
\right) \right) \text{ }\mathrm{I}_{2}\text{ ,\qquad }\mathrm{Tr}\mathbf{J}%
=0\qquad .
\end{equation}%
This method has several advantages mainly given by: \newline
\textbf{(i)} the similarity with the usual analysis of $\mathcal{N}=2$ black
holes in $4D$ \cite{MS,Wi}.\newline
\textbf{(ii)} the power of the spin $\frac{1}{2}$ representation of the $%
su\left( 2\right) $ algebra to deal with $su\left( 2\right) $ tensor
analysis. \newline
This method has also dis-advantages mainly associated with the fact that the
basic objects, in particular the moduli space variables, are non commuting
matrices. This property is not a technical difficulty; it captures in fact
the novelties brought by the hyperKahler geometry with respect to the
standard Kahler one.

\subsection{Quaternionized HyperKahler 2-Form}

\qquad The main lines of the matrix formulation towards the study of the
special hyperKahler \textit{d- geometry} may be summarized in the three
following points:\newline
\textbf{(1)} Represent the three closed Kahler 2-forms $J^{1},$ $J^{2}$ and $%
J^{3}$ by a traceless hermitian $2\times 2$ matrix as shown below 
\begin{equation}
\mathbf{J}=\left( 
\begin{array}{cc}
J^{3} & J^{1}+iJ^{2} \\ 
J^{1}-iJ^{2} & -J^{3}%
\end{array}%
\right) ,
\end{equation}%
or equivalently 
\begin{equation}
\mathbf{J}=\sum_{i=1}^{3}J_{i}\sigma ^{i}  \label{j}
\end{equation}%
with $\mathbf{J}^{+}\mathbf{=J}$, $d\mathbf{J}=0,$ and where we have set 
\begin{equation}
\Omega ^{\left( 2,0\right) }=J^{1}+iJ^{2},\qquad \Omega ^{\left( 0,2\right)
}=J^{1}-iJ^{2},\qquad \Omega ^{\left( 1,1\right) }=J^{3}.
\end{equation}%
The $\sigma ^{i}$'s in eq(\ref{j}) are the standard $2\times 2$ Pauli
matrices given by:%
\begin{equation}
\sigma ^{1}=\left( 
\begin{array}{cc}
0 & 1 \\ 
1 & 0%
\end{array}%
\right) ,\qquad \sigma ^{2}=\left( 
\begin{array}{cc}
0 & i \\ 
-i & 0%
\end{array}%
\right) ,\qquad \sigma ^{3}=\left( 
\begin{array}{cc}
1 & 0 \\ 
0 & -1%
\end{array}%
\right) .
\end{equation}%
\textbf{(2)} Quaternionify the isotriplet 1- form $\mathbf{J}$ by the
implementation of the NS-NS\ B- field on the 2-cycles. Setting 
\begin{equation}
B^{NS}\text{ }\mathrm{I}_{2\times 2}=\left( 
\begin{array}{cc}
B & 0 \\ 
0 & B%
\end{array}%
\right) ,
\end{equation}%
where $\mathrm{I}_{2\times 2}$ is the $2\times 2$ identity matrix, which
will be dropped now on for simplicity of notations, the \textit{%
quaternionified} closed hyperkahler 2-form reads as follows 
\begin{equation}
\mathcal{J}_{+}=B^{NS}+i\left( \sum_{i=1}^{3}J^{i}\sigma ^{i}\right) ,
\end{equation}%
or in a condensed form ($B^{NS}=B$) as 
\begin{equation}
\mathcal{J}_{+}=B+i\mathbf{\sigma J},  \label{pm}
\end{equation}%
with $d\mathcal{J}_{+}=0$. We also have 
\begin{equation}
\mathcal{J}_{-}=B-i\mathbf{\sigma J},\qquad \left( \mathcal{J}_{+}\right)
^{+}=\mathcal{J}_{-}
\end{equation}%
as well as 
\begin{equation}
B=\frac{1}{2}\mathrm{Tr}\left( \mathcal{J}_{\pm }\right) ,\qquad J^{i}=\mp 
\frac{i}{2}\mathrm{Tr}\left( \mathcal{J}_{\pm }\sigma ^{i}\right) .
\label{qa}
\end{equation}%
Note in passing that along with eqs(\ref{pm}), one may also quaternionify
the hyperKahler 2-form $\mathbf{J}$ by using the two following real
quantities,%
\begin{equation}
\mathcal{K}_{0}=B+\mathbf{\sigma J},\qquad \mathcal{L}_{0}=B-\mathbf{\sigma J%
},\qquad \mathbf{\sigma J=}\sum_{i=1}^{3}J^{i}\sigma ^{i},  \label{kl}
\end{equation}%
that is without the complex number $i$ in front of $\mathbf{\sigma J}$.
These two extra objects are self adjoints and lead basically to a \textit{%
vector like} theory with strong constraints on the Kahler potential. Though
interesting for a complete study, we will restrict our discussion to
considering only $\mathcal{J}_{+}$ and $\mathcal{J}_{-}$ for the following
reasons: \newline
\textbf{(i)} $\mathcal{J}_{\pm }$ are enough to define the components of the
quaternion (\ref{qa}). \newline
\textbf{(ii)} $\mathcal{J}_{\pm }$ are enough to have a reality condition
for building the (hyper) Kahler potential. \newline
\textbf{(iii)} $\mathcal{J}_{\pm }$ exhibit a striking parallel with the
complex Kahler geometry of 10D type IIA superstring on the Calabi-Yau
threefolds. There, the complexified Kahler 2-form is given by%
\begin{equation}
\boldsymbol{K}_{\pm }=B^{NS}\pm i\Omega ^{\left( 1,1\right) }.
\end{equation}%
(\textbf{iv}) $\mathcal{J}_{\pm }$ allow to define a "holomorphic"
prepotential $\mathcal{G}$ in same manner as in 10D type IIA superstring on
Calabi-Yau threefolds.\newline
However a complete analysis would also take into account the relations (\ref%
{kl}) as well. We will give a comment on the effect of implementing the
quantities (\ref{kl}) in the game later on.\newline
\textbf{(3)} Then require that observable quantities such as hyperkahler
potential $\mathcal{H}$, and the "holomorphic" prepotential $\mathcal{G}$,
to be invariant under the following $SU\left( 2\right) $ gauge
transformations,%
\begin{equation}
\mathcal{J}_{\mathcal{\pm }}^{\prime }=\mathcal{U}^{+}\text{ }\mathcal{J}%
_{\pm }\text{ }\mathcal{U},  \label{l}
\end{equation}%
where $\mathcal{U}$ is an arbitrary unitary $2\times 2$ matrix of $SU\left(
2\right) $. In other words, thinking about the hyperkahler potential $%
\mathcal{H}$ as a hermitian function $\mathcal{H}\left( w^{+},w^{-}\right) $
on the quaternionic moduli 
\begin{equation}
w^{\pm }=\int_{C_{2}}\mathcal{J}_{\pm },
\end{equation}%
to be introduced with details later on,\ we should have%
\begin{equation}
\mathcal{H}\left( w^{+\prime },w^{-\prime }\right) =\mathcal{H}\left(
w^{+},w^{-}\right) ,  \label{k}
\end{equation}%
where $w^{\pm \prime }$ are the transform of $w^{\pm }$ under $SU\left(
2\right) $ group. A similar statement can be said about the holomorphic
prepotential $\mathcal{G}$.

Notice that, though very interesting, this method has also some weak points.
One of the difficulties of this method is that the derivatives $\partial
_{-I}=\frac{\partial }{\partial w^{+I}}$ and $\partial _{+I}=\frac{\partial 
}{\partial w^{-I}}$ with respect to $w^{+}$ and $w^{-}$ have non trivial
torsions. They make the building of the hyperkahler metric following from
the potential $\mathcal{H}$ a non easy task. The point is that by defining
the derivatives with respect to $w^{+}$ and $w^{-}$ as, 
\begin{eqnarray}
\partial _{-I} &=&\frac{\partial }{\partial w^{+I}}=\frac{1}{2}\left( \frac{%
\partial }{\partial y^{I}}-\frac{i}{3}\mathbf{\sigma .\nabla }_{I}\right) , 
\notag \\
\partial _{+I} &=&\frac{\partial }{\partial w^{-I}}=\frac{1}{2}\left( \frac{%
\partial }{\partial y^{I}}+\frac{i}{3}\mathbf{\sigma .\nabla }_{I}\right) ,
\end{eqnarray}%
satisfying 
\begin{equation}
\frac{\partial w^{+J}}{\partial w^{+I}}=\delta _{I}^{J},\qquad \frac{%
\partial w^{-J}}{\partial w^{+I}}=0
\end{equation}%
and so on, we can check that their commutators $\left[ \partial
_{-I},\partial _{+J}\right] $ and $\left[ \partial _{\pm I},\partial _{\pm J}%
\right] $ are non zero. This property puts strong restrictions on the
differential analysis using $w^{+}$ and $w^{-}$ matrices as basic variables.
This difficulty, which is due to the use of Pauli matrices that do not
commute, 
\begin{equation}
\sigma ^{i}\sigma ^{j}-\sigma ^{j}\sigma ^{i}=i\sum_{k=1}^{3}\epsilon
^{ijk}\sigma ^{k},
\end{equation}%
can be overcome by using a geometric representation of $SU\left( 2\right) $
relying on the duality $SU\left( 2\right) \sim S^{3}$. However, this method
is beyond the scope of present study. We refer to \textrm{\cite{BS2}} for
more details.

Before going ahead, we would like to comment on the parallel between the
Kahler and hyperKahler geometries. This study will be helpful when we
consider the building of the potential $\mathcal{H}$ and the \emph{%
"holomorphic"} prepopotential $\mathcal{G}\left( w\right) $.

\subsection{Kahler/HyperKahler correspondence}

\qquad We start by recalling that hyperKahler manifolds form a particular
subset of complex $2n$ dimensional Ricci flat Kahler ones. To study special
hyperKahler \textit{d-geometry} of the moduli space of 10D type IIA
superstring on the K3 surface, we shall then use the matrix formulation
presented above and mimic the method made for the case of special Kahler 
\textit{d-geometry} of type IIA superstring on Calabi-Yau threefolds. Here
we first give a general correspondence between Kahler and hyperKahler
geometries. Then we make comments on their geometric interpretations, in
particular the issue regarding the realization of the central charges of the
4D/6D $\mathcal{N}=2$ superalgebras, the D2-brane wrapping 2-cycles and
corresponding potentials.

\subsubsection{Type IIA superstring on CY3}

\qquad In studying the moduli space of 10D type IIA superstring on
Calabi-Yau threefolds, we have few geometric objects that play a central
role. Some of these basic quantities and their corresponding physical
interpretations are collected in the following table:

\begin{equation}
\begin{tabular}{lll}
$\left\{ 
\begin{array}{c}
\text{{\small Kahler 2- form}: }{\small \Omega }^{\left( 1,1\right) }=%
\overline{{\small \Omega }^{\left( 1,1\right) }} \\ 
\text{{\small Complexified}: }\boldsymbol{K}_{\pm }={\small B}^{NS}{\small %
\pm i\Omega }^{\left( 1,1\right) } \\ 
\text{{\small Volume form }}{\small \Omega }^{\left( 3,0\right) }\wedge 
{\small \Omega }^{\left( 0,3\right) }\text{{\small \ }}%
\end{array}%
\right. $ & $\leftrightarrow $ & $\left\{ 
\begin{array}{c}
\text{{\small real moduli:} }x=\int_{C_{2}}\Omega ^{\left( 1,1\right) } \\ 
\text{{\small complex: }}z=\int_{C_{2}}\boldsymbol{K}_{+} \\ 
\text{{\small Kahler potential:} }\mathcal{K}\left( z,\overline{z}\right)%
\end{array}%
\right. $%
\end{tabular}
\label{cy}
\end{equation}

$\ \ $\newline
Here $\boldsymbol{K}_{+}=B^{NS}+i\Omega ^{\left( 1,1\right) }$ is the
complexification of the Kahler 2- form $\Omega ^{\left( 1,1\right) }$ and $%
\boldsymbol{K}_{-}$ is its complex conjugate $\overline{\left( \boldsymbol{K}%
_{+}\right) }$. We also have the usual globally defined complex holomorphic
3- form of the CY3 $\Omega ^{\left( 3,0\right) }$ together with its
antiholomorphic partner $\Omega ^{\left( 0,3\right) }$. These quantities
play an important role in type IIB superstring on the mirror of the
Calabi-Yau threefold \cite{AMS}.

In the type IIA picture we are considering here, the Kahler deformations of
the CY3 metric are parameterized by complex numbers 
\begin{equation}
z^{I}=y^{I}+ix^{I}
\end{equation}%
which on the physical side describe the vevs of the scalar fields in the 4D $%
\mathcal{N}=2$ Maxwell gauge supermultiplets. The real numbers $x^{I}$ are
the area of the 2- cycle $C_{2}^{I}$ inside the CY3 and $y^{I}$ are the real
fluxes of the NS-NS B- field through the $C_{2}^{I}$. Notice that on the
moduli space of 10D type IIA superstring on CY3, the Kahler form $%
\boldsymbol{J}=\Omega ^{\left( 1,1\right) }$, the NS-NS B- field $%
\boldsymbol{B}^{NS}$ and the complexified Kahler form $\boldsymbol{K}_{+}$
can be respectively defined as follows: 
\begin{equation}
\boldsymbol{J}=\sum_{I=1}^{h_{\text{{\tiny CY3}}}^{\left( 1,1\right)
}}x^{I}J_{I},\qquad \boldsymbol{B}^{NS}=\sum_{I=1}^{h_{\text{{\tiny CY3}}%
}^{\left( 1,1\right) }}y^{I}J_{I},\qquad \boldsymbol{K}_{+}=\sum_{I=1}^{h_{%
\text{{\tiny CY3}}}^{\left( 1,1\right) }}z^{I}J_{I},
\end{equation}%
where $\left\{ J_{I}\right\} $ is a canonical basis of real 2-forms
normalized as 
\begin{equation}
\int_{C_{2}^{I}}J_{K}=\delta _{K}^{I},\qquad I,\text{ }K=1,...,h_{\text{%
{\tiny CY3}}}^{\left( 1,1\right) }.
\end{equation}%
Along with these objects, we also define three more interesting quantities: 
\newline
\textbf{(i)} the particular real 2-cycle $C_{2}$ given by the following
integral linear combination 
\begin{equation}
C_{2}=\sum_{I}q_{I}C_{2}^{I},  \label{c2}
\end{equation}%
with $q_{I}$ integers. A D2-brane wrapping $C_{2}$ splits in general as the
sum of $q_{I}$ components wrapping the basis $C_{2}^{I}$. The charge $q_{I}$
is then interpreted as the number of D2-branes wrapping $C_{2}^{I}$. \newline
\textbf{(ii)} the real geometric area of $C_{2}$ given by 
\begin{equation}
Z_{e}=\int_{C_{2}}\Omega ^{\left( 1,1\right) },
\end{equation}%
which, by using above relations, can be put in the explicit form 
\begin{equation}
Z_{e}=\sum_{I=1}^{h_{\text{{\small CY3}}}^{\left( 1,1\right) }}q_{I}x^{I}.
\label{ze}
\end{equation}%
Up to an overall factor of the tension, this $Z_{e}$ is just \textit{the mass%
} of a D2-brane wrapping $C_{2}$ eq(\ref{c2}) and is interpreted as the
electric central charge of the 4D $\mathcal{N}=2$ superalgebra.\newline
\textbf{(iii)} in type IIB superstring set up, the Kahler potential $%
\mathcal{K}\left( \mathrm{x,}\overline{\mathrm{x}}\right) $ reads in terms
of the holomorphic 3- forms as follows 
\begin{equation}
\mathcal{K}\left( \mathrm{x,}\overline{\mathrm{x}}\right) =i\int_{CY3}%
{\small \Omega }^{\left( 3,0\right) }\wedge {\small \Omega }^{\left(
0,3\right) }.
\end{equation}%
By using the usual symplectic basis $\left( \mathrm{A}^{\mathrm{\mu }},%
\mathrm{B}_{\mathrm{\mu }}\right) $ of real 3-cycles within the Calabi-Yau
threefold and the complex moduli 
\begin{equation}
\mathrm{x}^{\mathrm{\mu }}=\int_{\mathrm{A}^{\mathrm{\mu }}}{\small \Omega }%
^{\left( 3,0\right) },\qquad \mathrm{F}_{\mathrm{\mu }}=\int_{\mathrm{B}_{%
\mathrm{\mu }}}{\small \Omega }^{\left( 3,0\right) },
\end{equation}%
together with their complex conjugates $\overline{\left( \mathrm{x}^{\mathrm{%
\mu }}\right) }$ and $\overline{\left( \mathrm{F}_{\mathrm{\mu }}\right) }$,
we can put $\mathcal{K}\left( \mathrm{x,}\overline{\mathrm{x}}\right) $ in
the following explicit form%
\begin{equation}
\mathcal{K}\left( \mathrm{x,}\overline{\mathrm{x}}\right) =i\sum_{\mathrm{%
\mu }=1}^{h_{\text{{\tiny CY3}}}^{2,1}}\left( \mathrm{x}^{\mathrm{\mu }}%
\overline{\left( \mathrm{F}_{\mathrm{\mu }}\right) }-\overline{\left( 
\mathrm{x}^{\mathrm{\mu }}\right) }\mathrm{F}_{\mathrm{\mu }}\right) .
\end{equation}%
In type IIA superstring on CY3, the Kahler potential $\mathcal{K}\left( 
\mathrm{z,}\overline{\mathrm{z}}\right) $ reads in terms of the Kahler form 
\begin{equation}
\mathcal{K}\left( \mathrm{z,}\overline{\mathrm{z}}\right) =\int_{\mathrm{CY3}%
}\Omega ^{\left( 1,1\right) }\wedge \Omega ^{\left( 1,1\right) }\wedge
\Omega ^{\left( 1,1\right) },
\end{equation}%
where now the $\mathrm{z}$ 's stand for the 2-cycles area. We also have $%
2i\Omega ^{\left( 1,1\right) }=\left( \boldsymbol{K}_{+}-\boldsymbol{K}%
_{-}\right) $.

\subsubsection{Type IIA superstring on K3}

\qquad The relations given above have quite similar partners for the case
the moduli space of 10D type IIA superstring on the K3 surface. For the
analog of the table eq(\ref{cy}), we have

\begin{equation}
\begin{tabular}{lll}
$\left\{ 
\begin{array}{c}
\text{{\small HyperKahler- form}:}\overline{J^{\left( ij\right) }}{\small %
=\epsilon }_{ik}{\small \epsilon }_{jl}{\small J}^{\left( kl\right) } \\ 
\text{{\small Quaternionified }:}{\small J=B}^{NS}{\small +i\sigma J} \\ 
\text{{\small Volume form: }}{\small \sigma J}\wedge {\small \sigma J}\text{%
{\small \ }}%
\end{array}%
\right. $ & $\quad \leftrightarrow \quad $ & $\left\{ 
\begin{array}{c}
\text{{\small real moduli:} }{\small x\sigma =}\int_{C^{2}}{\small \sigma J}
\\ 
\text{{\small Quaternion}: }{\small w=}\int_{C^{2}}{\small J} \\ 
\text{{\small Hyperpotential} }\mathcal{H}\left( w,\overline{w}\right)%
\end{array}%
\right. $%
\end{tabular}%
,  \label{cx}
\end{equation}

\ \ \newline
where we have used the $2\times 2$ matrix representation 
\begin{equation}
\mathbf{\sigma .J}=\sigma ^{1}J^{1}+\sigma ^{2}J^{2}+\sigma ^{3}J^{3},
\label{si}
\end{equation}%
to represent the real isotriplet 2-form $J^{\left( ij\right) }$ and where $%
B^{NS}+i\mathbf{\sigma .J}$ is the quaternionified hyperKahler 2-form.
Notice that $B^{NS}+i\mathbf{\sigma .J}$ can in general be written as a spin 
$\left( \frac{1}{2},\frac{1}{2}\right) $ \ representation $J^{ia}$ of the
group $SU\left( 2\right) \times SU^{\prime }\left( 2\right) $. Notice also
that on the moduli space (\ref{ms}) of 10D type IIA superstring on K3, the
hyperKahler form $\mathbf{\sigma .J}$, the NS-NS B-field $\boldsymbol{B}%
^{NS} $ and the quaternionified hyperKahler form $\mathcal{J}$ can be
defined as follows: 
\begin{eqnarray}
\mathbf{\sigma .J} &=&\sum_{I=1}^{h_{\text{{\tiny K3}}}^{\left( 1,1\right)
}}J_{I}\mathbf{x}^{I}.\mathbf{\sigma },\qquad  \notag \\
\boldsymbol{B}^{NS} &=&\sum_{I=1}^{h_{\text{{\tiny CY3}}}^{\left( 1,1\right)
}}y^{I}J_{I},\qquad \\
\mathcal{J}_{+} &=&\sum_{I=1}^{h_{\text{{\tiny CY3}}}^{\left( 1,1\right)
}}\left( y^{I}+i\mathbf{\sigma .x}^{I}\right) J_{I},  \notag
\end{eqnarray}%
where $\left\{ J_{I}\right\} $ is a canonical basis of real 2-forms
normalized as: 
\begin{equation}
\int_{C_{2}^{I}}J_{K}=\delta _{K}^{I},\qquad I,\text{ }K=1,\ldots ,h_{\text{%
{\tiny K3}}}^{\left( 1,1\right) }.
\end{equation}%
Like for eqs(\ref{c2}-\ref{ze}), we define the integral 2-cycle $C_{2}$ of
the complex surface K3, 
\begin{equation}
C_{2}=\sum_{I}q_{I}C_{2}^{I},
\end{equation}%
where $q_{I}$ is the number of a D2-brane wrapping $C_{2}^{I}$. This special
cycle allows to compute the central charge isotriplet $Z_{e}^{\left(
ij\right) }$. Indeed, notice first that like the $J^{\left( ij\right) }$
hyperKahler 2-form, we can use the Pauli matrices to define $Z^{\left(
ij\right) }$ isotriplet as a hermitian traceless $2\times 2$ matrix $\mathbf{%
\sigma .Z}$ eq(\ref{si}). So we have 
\begin{equation}
Z_{e}^{\left( ij\right) }=\int_{C_{2}}J^{\left( ij\right) }\mathbf{,\qquad
\Leftrightarrow \qquad \sigma .Z}_{e}=\int_{C_{2}}\mathbf{\sigma J,}
\label{cch}
\end{equation}%
which up on integration leads to 
\begin{equation}
\sum_{m=1}^{3}Z_{e}^{m}\mathbf{\sigma }^{m}=\sum_{m=1}^{3}\left(
\sum_{I=1}^{h_{\text{{\tiny K3}}}^{\left( 1,1\right) }}q_{I}x^{mI}\right) 
\mathbf{\sigma }^{m},  \label{ch}
\end{equation}%
from which one reads the explicit expression of $\mathbf{Z}_{e}^{m}$ namely%
\begin{equation}
Z_{e}^{m}=\sum_{I=1}^{h_{\text{{\tiny K3}}}^{\left( 1,1\right)
}}q_{I}x^{mI},\qquad \Leftrightarrow \qquad \mathbf{Z}_{e}=\sum_{I=1}^{h_{%
\text{{\tiny K3}}}^{\left( 1,1\right) }}q_{I}\mathbf{x}^{I}.
\end{equation}%
Notice that in presence of the \textit{NS-NS} B-field, one has extra
contribution associated with the isosinglet $Z^{0}$ as shown below: 
\begin{equation}
Z^{0}=\sum_{I=1}^{h_{\text{{\tiny K3}}}^{\left( 1,1\right) }}q_{I}y^{I}.
\label{chh}
\end{equation}%
The isosinglet $Z_{0}$ and the isotriplet $Z^{\left( ij\right) }$ can be
combined into the real four dimensional quantity 
\begin{equation}
\mathcal{Z}=\int_{C_{2}}\mathcal{J}.
\end{equation}%
Using the moduli 
\begin{equation}
\mathrm{w}^{ia\mathrm{I}}=\int_{\mathrm{C}_{2}^{\mathrm{I}}}J^{ia},\qquad 
\mathrm{G}_{\mathrm{I}}^{ia}=\int_{\mathrm{K3}}J^{ia}\wedge J_{I},\qquad i,%
\text{ }a=1,2,
\end{equation}%
where the $\mathrm{C}_{2}^{\mathrm{I}}$'s is a real basis of real 2-cycles
of the K3 surface, one can write down the basic relations for the special
hyperKahler geometry. \newline
Because of the $SU\left( 2\right) \times SU^{\prime }\left( 2\right) $
tensor structure, the generalized hyperKahler 2- form is captured by the
real 2-form $J^{ia}$ and so the general object we can write down is given by 
\begin{equation}
\mathcal{H}^{ijab}\left( \mathrm{w,G}\right) =\int_{K3}\left( J^{ia}\wedge
J^{jb}\right)
\end{equation}%
or equivalently%
\begin{equation}
\mathcal{H}^{ijab}\left( \mathrm{w,G}\right) =\sum_{I=1}^{h_{\text{{\tiny K3}%
}}^{\left( 1,1\right) }}\left( \mathrm{w}^{ia\mathrm{I}}\mathrm{G}_{\mathrm{I%
}}^{jb}+\mathrm{w}^{jb\mathrm{I}}\mathrm{G}_{\mathrm{I}}^{ia}\right) .
\end{equation}%
Notice that $\mathcal{H}^{ijab}$ is a reducible tensor and can be decomposed
in four components as follows:%
\begin{equation}
\mathcal{H}^{ijab}=\epsilon ^{ij}\epsilon ^{ab}\mathcal{H}_{0}+\epsilon ^{ij}%
\mathcal{H}_{1}^{\left( ab\right) }+\epsilon ^{ab}\mathcal{H}_{2}^{\left(
ij\right) }+\mathcal{H}_{3}^{\left( ij\right) \left( ab\right) }  \label{h0}
\end{equation}%
with $\mathcal{H}_{0}=2\sum_{\mathrm{I}=1}^{h_{\text{{\tiny K3}}}^{1,1}}%
\mathrm{w}^{ia\mathrm{I}}\mathrm{G}_{ia\mathrm{I}}$ and,%
\begin{eqnarray}
\mathcal{H}_{1}^{\left( ab\right) } &=&\sum_{\mathrm{I}=1}^{h_{\text{{\tiny %
K3}}}^{1,1}}\left( \mathrm{w}^{ia\mathrm{I}}\mathrm{G}_{i\mathrm{I}}^{b}+%
\mathrm{w}^{ib\mathrm{I}}\mathrm{G}_{i\mathrm{I}}^{a}\right) ,  \notag
\label{h00} \\
\mathcal{H}_{2}^{\left( ij\right) } &=&\sum_{\mathrm{I}=1}^{h_{\text{{\tiny %
K3}}}^{1,1}}\left( \mathrm{w}^{ia\mathrm{I}}\mathrm{G}_{a\mathrm{I}}^{j}+%
\mathrm{w}^{ja\mathrm{I}}\mathrm{G}_{a\mathrm{I}}^{i}\right) , \\
\mathcal{H}_{3}^{\left( ij\right) \left( ab\right) } &=&\sum_{\mathrm{I}%
=1}^{h_{\text{{\tiny K3}}}^{1,1}}\left( \mathrm{w}^{ia\mathrm{I}}\mathrm{G}_{%
\mathrm{I}}^{jb}+\mathrm{w}^{ib\mathrm{I}}\mathrm{G}_{\mathrm{I}}^{ja}+%
\mathrm{w}^{ja\mathrm{I}}\mathrm{G}_{\mathrm{I}}^{ib}+\mathrm{w}^{jb\mathrm{I%
}}\mathrm{G}_{\mathrm{I}}^{ia}\right) ,  \notag
\end{eqnarray}%
By identifying the two SU$\left( 2\right) $ group factors, $\mathcal{H}%
^{ijab}$\ becomes $\mathcal{H}^{ijkl}$\ and can be reduced as follows%
\begin{equation}
\mathcal{H}^{ijkl}=\epsilon ^{i(j}\epsilon ^{kl)}\mathcal{H}_{0}+\left(
\epsilon ^{i(j}\mathcal{H}_{1}^{kl)}+\epsilon ^{i(j}\mathcal{H}%
_{1}^{lk)}\right) +\mathcal{H}_{3}^{\left( ijkl\right) }  \label{h2}
\end{equation}%
We will develop further these issues later by using the 2$\times $2 matrix
formulation; see section 5, all we need is to keep in mind the two
following: \newline
\textbf{(i)} the correspondence between Kahler and hyperKahler manifolds,
eqs(\ref{cy}-\ref{cx}), \newline
\textbf{(ii)} the role played by the leading $SU\left( 2\right) $
representations in specifying the hyperKaler geometry, eq(\ref{h2}).

We end this subsection by noting that in the case of moduli space of $10D$
type IIA superstring compactification on Calabi-Yau manifolds, the
connection between supersymmetry and complex geometry reads for the Higgs
and Coulomb branches of $4D$ and $6D$ supersymmetric field theory limits as
follows:

\begin{equation}
\begin{tabular}{|l|l|l|l|l|}
\hline
{\small Moduli space} & {\small 4D} $\mathcal{N}${\small =1} & {\small 4D} $%
\mathcal{N}${\small =2} & {\small 6D} $\mathcal{N}${\small =1} & {\small 6D} 
$\mathcal{N}$=$\left( {\small 1,1}\right) $ \\ \hline
{\small Higgs Branch} & {\small Kahler} & {\small hyperkahler} & {\small %
hyperkahler} & {\small -} \\ \hline
{\small Coulomb Branch} & {\small -} & {\small Kahler} & {\small -} & 
{\small hyperkahler} \\ \hline
\end{tabular}%
\end{equation}

\ \newline
Notice in passing that in present study we are interested in the hyperKahler
structure of $6D$ $\mathcal{N}=2$ theory. This is also given by the Higgs
branch of $6D$ $\mathcal{N}=1$ supersymmetric theory in the same spirit as
does the Kahler geometry in $4D$ $\mathcal{N}=2$ and $4D$ $\mathcal{N}=1$
theories.

\section{Quaternionic Geometry in Matrix Formulation}

\qquad We start by setting up the problem of \textit{special hyperkahler
d-geometry }for the moduli space of type IIA superstring on the K3 surface.
Then we consider the way to solve it by using the $2\times 2$ matrix
formulation.

\subsection{Moduli Space $\boldsymbol{M}_{\text{IIA/K3}}$}

\qquad We first give some general on the moduli space $\boldsymbol{M}_{\text{%
{\small IIA/K3}}}=\boldsymbol{M}$ of the type IIA superstring on the K3
surface with fixed string coupling constant \cite{W,V,A,A1,HT}.\textbf{\ }%
Then we describe $\boldsymbol{M}$ by using the quaternionic moduli.

\subsubsection{Spectrum 10D Type IIA on the K3 surface}

\qquad To begin recall that the spectrum of 10D Type IIA superstring has two
sectors: perturbative (fields) and non perturbative (D2$n$-branes, $%
n=0,1,2,3 $ sources of RR fields).

\paragraph{10D Supermultiplet \ \qquad\ \newline
}

In the perturbative massless sector, the bosonic fields of 10D type IIA
superstring spectrum are given by 
\begin{equation}
\phi _{dil}\text{, }\quad G_{\left( MN\right) },\text{ }\quad B_{\left[ MN%
\right] },\quad \mathcal{A}_{M},\quad \mathcal{C}_{MNK},
\end{equation}%
where $M,N,K=0,\cdots ,9$ are vector indices of $SO\left( 1,9\right) $.
Altogether have a total number of $128$ on shell degrees of freedom. Their
fermionic partners involve two 10D-gravitinos and two 10D-gauginos. These
are 10D Majorana-Weyl spinors. For simplicity, we shall deal with the
bosonic sector only.

The compactification of type IIA superstring on the K3 surface is obtained
by breaking space-time symmetry $SO\left( 1,9\right) $ down to the subgroup $%
SO\left( 1,5\right) \times SU\left( 2\right) $ which is contained in $%
SO\left( 1,5\right) \times SO\left( 4\right) $. Degrees of freedom of type
IIA string on the K3 surface are determined by retaining only half of the
original ones since K3 preserves only half of the 32 original
supersymmetries.

\paragraph{6D Supermultiplets \ \qquad\ \newline
}

In six dimensions, the spectrum of bosonic fields with non zero spin is
formally given by,%
\begin{equation}
G_{\mu \nu },\text{ }B_{\mu \nu },\text{ }\mathcal{A}_{\mu },\text{ }%
\mathcal{C}_{\mu \nu \rho },\text{ }\mathcal{C}_{\mu mn},\qquad 0\leq \mu
,\nu ,\rho \leq 5,
\end{equation}%
where $G_{\mu \nu }$ is the 6D metric, $B_{\mu \nu }$ and $\mathcal{C}_{\mu
\nu \rho }$ are the 6D antisymmetric gauge fields. We also have the 6D
gravi-photon $\mathcal{A}_{\mu }$ and the Maxwell gauge fields $\mathcal{C}%
_{\mu mn}$ following from the compactification of the RR 3-form on the real
2- cycles of K3 surface. These real 6D 1-forms have two indices on the K3
surface and should be thought of as 
\begin{equation}
\mathcal{C}_{\mu mn}\equiv \mathcal{C}_{\mu }^{I}\qquad I=1,\cdots ,22.
\end{equation}%
For the spectrum of the $6D$ scalars, we have in addition to the $6D$
dilaton $\phi _{dil}$, the following 
\begin{equation}
\phi _{dil},\text{ }B_{mn},\text{ }G_{mn}^{\text{{\scriptsize kahler}}},%
\text{ }G_{mn}^{\text{{\scriptsize complex}}},\text{ }\overline{G}_{mn}^{%
\text{{\scriptsize complex}}}.
\end{equation}%
They describe the scalars $B_{mn}$ resulting from the compactification of
the NS-NS 2-form on the real 2-cycles of the K3 surface. These $B_{mn}$'s
and the field moduli $G_{mn}=\left\{ G_{mn}^{\text{{\scriptsize kahler}}},%
\text{ }G_{mn}^{\text{{\scriptsize complex}}}\right\} $ following from the
compactification of metric field should be understood as follows:%
\begin{eqnarray}
G_{mn}^{\text{{\scriptsize kahler}}} &\equiv &G_{\text{{\scriptsize kahler}}%
}^{I},\text{ }\qquad I=1,\cdots ,20,  \notag \\
G_{mn}^{\text{{\scriptsize complex}}} &\equiv &G_{\text{{\scriptsize complex}%
}}^{J},\qquad J=1,\cdots ,19, \\
B_{mn} &\equiv &B^{K},\qquad \qquad K=1,\cdots ,22.  \notag
\end{eqnarray}%
Notice that the field moduli $G_{\text{{\small kahler}}}^{I}$ and the
complex $G_{\text{{\small complex}}}^{J}$ stand for Kahler and complex
deformations of \ K3 metric with SU(2) holonomy group. \newline
On the other hand, viewed from $6D$ $\mathcal{N}=2$ supergravity low energy
limit, these fields combine as follows: \newline
(\textbf{i}) \textbf{Gravity}:$\qquad \ \ $\newline
$\left( 32+32\right) $ on shell degrees of freedom for the $6D$ $\mathcal{N}%
=2$ gravity supermultiplet whose bosonic sector is as follows:%
\begin{equation}
G_{\mu \nu },\text{ }B_{\mu \nu },\text{ }\mathcal{A}_{\mu }^{\left(
ij\right) },\text{ }\mathcal{C}_{\mu \nu \rho },\text{ }\phi _{dil},
\end{equation}%
involving, besides the 3-form $H=dB$, the gauge field strengths: 
\begin{eqnarray}
\mathcal{F}_{2}^{\left( ij\right) } &=&d\mathcal{A}_{1}^{\left( ij\right)
},\qquad \mathcal{F}_{4}^{\left( ij\right) }=\text{ }^{\ast }\mathcal{F}%
_{2}^{\left( ij\right) }  \notag \\
\mathcal{F}_{4}^{0} &=&d\mathcal{C}_{3},\qquad \mathcal{F}_{2}^{0}=\text{ }%
^{\ast }\mathcal{F}_{4}^{0}.
\end{eqnarray}%
Notice that $\mathcal{A}_{\mu }^{\left( ij\right) }$, $\mathcal{F}%
_{2}^{\left( ij\right) }$ and $^{\ast }\mathcal{F}_{2}^{\left( ij\right) }$
are isotriplets and all remaining others singlets. \newline
\textbf{(ii) Coulomb:\qquad\ }\newline
$20\times \left( 8+8\right) $ on shell degrees of freedom for the six
dimensional $\mathcal{N}=2$ supersymmetric Maxwell multiplets with bosonic
sector%
\begin{equation}
\mathcal{C}_{\mu }^{I},\text{ \ }w^{\pm I},\qquad I=1,\cdots ,20.
\end{equation}%
Setting $\mathcal{C}_{1}^{I}=dx^{\mu }\mathcal{C}_{\mu }^{I}$, the
corresponding gauge field strengths and their 6D duals are as follows:%
\begin{equation}
\mathcal{F}_{2}^{I}=d\mathcal{C}_{1}^{I},\qquad \mathcal{F}_{4}^{I}=\text{ }%
^{\ast }\mathcal{F}_{2}^{I},\qquad I=1,\cdots ,20.
\end{equation}%
These $6d$ $\mathcal{N}=2$ supermultiplets are $SU\left( 2\right) $ singlets
but transforms as a vector under $SO\left( 20\right) $. They can be nicely
described in the $6d$ $\mathcal{N}=1$ superspace formalism by still
maintaining $SU\left( 2\right) $ isometry. Below, we give some details.

\subsubsection{Supersymmetry in 6D}

\qquad In six dimensional space-time, one distinguishes two supersymmetric
algebras: (i) the 6D $\mathcal{N}=\left( 2,0\right) $ chiral superalgebra
and (ii) the 6D $\mathcal{N}=\left( 1,1\right) $ (6D $\mathcal{N}=2$) non
chiral one given by eq(\ref{n11}). More details on the graded commutation
relations of these superalgebras including central extensions are exhibited
in the appendix. What we need here is the physical representations of the 6D 
$\mathcal{N}=2$ superalgebra and the field theoretical way to deal with them.

In the language of $6D$ $\mathcal{N}=1$ supersymmetric representations, $6D$ 
$\mathcal{N}=2$ supermultiplets $\left( R_{6D}^{\mathcal{N}=2}\right) $
split into pairs of $6D$ $\mathcal{N}=1$ representations \textrm{\cite{S}}
as given below, 
\begin{equation}
R_{6D}^{\mathcal{N}=2}=R_{6D}^{\mathcal{N}=1}\oplus R_{6D}^{\prime \mathcal{N%
}=1}.
\end{equation}%
To fix the idea, we consider here after the $6D$ $\mathcal{N}=2$ gauge
multiplet $V_{6D}^{\mathcal{N}=2}$ which, as given above, consists of the
following on shell degrees of freedom,%
\begin{equation}
V_{6D}^{\mathcal{N}=2}=\left( 1,\frac{1}{2}^{2},0^{4}\right) ,
\end{equation}%
where $1,$ $\frac{1}{2}$ and $0$ stand for the spin of the component field
content and the powers for their numbers. The existence of four scalar gauge
fields (\textit{sgauginos}) within the $6D$ $\mathcal{N}=2$ reflects in some
sense the four central charges $Z^{ij}$ of the underlying superalgebra eq(%
\ref{n11}). In the $6D$ $\mathcal{N}=1$ formalism, the $V_{6D}^{\mathcal{N}%
=2}$ vector multiplet splits into a vector $V_{6D}^{\mathcal{N}=1}$ and a
hypermultiplet $H_{6D}^{\mathcal{N}=1}$ as follows%
\begin{equation}
V_{6D}^{\mathcal{N}=2}=V_{6D}^{\mathcal{N}=1}\oplus H_{6D}^{\mathcal{N}=1},
\end{equation}%
where%
\begin{equation}
V_{6D}^{\mathcal{N}=1}=\left( 1,\frac{1}{2}\right) ,\qquad H_{6D}^{\mathcal{N%
}=1}=\left( \frac{1}{2},0^{4}\right) .
\end{equation}%
Notice that $V_{6D}^{\mathcal{N}=1}$ has no scalar in six dimensions while $%
H_{6D}^{\mathcal{N}=1}$ has four scalars capturing the hyperkahler structure
of the Coulomb branch of $6D$ $\mathcal{N}=2$ supersymmetry. Notice also
that in the $4D$ $\mathcal{N}=2$ language, these multiplets reduce generally
to 
\begin{equation}
V_{4D}^{\mathcal{N}=2}=\left( 1,\frac{1}{2}^{2},0^{2}\right) ,\qquad H_{4D}^{%
\mathcal{N}=2}=\left( \frac{1}{2}^{2},0^{4}\right) .
\end{equation}%
Notice moreover that in the case of toroidal compactification from 6D down
to 4D, $6D$ $\mathcal{N}=2$ supersymmetry leads to $4D$ $\mathcal{N}=4$
supersymmetry. The gauge multiplet $V_{6D}^{\mathcal{N}=2}$ gives $V_{4D}^{%
\mathcal{N}=4}$ with on shell degrees of freedom, 
\begin{equation}
V_{4D}^{\mathcal{N}=4}=\left( 1,\frac{1}{2}^{4},0^{6}\right) ,
\end{equation}%
corresponding precisely to the combination of $V_{4D}^{\mathcal{N}=2}$ and $%
H_{4D}^{\mathcal{N}=2}$ multiplets. The six scalars involved in $V_{4D}^{%
\mathcal{N}=4}$ are exactly the six dimensional vector moduli that we
encounter in the study of the moduli space 
\begin{equation}
\boldsymbol{M}_{het/T^{6}}=\frac{SO\left( 6,22\right) }{SO\left( 6\right)
\times SO\left( 22\right) },
\end{equation}%
of the toroidal compactification of 10D heterotic superstring on T$^{6}$.

The splitting of $6D$ $\mathcal{N}=2$ representations into pairs of $6D$ $%
\mathcal{N}=1$ ones has a remarkable parallel with the reduction $4D$ $%
\mathcal{N}=2$ into $4D$ $\mathcal{N}=1$ representations. This is a crucial
technical point that can be used for the study of the Coulomb branch $4D$ $%
\mathcal{N}=2$ theories and black holes in type II superstrings on CY3. This
parallel can be learnt on the following correspondence,

\begin{equation*}
\begin{tabular}{|l|l|l|}
\hline
vector multiplet in {\small 4D} $\mathcal{N}${\small =2} & $\rightarrow $ & 
chiral multiplet in {\small 4D} $\mathcal{N}${\small =1} \\ \hline
vector multiplet in {\small 6D} $\mathcal{N}$=$\left( {\small 1,1}\right) $
& $\rightarrow $ & hypermultiplet in {\small 6D} $\mathcal{N}$=${\small 1}$
\\ \hline
\end{tabular}%
\end{equation*}

\ \newline
It is interesting to note that one can still exhibit manifestly the $%
SU\left( 2\right) $ symmetry within $6D$ $\mathcal{N}=1$ supersymmetry. This
property follows from the use of by the supercharges $Q_{\alpha }^{i}$ eq(%
\ref{mw}) as the generators of the $6D$ $\mathcal{N}=1$ algebra whose
anticommutation relations can be directly read from eq(\ref{n11}) as shown
below,%
\begin{eqnarray}
\left\{ Q_{\alpha }^{i},Q_{\beta }^{j}\right\} &=&\epsilon ^{ij}P_{\left[
\alpha \beta \right] },\qquad \alpha ,\beta =1,...,4  \notag \\
\left[ P_{\left[ \alpha \beta \right] },Q_{\gamma }^{i}\right] &=&0,\qquad
i,j=1,2.
\end{eqnarray}%
Here $P_{\left[ \alpha \beta \right] }=\sum_{\mu =0}^{5}\left( \Gamma ^{\mu
}\right) _{\alpha \beta }P_{\mu }$ is the 6D- energy momentum vector and the 
$\Gamma ^{\mu }$'s are the 6D gamma matrices.

One can also develop a $6D$ $\mathcal{N}=1$ superspace formalism to describe
supersymmetric multiplets in terms of superfields $\Phi =\Phi \left( x^{\mu
},\theta ^{\alpha i}\right) $. For instance the $6D$ $\mathcal{N}=1$ Maxwell
superfield in the Wess- Zumino gauge is defined, using 6D Grassmann
variables $\theta ^{\alpha i}$, by a real isotriplet superfield $V^{\left(
ij\right) }$ with $\theta $- expansion given, in the Wess-Zumino gauge, by 
\begin{equation}
V^{\left( ij\right) }=\theta ^{\alpha (i}\theta ^{j)\beta }A_{\left[ \alpha
\beta \right] }+\theta ^{\alpha (i}\lambda _{\alpha }^{j)}+\left( \theta
^{4}D\right) ^{\left( ij\right) },
\end{equation}%
where $\left( \theta ^{4}D\right) ^{\left( ij\right) }$ stands for $\epsilon
_{\alpha \beta \gamma \delta }\theta ^{\alpha (i}\theta ^{j\beta }\theta
^{\gamma k}\theta ^{l)\delta }D_{\left( kl\right) }$. Here $A_{\left[ \alpha
\beta \right] }\sim A_{\mu }$ is the 6D gauge field, $\lambda _{\alpha }^{i}$%
\ its gaugino partners and $D^{\left( kl\right) }$ an isotriplet of
auxiliary fields. Quite similar quantities can be written down for the
hypermultiplets which are also described by superfields in $SU\left(
2\right) $ representations; for details see \textrm{\cite{HST,HS,S}}. Note
finally that a more convenient way to deal with $6D$ $\mathcal{N}=1$ vector
and hypermultiplet representations ($6D$ $\mathcal{N}=2$ vector multiplet)
is to use $4D$ $\mathcal{N}=2$ superspace obtained by decomposing 6D vectors
in 4D vectors and 2 scalars and 6D spinors $\theta ^{\alpha i}$ into a 4D
Weyl spinor $\theta ^{ai}$ and its complex conjugate $\overline{\theta }_{%
\dot{a}i}$,%
\begin{equation}
\left( \theta ^{\alpha i}\right) _{1\leq \alpha \leq 4}\qquad \rightarrow
\qquad \left( \theta ^{ai},\overline{\theta }_{\dot{a}i}\right) _{a=1,2}.
\end{equation}%
It is an interesting task to study this reduction in the framework of the
harmonic superspace formalism \textrm{\cite{HS}} where hyperkahler geometry
is nicely described in terms of $4D$ $\mathcal{N}=2$ hypermultiplets
couplings.

\subsection{Quaternionic Moduli in 6D $\mathcal{N}=2$ Gauge Theory}

\qquad From the set up of 10D type IIA superstring on a generic K3 surface,
the real \emph{80} degrees of freedom of $\boldsymbol{M}$ (\ref{ms}) are
arranged as follows: \newline
(\textbf{i}) Twenty (\emph{20}) Kahler deformations to be denoted as $x^{0I}$%
. They can be expressed in terms of the Kahler structure of the K3 surface
captured by the Kahler 2-form $\Omega ^{\left( 1,1\right) }=J^{0}$, as
follows,%
\begin{equation}
x^{0I}=\int_{C_{2}^{I}}J^{0},\qquad I=1,...,20.  \label{x0}
\end{equation}%
Here $C_{2}^{I}\in H_{2}\left( K3,R\right) $ is a real basis of 2-cycles of
\ the K3 surface. So $x^{0I}$ can be thought of as the real area of the 2-
cycle $C_{2}^{I}$. Notice that using the basis $\left\{ J_{I}^{0}\right\} $
of real 2-forms normalized as 
\begin{equation}
\delta _{J}{}^{I}=\int_{C_{2}^{I}}J_{J}^{0},
\end{equation}%
one can invert the previous relation as follows 
\begin{equation}
J^{0}=\sum x^{0I}J_{I}^{0}.
\end{equation}%
Note in passing that the \emph{20} Kahler moduli should be split as $\emph{%
20=1+19}$. From $SU\left( 2\right) $ representation theory, only one of
these degrees of freedom is an isosinglet. The remaining \emph{19} ones are
components belonging to \emph{19} isotriplets. The missing $2\times 19$
moduli come from the complex deformations to be considered in moment. Notice
also that one can express the real volume of the K3 surface as follows 
\begin{equation}
\mathcal{V}_{K3}=\int_{K3}J^{0}\wedge J^{0}=d_{IJ}x^{0I}x^{0J},\qquad
d_{IJ}=\int_{K3}J_{I}^{0}\wedge J_{J}^{0}  \label{vk3}
\end{equation}%
(\textbf{ii}) Nineteen (\emph{19}) complex deformations to be denoted as $%
x^{+I}$ and $x^{-I}$, that is thirty eight (\emph{38}) real moduli, 
\begin{equation}
x^{+I}=x^{1I}+ix^{2I},\qquad x^{-I}=x^{1I}-ix^{2I},\qquad I=1,...,19.
\end{equation}%
They can be expressed in terms of deformations of the holomorphic and
antiholomorphic 2- forms $J^{+}=\Omega ^{\left( 2,0\right) }$ and $%
J^{-}=\Omega ^{\left( 0,2\right) }$ as follows: 
\begin{equation}
x^{+I}=\int_{C_{2}^{I}}J^{+},\qquad x^{-I}=\int_{C_{2}^{I}}J^{-},\qquad
I=1,...,19.  \label{x-}
\end{equation}%
Using the 2-form basis $\left\{ J_{I}\right\} $, these relations can be also
inverted as 
\begin{equation}
J^{+}=\sum x^{+I}J_{I},\qquad J^{-}=\sum x^{-I}J_{I}.\qquad
\end{equation}%
The $x^{\pm I}$ moduli may be interpreted as holomorphic volumes of the
2-cycles $C_{2}^{I}$. Indeed computing the volume of the K3 surface but now
using the relation 
\begin{equation}
\mathcal{V}_{K3}=\int_{K3}J^{+}\wedge
J^{-}=2\sum_{I,J=1}^{2}d_{IJ}x^{+I}x^{-J}.
\end{equation}%
Then equating it with (\ref{vk3}), we learn that the volume of $\mathcal{V}%
_{K3}$ can be rewritten as, 
\begin{equation}
\mathcal{V}_{K3}=d_{IJ}\left( \frac{1}{2}x^{+I}x^{-J}+\frac{1}{2}%
x^{-I}x^{+J}+x^{0I}x^{0J}\right)
\end{equation}%
showing that $\mathcal{V}_{K3}$ is an $su\left( 2\right) $ invariant and can
be put in the form $\mathcal{V}_{K3}=d_{IJ}\left( \mathbf{x}^{I}\mathbf{%
\cdot x}^{J}\right) $. \newline
\textbf{(iii)} $22=\left( 20+2\right) $ real moduli coming from the values
of the B$_{NS}$ field on 2-cycles of the K3 surface. It is given by the
Betti number $b_{2}\left( K3\right) =h^{\left( 2,0\right) }+h^{\left(
0,2\right) }+h^{\left( 1,1\right) }$ which is equal to $1+1+\left(
4+16\right) $. The first $\left( 4+16\right) =20$ moduli have same nature as
Kahler deformations and are written as%
\begin{equation}
y^{0I}=\int_{C_{2}^{I}}B_{NS}^{0},\qquad I=1,...,20.
\end{equation}%
The two extra ones should viewed as the 20-th complex moduli, that is 
\begin{equation}
y_{21}^{0}+iy_{22}^{0}=x^{+20},\qquad y_{21}^{0}-iy_{22}^{0}=x^{-20}.
\label{tr}
\end{equation}%
Notice that nineteen of the $y^{0I}$, say $I=1,...,19$, are isosinglets, and
the remaining three namely $\left( y_{20}^{0},y_{21}^{0},y_{22}^{0}\right) $
can be combined as in eq(\ref{tr}) to form altogether an isotriplet; see
also previous discussion concerning the Kahler form. This is a crucial point
which we will encounter when we consider the uplifting to 7D.

To exhibit the hyperkahler structure of the moduli space of 10D type IIA
superstring on the K3 surface and keeping in mind the above discussion, we
can use the $SU\left( 2\right) $ spin $\frac{1}{2}\times \frac{1}{2}$
representations to split the real \emph{80} moduli as \emph{20} isosinglets $%
\left\{ y^{0I}\right\} $ plus \emph{20} isotriplets $\left\{
x^{-I},x^{0I},x^{+I}\right\} $: 
\begin{equation}
80=20\times 1+20\times 3.
\end{equation}%
As such the real \emph{80} moduli can be grouped into \emph{20} quaternions $%
w^{+I}$ as shown below 
\begin{equation}
w_{I}^{+}=y_{I}^{0}+i\mathbf{x}_{I}\cdot \mathbf{\sigma },\qquad I=1,...,20,
\label{yx}
\end{equation}%
together with there adjoint conjugates $w^{-I}=\left( w^{+I}\right) ^{+}$.
Here $y_{I}^{0}$ should be thought of as $y_{I}^{0}$ times the identity
matrix $I_{2\times 2}$ and $\mathbf{x}_{I}.\mathbf{\sigma }$ as%
\begin{equation}
\mathbf{x}_{I}.\mathbf{\sigma }=x_{I}^{-}\sigma ^{+}+x_{I}^{0}\sigma
^{0}+x_{I}^{+}\sigma ^{-},
\end{equation}%
with $\sigma ^{0,\pm }$ being the usual $2\times 2$ Pauli matrices
satisfying both $SU\left( 2\right) $ commutation relations and the 2D
Clifford algebra anticommutations. In particular we have:%
\begin{eqnarray}
\left( i\sigma ^{1}\right) ^{2} &=&\left( i\sigma ^{2}\right) ^{2}=\left(
i\sigma ^{3}\right) ^{2}=-1,  \notag \\
\sigma ^{\pm } &=&\sigma ^{1}\pm i\sigma ^{2},\qquad \sigma ^{0}=\sigma ^{3}.
\end{eqnarray}%
For later use, we define the adjoint conjugate\footnote{%
Notice that quaternions have three complex structures $\mathbf{i}$, $\mathbf{%
j}$ and $\mathbf{k}$ related as $\mathbf{i}\wedge \mathbf{j}=\mathbf{k}$. As
such, one should have three kinds of complex conjugations, say one for $%
\mathbf{i}$ ($\overline{\mathbf{i}}=-\mathbf{i}$), an other for $\mathbf{j}$
($\mathbf{j}^{\ast }\mathbf{=-j}$) and the combined one for $\mathbf{k}$ ($%
\overline{\mathbf{k}}=-\mathbf{k}$, $\mathbf{k}^{\ast }\mathbf{=-k}$, $%
\overline{\mathbf{k}}^{\ast }\mathbf{=k}$). An aspect of this feature will
be considered when we introduce harmonic space.} of the $w_{I}$ quaternions
as $w_{I}^{+}=y_{I}^{0}-i\mathbf{x}_{I}.\mathbf{\sigma }$. So the "real" and
"imaginary" parts (see also footnote) of the $w_{I}^{\pm }$ quaternions as
given by,%
\begin{eqnarray}
y_{I}^{0} &=&\frac{w_{I}+w_{I}^{+}}{2},  \notag \\
\mathbf{x}_{I}.\mathbf{\sigma } &=&\frac{w_{I}-w_{I}^{+}}{2i}.
\end{eqnarray}%
With these tools, it is not difficult to see that the moduli (\ref{x0}-\ref%
{x-}) can be rewritten as 
\begin{equation}
\mathbf{x}^{I}.\mathbf{\sigma =}\int_{C_{2}^{I}}\mathbf{\sigma .J,\qquad
\sigma .J=}\sum_{m=1}^{3}\sigma ^{-m}J^{m}.
\end{equation}%
Introducing the following "quaternionified" 2-form, in analogy with the
usual complexified 2- form in the Kahler geometry 
\begin{equation}
\mathcal{J}_{+}=B_{NS}+i\mathbf{\sigma .J,\qquad }\mathcal{J}_{-}=B_{NS}-i%
\mathbf{\sigma .J,\qquad }\mathcal{J}_{-}=\left( \mathcal{J}_{+}\right)
^{\dagger }  \label{q2f}
\end{equation}%
we see that the $\left\{ y^{0I},x^{-I},x^{0I},x^{+I}\right\} $ moduli can be
grouped altogether in quaternions as shown below 
\begin{equation}
w^{I}=\int_{C_{2}^{I}}\mathcal{J}_{+},\qquad \overline{w}^{I}=%
\int_{C_{2}^{I}}\mathcal{J}_{-},
\end{equation}%
where $w^{I}$ stand for $w^{+I}$ and $\overline{w}^{I}=w^{-I}$\ for its
adjoint conjugate. For later use, it is intersesting to introduce the basis
of quaternionic 2-forms $\left\{ \mathcal{J}_{I}\right\} $ on the $C_{2}^{I}$%
- cycles. With these objects, the above relations may be rewritten as%
\begin{equation}
\mathcal{J}_{+}=\sum w^{+I}\mathcal{J}_{I},\qquad \mathcal{J}_{-}=\sum w^{-I}%
\mathcal{J}_{I},  \label{dcj}
\end{equation}%
where we have used $\int_{C_{2}^{I}}\mathcal{J}_{J}=\delta _{J}^{I}$.

\subsection{7D Uplifting}

$\qquad $Large distance $7D$ $\mathcal{N}=2$ supergravity theory limit of
10D superstring and 11D M- theory compactifications can be obtained in
different, but equivalent, ways. In particular, they are obtained by the
three following routes: \newline
\textbf{(1)} compactification of eleven dimensional M-theory on the K3
surface. \newline
\textbf{(2)} compactification of 10D heterotic superstring on a real
3-torus. \newline
\textbf{(3)} uplifting $6D$ $\mathcal{N}=2$ theory to seven dimensions.

The last method is useful for studding $\mathcal{N}=2$ extremal black
objects in seven dimensions from six dimensional view. This way is in the
same spirit used in the uplifting of 4D $\mathcal{N}=2$ theory to five
dimensions. The 7D uplifting moduli space is obtained by putting appropriate
constraint eqs on the 6D one. These constraint eqs, that remain to be worked
out, have to reduce the real \emph{eighty one} dimension moduli space of
type IIA superstring on the K3 surface, namely 
\begin{equation}
\frac{SO\left( 4,20\right) }{SO\left( 4\right) \times SO\left( 20\right) }%
\times SO\left( 1,1\right) ,
\end{equation}%
down to 
\begin{equation}
\frac{SO\left( 3,19\right) }{SO\left( 3\right) \times SO\left( 19\right) }%
\times SO\left( 1,1\right) .
\end{equation}%
This reduction corresponds to fix $\left( 81-58\right) =23$ real moduli.

In our matrix formulation, the 7D uplifting corresponds to put adequate
constraint eqs on the quaternionic moduli $w^{\pm I}$. Using the analysis of 
\emph{subsection 4.2}, it is not difficult to see that the real \emph{23}
constraint equations are obtained by:\newline
\textbf{(i)} killing the real \emph{22} moduli coming from the NS-NS B-field
on the 2-cycles of K3.\newline
\textbf{(ii)} fixing the \emph{real volume} of the K3 surface.\newline
\textbf{(iii)} breaking the $SO\left( 4\right) \sim SU\left( 2\right) \times
SU^{\prime }\left( 2\right) $ down to $SO\left( 3\right) \sim SU\left(
2\right) $.\newline
This is achieved as follows: Since \emph{nineteen} of the real \emph{22}
moduli the NS-NS B-field come as isosinglets, see discussion after eq(\ref%
{tr}), they are killed by the conditions 
\begin{equation}
\mathrm{Tr}\left( w^{\pm I}\right) =0,\qquad I=1,...,19.
\end{equation}%
The remaining three moduli appear as a isotriplet and are then killed by the
vectorial condition 
\begin{equation}
\mathrm{Tr}_{{\tiny SU}\left( 2\right) }\left( \int_{K3}\left( \mathbf{%
\sigma }w^{+}w^{-}\right) \right) =0,  \label{c1}
\end{equation}%
which, up on integration, gives:%
\begin{equation}
\sum_{I,J=1}^{20}d_{IJ}\mathrm{Tr}\left( \sigma ^{i}w^{+I}w^{-I}\right) =0,%
\text{\qquad }i=1,2,3.
\end{equation}%
The \textit{23-rd} modulus is fixed by imposing a condition on the volume of
the K3 surface as shown below,%
\begin{equation}
\sum_{I,J=1}^{20}\mathrm{Tr}\left( w^{+I}d_{IJ}w^{-I}\right) =\mathcal{V}%
_{K3}=\text{ {\small constant}}{\small ,}  \label{c3}
\end{equation}%
where the constant can be taken as $\mathcal{V}_{K3}=1$. Notice that the
constraint relations (\ref{c1}-\ref{c3}) can be equivalently stated as
follows: 
\begin{equation}
\sum_{I,J}w^{+I}d_{IJ}w^{-I}=\frac{\mathcal{V}_{K3}}{2}I_{2},
\end{equation}%
that is as a hermitian $2\times 2$ matrix condition fixing four real degrees
of freedom.

\section{Computing the Potentials}

\qquad The special hyperkahler potential $\mathcal{H}$, whose $SU\left(
2\right) $ tensor structure has been obtained earlier eqs(\ref{h0}-\ref{h2}%
), and the prepotential $\mathcal{G}\left( w\right) $ will be obtained as
follows: \newline
\textbf{(1)} Using the matrix formulation developed above; in particular the
quaternionified 2- forms $\mathcal{J}_{\pm }$ and the corresponding
quaternionic moduli w$^{\pm I}$. \newline
\textbf{(2)} Mimicking the analysis for computing the Kahler potential of
the moduli space of the 10D type IIA superstring on Calabi-Yau threefolds.

\subsection{Special hyperkahler potential $\mathcal{H}$}

\qquad Extending the analysis made for the special Kahler geometry, one
discovers that the special hyperKahler potential $\mathcal{H}$ of the 10D\
type IIA superstring on the K3 surface can be given in term of the integral
on the volume 4-form 
\begin{equation}
\mathbf{J}\wedge \mathbf{J,}
\end{equation}%
on the K3 surface. A priori, $\mathcal{H}$ is not a simple real number
since, from SU$\left( 2\right) $ group theoretical view, the tensor product $%
\mathbf{J}\otimes \mathbf{J}$ has real nine dimensions which decomposes in
terms of SU$\left( 2\right) $ irreducible representations as 
\begin{equation}
J^{n}\otimes J^{m}=\delta ^{nm}\left( J^{r}J_{r}\right) \oplus
J^{[n}J^{m]}\oplus J^{(n}J^{m)},\qquad n,\text{ }m=1,2,3.
\end{equation}%
These factors are precisely the components given by eqs(\ref{h0}-\ref{h2}).
Moreover, notice that since the wedge product $\mathbf{J}\wedge \mathbf{J}$
is symmetric in the interchange of the 2- form isotriplets $\mathbf{J}$, the
antisymmetric part in the above decomposition should not contribute. As such 
$\mathbf{J}\wedge \mathbf{J}$ contains an isosinglet which we set as $V_{0}$
and a quintet which we represent by a traceless symmetric 3$\times $3 matrix 
$S$. With these objects, we can show that the hyperKahler potential reads as 
\begin{equation}
\mathcal{H}=Tr\left[ \ln \left( V_{0}-S\right) \right] .
\end{equation}%
Let us give details on the way of building $\mathcal{H}$.

Using Pauli matrices, the volume 4- form $\mathbf{J}\wedge \mathbf{J}$ can
be also written like $\mathbf{\sigma .J}\wedge \mathbf{\sigma .J}$.
Introducing the hermitian \emph{"volume"} matrix, 
\begin{equation}
\boldsymbol{V}=\int_{K3}\mathbf{\sigma .J}\wedge \mathbf{\sigma .J,}
\label{vo}
\end{equation}%
we can state the problem of the determination of the potential $\mathcal{H}$
as follows, 
\begin{equation}
\mathcal{H}=f\left( \boldsymbol{V}\right) ,
\end{equation}%
where $f$ is some function that remains to be specified. It should be
invariant under $SO\left( 4\right) \times SO\left( 20\right) $. Before
proceeding further, let us make three comments: \newline
\textbf{(1)} The hyperKahler 2-form $\mathbf{J}$ on the K3 surface is an
isotriplet. The hermitian quantity $\boldsymbol{V}$ is valued in the tensor
algebra of two SU$\left( 2\right) $ isotriplets which generally decompose as
follows 
\begin{equation}
3\times 3=1+3+5.  \label{fo}
\end{equation}%
These irreducible components correspond, in the language of the tensor
algebra of Pauli matrices, to 
\begin{equation}
\mathbf{\sigma \otimes \sigma =}\left( \mathbf{1}\right) \oplus \left( 
\mathbf{\sigma }\right) \oplus \left( \mathbf{\sigma \otimes \sigma }\right)
_{sym}
\end{equation}%
where $\left( \mathbf{\sigma \otimes \sigma }\right) _{sym}$stands for
traceless symmetric product. They are respectively associated with
isosinglet, isotriplet and isoquintet.\newline
\textbf{(2)} Using eq(\ref{q2f}), 
\begin{equation}
\mathbf{\sigma .J=}\frac{1}{2i}\left( \mathcal{J}_{+}-\mathcal{J}_{-}\right)
,
\end{equation}%
we can express $\boldsymbol{V}$ as 
\begin{equation}
\boldsymbol{V}=\frac{1}{4}\left( \boldsymbol{H}_{0}-\boldsymbol{H}_{++}-%
\boldsymbol{H}_{--}\right)
\end{equation}%
with%
\begin{eqnarray}
\boldsymbol{H}_{--} &=&\int_{K3}\left( \mathcal{J}_{-}\wedge \mathcal{J}%
_{-}\right) ,  \notag \\
\boldsymbol{H}_{0} &=&\int_{K3}\left( \mathcal{J}_{+}\wedge \mathcal{J}_{-}+%
\mathcal{J}_{-}\wedge \mathcal{J}_{+}\right) , \\
\boldsymbol{H}_{++} &=&\int_{K3}\left( \mathcal{J}_{+}\wedge \mathcal{J}%
_{+}\right) .  \notag
\end{eqnarray}%
Then using the expansions $\mathcal{J}_{\pm }=\sum w^{\pm I}J_{I}$ and
integrating over the 2- cycles of K3, we get on one hand 
\begin{eqnarray}
\boldsymbol{H}_{++} &=&\sum_{I,J}w^{+I}d_{IJ}w^{+J},  \notag \\
\boldsymbol{H}_{0} &=&\sum_{I,J}\left(
w^{+I}d_{IJ}w^{-J}+w^{-I}d_{IJ}w^{+J}\right)  \label{he} \\
\boldsymbol{H}_{--} &=&\sum_{I,J}w^{-I}d_{IJ}w^{-J},  \notag
\end{eqnarray}%
where $d_{IJ}=\int_{K3}\left( \mathcal{J}_{I}\wedge \mathcal{J}_{J}\right) $
stands for the intersection matrix numbers of the 2-cycles of the K3 surface
encountered earlier. It reads for a generic K3 surface as follows: 
\begin{equation}
d_{IJ}=\left( 
\begin{array}{cccc}
K\left( E_{8}\right) & 0_{8\times 8} & 0_{8\times 2} & 0_{2\times 2} \\ 
0_{8\times 8} & K\left( E_{8}\right) & 0_{8\times 2} & 0_{2\times 2} \\ 
0_{2\times 8} & 0_{2\times 8} & \Delta _{2} & 0_{2\times 2} \\ 
0_{2\times 2} & 0_{2\times 2} & 0_{2\times 2} & \Delta _{2}%
\end{array}%
\right) _{20\times 20}
\end{equation}%
where $K\left( E_{8}\right) $ is the Cartan matrix of ordinary $E_{8}$\ Lie
algebra and where%
\begin{equation}
\Delta _{2}=\left( 
\begin{array}{cc}
0 & 1 \\ 
1 & 0%
\end{array}%
\right) ,
\end{equation}%
generates the intersections of the $T^{2}$-cycles within $T^{4}$ in the
orbifold limit construction of K3 of the K3 surface.\newline
On the other hand, using the moduli $w^{\pm I}=\left( \int_{C_{2}^{I}}%
\mathcal{J}_{\pm }\right) $ we have:%
\begin{equation}
\boldsymbol{V}=\frac{1}{2}\sum_{I,J}\left( w^{+I}-w^{-I}\right) d_{IJ}\left(
w^{+J}-w^{-J}\right) .  \label{v}
\end{equation}%
This relation can be also rewritten as follows%
\begin{equation}
\boldsymbol{V}=\frac{1}{2}\sum_{I,J}\left( \sum_{n,m=1}^{3}\sigma _{n}\sigma
_{m}x^{nI}d_{IJ}x^{mJ}\right) =\sum_{n,m=1}^{3}\sigma _{n}\sigma _{m}%
\boldsymbol{V}^{nm}.
\end{equation}%
Notice that since $d_{IJ}$ is symmetric, the $SU\left( 2\right) $ tensor 
\begin{equation}
\boldsymbol{V}^{nm}=\frac{1}{2}\sum d_{IJ}x^{nI}x^{mJ},
\end{equation}%
which reads as well as 
\begin{equation}
\frac{1}{4}\sum d_{IJ}\left( x^{nI}x^{mJ}+x^{nJ}x^{mI}\right) ,
\end{equation}%
is also symmetric; $\boldsymbol{V}^{nm}=\boldsymbol{V}^{mn}$. So there is no
contribution coming from the isotriplet component of the formal expansion (%
\ref{fo}) since 
\begin{equation}
\boldsymbol{V}^{\left[ nm\right] }=\frac{1}{4}\sum d_{IJ}\left(
x^{nI}x^{mJ}-x^{mI}x^{nJ}\right) ,
\end{equation}%
vanishes identically. As such, it is interesting to split the $3\times 3$
matrix $\boldsymbol{V}^{nm}$\ as follows%
\begin{equation}
\boldsymbol{V}^{nm}=\boldsymbol{V}_{0}\delta ^{nm}-\boldsymbol{S}^{nm},
\label{vm}
\end{equation}%
where 
\begin{equation}
\boldsymbol{V}_{0}=\frac{1}{3}\mathrm{Tr}\left( \boldsymbol{V}^{nm}\right) ,
\end{equation}%
which reads in terms of 
\begin{equation}
\boldsymbol{V}_{K3}=\sum_{I,J}\mathbf{x}^{I}d_{IJ}\mathbf{x}^{J},
\end{equation}%
the volume of K3, as%
\begin{equation}
\boldsymbol{V}_{0}=\frac{1}{3}\sum_{n=1}^{3}\left( \frac{1}{2}%
\sum_{I,J}x^{nI}d_{IJ}x^{nJ}\right) =\frac{1}{6}\sum_{I,J}\mathbf{x}%
^{I}d_{IJ}\mathbf{x}^{J},
\end{equation}%
and where the traceless matrix 
\begin{equation}
\boldsymbol{S}^{nm}=\boldsymbol{V}^{nm}+\boldsymbol{V}_{0}\delta ^{nm}.
\end{equation}%
By substituting eq(\ref{vm}) back into 
\begin{equation}
\boldsymbol{V}=\sum_{n,m=1}^{3}\sigma _{n}\sigma _{m}\boldsymbol{V}^{nm},
\end{equation}%
we can put the volume matrix as follows 
\begin{equation}
\boldsymbol{V}=\boldsymbol{V}_{0}\text{ }\mathrm{I}_{3}-\boldsymbol{S}.
\end{equation}%
$\mathrm{I}_{3}$ stands for the $3\times 3$ identity matrix which will be
dropped now on and is given by, 
\begin{equation}
\boldsymbol{S}=\sum_{n,m=1}^{3}\sigma _{n}\sigma _{m}\boldsymbol{S}^{nm}.
\end{equation}%
Notice that while $\boldsymbol{V}_{0}$ is invariant under $SO\left( 4\right)
\times SO\left( 20\right) $, the matrix $\boldsymbol{S}$ is invariant under $%
SO\left( 4\right) $ but still transforms as quintet under $SU\left( 2\right)
\subset SO\left( 4\right) $. \newline
\textbf{(3)} Pushing further the similarity with the Kahler geometry, one
can then define the hyperKahler potential $\mathcal{H}$ in term of the
logarithm of the volume form $\boldsymbol{V}$ as follows%
\begin{equation}
\mathcal{H}=\mathrm{Tr}\left( \ln \boldsymbol{V}\right) =\mathrm{Tr}\left[
\ln \left( \boldsymbol{V}_{0}-\boldsymbol{S}\right) \right] ,
\end{equation}%
which can be put in the form 
\begin{equation}
\mathcal{H}=\mathrm{Tr}\left[ \ln \boldsymbol{V}_{0}\right] +\mathrm{Tr}%
\left[ \ln \left( 1-\boldsymbol{V}_{0}^{-1}\boldsymbol{S}\right) \right]
\label{hyp}
\end{equation}%
and then expanded into in power series in the inverse of the volume $%
\boldsymbol{V}_{0}$ of the K3 surface. Notice that the leading term $\mathrm{%
Tr}\left[ \ln \boldsymbol{V}_{0}\right] $ appears as the Kahler component
which is independent of $\boldsymbol{S}$. The next leading term given by the
expansion of $\ln \left( 1-\boldsymbol{V}_{0}^{-1}\boldsymbol{S}\right) $
namely 
\begin{equation}
\mathrm{Tr}\left[ \boldsymbol{V}_{0}^{-1}\boldsymbol{S}\right] =\boldsymbol{V%
}_{0}^{-1}\mathrm{Tr}\left[ \boldsymbol{S}\right]
\end{equation}%
vanishes identically due to the property $\mathrm{Tr}\left[ \boldsymbol{S}%
\right] =0$. This feature may be associated with the Ricci flat property of
hyperKahler manifolds.

\subsection{Matrix prepotential $\mathcal{G}\left( w\right) $}

\qquad In this subsection, we want to derive the \emph{"holomorphic"}\
matrix prepotential $\mathcal{G}\left( w\right) $ of type IIA superstring on
K3 surface. Holomorphicity should be understood in terms of the matrix
formulation. In other words $\mathcal{G}\left( w\right) $ is a holomorphic 2$%
\times $2 matrix which do not depend on $\overline{w}$,%
\begin{equation}
\partial \mathcal{G}\left( w\right) /\partial \overline{w}=0.
\end{equation}%
As we will show, $\mathcal{G}\left( w\right) $ and its adjoint conjugate $%
\mathcal{G}\left( \overline{w}\right) $ are prepotentials involved in the
building of the volume matrix V eq(\ref{vo}). Notice that we have used the
terminology \emph{holomorphic} because of the analog role of $\mathcal{G}%
\left( w\right) $ with the usual classical holomorphic prepotential $%
\mathcal{F}\left( z\right) $ of 10D type IIA superstring on CY3.

\subsubsection{Prepotential $\mathcal{F}\left( z\right) $ in type IIA on CY3}

\qquad Recall that in type IIA superstring on CY3, the classical expression
of the prepotential $\mathcal{F}\left( z\right) $ is given in term of the
local complex coordinates 
\begin{equation}
z^{I}=\int_{C_{2}^{I}}K_{+},\qquad I=1,...,h^{\left( 1,1\right) }\left(
CY3\right) ,
\end{equation}%
as follows:%
\begin{equation}
\mathcal{F}\left( z\right) =\frac{1}{3!}\sum_{I,J,K=1}^{h_{\text{{\tiny CY3}}%
}^{\left( 1,1\right) }}d_{IJK}z^{I}z^{J}z^{K}.
\end{equation}%
In these relations $K_{\pm }=B^{NS}\pm i\Omega ^{\left( 1,1\right) }$ is the
complexified Kahler 2- form and 
\begin{equation}
d_{IJK}=\int_{CY3}\left( \mathcal{J}_{I}\wedge \mathcal{J}_{J}\wedge 
\mathcal{J}_{K}\right) ,
\end{equation}%
is the triple intersection of 2-cycles within the Calabi-Yau treefold. The
prepotential $\mathcal{F}\left( z\right) $ is a cubic holomorphic function
playing a crucial role in the characterization of the Kahler potential of
the Kahler special {$d$-geometry} for type IIA superstring on CY3. We also
have for the Kahler potential 
\begin{equation}
\mathcal{K}\left( z,\overline{z}\right) =\ln \left( \int_{CY3}\left(
K_{+}-K_{-}\right) \wedge \left( K_{+}-K_{-}\right) \wedge \left(
K_{+}-K_{-}\right) \right) .
\end{equation}%
By integration, we get precisely 
\begin{equation}
\mathcal{K}\left( z,\overline{z}\right) =\ln \left( \sum_{I,J,K=1}^{h_{\text{%
{\tiny CY3}}}^{\left( 1,1\right) }}d_{IJK}\left( z^{I}-\overline{z}%
^{I}\right) \left( z^{J}-\overline{z}^{J}\right) \left( z^{K}-\overline{z}%
^{K}\right) \right) ,  \label{ka}
\end{equation}%
which can be also rewritten as 
\begin{equation}
\mathcal{K}\left( z,\overline{z}\right) =\ln \left( \sum_{I,J,K=1}^{h_{\text{%
{\tiny CY3}}}^{\left( 1,1\right) }}d_{IJK}\left( z^{I}-\overline{z}%
^{I}\right) \left( z_{I}-\overline{z}_{I}\right) \right) ,
\end{equation}%
where we have set 
\begin{equation}
z_{I}=\int_{CY3}K_{+}\wedge K_{+}\wedge J_{I},
\end{equation}%
and similarly for its complex conjugates $\overline{z}_{I}$.

\subsubsection{Prepotential $\mathcal{G}\left( w\right) $ in type IIA on the
K3 surface}

\qquad In the case of 10D type\ IIA superstring on the K3 surface, the
situation is a little bit subtle but we do still have quite similar
relations. More precisely, using the local quaternionic coordinates 
\begin{eqnarray}
w^{+I} &=&\int_{C_{2}^{I}}\mathcal{J}_{+},\qquad I=1,...,20,  \notag \\
w_{I}^{+} &=&\int_{C_{2}^{I}}\mathcal{J}_{+}\wedge \mathcal{J}_{I},
\end{eqnarray}%
with $\mathcal{J}_{+}=\left( B^{NS}+i\mathbf{\sigma .J}\right) $ which can
be expanded as 
\begin{equation}
\mathcal{J}_{+}=\sum_{I=1}^{20}w^{+I}\mathcal{J}_{I},
\end{equation}%
and%
\begin{equation}
\int_{C_{2}^{I}}\mathcal{J}_{K}=\delta _{K}^{I}\mathbf{,}
\end{equation}%
as well as 
\begin{equation}
w_{I}^{+}=d_{IJ}w^{+J},
\end{equation}%
the matrix prepotential $\mathcal{G}\left( w^{+}\right) $ is given by%
\begin{equation}
G^{++}=\frac{1}{2}\sum_{I,J}w^{+I}d_{IJ}w^{+J}.
\end{equation}%
Notice that $G^{++}$, which is invariant under $SO\left( 20\right) $
transformations, is \emph{holomorphic} in $w^{+}$ in the sense that it does
not depend on $w^{-}$. We also have 
\begin{equation}
G^{--}=\frac{1}{2}w^{-I}d_{IJ}w^{-J},\qquad G^{++}=\frac{1}{2}%
w^{+I}d_{IJ}w^{+J},
\end{equation}%
and 
\begin{eqnarray}
G_{I}^{+} &=&\frac{\partial G^{++}}{\partial w^{+I}}=d_{IJ}w^{+J},  \notag \\
G_{I}^{-} &=&\frac{\partial G^{--}}{\partial w^{-I}}=d_{IJ}w^{-J}.
\end{eqnarray}%
Thinking about $d_{IJ}$ as a tensor metric, we can also set $%
G_{I}^{+}=w_{I}^{+}$ and $G_{I}^{-}=w_{I}^{-}$. Using these relations, we
can rewrite eq(\ref{v}) as follows: 
\begin{equation}
\boldsymbol{V}=\frac{1}{2}\sum_{I}\left( w^{+I}-w^{-I}\right) \left(
w_{I}^{+}-w_{I}^{-}\right) ,
\end{equation}%
which should be compared with eq(\ref{ka}).

\section{Black objects in 6D/7D}

\qquad We first study black objects in 6D space time in the context of 10D
type IIA superstring compactified on the K3 surface. Then we consider the
uplifting of these black objects to 7D.

\subsection{Black objects in 6D}

\qquad We start by noting that in six dimensional space-time, the
electric/magnetic duality requires that if we have an electrically charged $%
D_{1}$-brane, the magnetically charged dual object is a $D_{2}$-brane such
that 
\begin{equation}
D_{1}+D_{2}=2.
\end{equation}%
It follows that there are essentially three kinds of black objects in 6D
with the following horizon geometries: \newline
\textbf{(1)} $D_{1}=2$ and $D_{2}=0$ with an $AdS_{2}\times \ S^{4}$ horizon
geometry describing a magnetically charged 6D black hole. It is associated
with the $6D$\ $\mathcal{N}=2$ supergravity limit of\ 10D type IIA
superstring on the K3\ surface or equivalently to 10D hererotic superstring
on 4-torus $T^{4}$. Notice also that such extremal 6D black holes can be
also recovered from the flux compactification of 11D M-theory on $K3\times
S^{1}$. The explicit D-brane\ configurations of these black holes will be
described later. \newline
\textbf{(2)} $AdS_{4}\times S^{2}$ horizon geometry describing an
electrically charged 6D black 2-brane. This black object, which solves the
above eq as $D_{1}=0$ and $D_{2}=2$, is dual to the previous 6D $\mathcal{N}%
=2$ black hole with $D_{1}=2$ and $D_{2}=0$. \newline
\textbf{(3)} $AdS_{3}\times \ S^{3}$ horizon geometry describing a dyonic
black F-string. Black D-string should, a priori, be described in the context
of 10D\ type IIB superstring \ moving on \ the K3 surface. \newline
Below, we consider the first two configurations; their attractor mechanism
will be considered in section 7. Dyonic black F-string ant its attractor
mechanism will be studied in section 7 and black D-string will be considered
elsewhere \cite{BDSS}.

\subsubsection{GVW potential for the 6D black hole}

\qquad By 6D black hole, we mean the background of the $6D$\ $N=2$\
supergravity describing the large distance limit of type IIA superstring on
the K3 surface with \ the near horizon geometry $AdS_{2}\times S^{4}$. This
black hole\ can produced \ by a system of \ D0-D2-D4-branes wrapping the
appropriate cycles of \ the K3 surface with a Gukov-Vafa-Witten (GVW) type
prepotential\footnote{%
The Gukov-Vafa-Witten (GVW) type superpotential G$_{BH}$ should not be
confused with the Weinhold scalar potential $\mathcal{V}_{eff}$ that
describe the attractor mechanism of the 6D $\mathcal{N}=2$ black objects;
see section 7 for details. The GVW superpotential deals with type II A
compactification on K3 in presence of fluxes.} $G_{BH}$\ induced by RR
fluxes on it. Below, we determine the prepotential $G_{BH}$\ which is given
by flux contributions in the K3 compactification. To that purpose, we shall
first consider the non zero flux compactification of type IIA superstring on
the K3 surface. Then, we consider switching on the fluxes.

\paragraph{\textbf{Compactification with non zero fluxes \ }\newline
}

\qquad To start recall that to fully specify the vacuum background of 10D
type IIA superstring on complex $n$-dimensional Calabi-Yau manifold $X_{n}$ (%
$n=2,4$; but in present discussion $n=2$), one must specify not just the
geometry of $X_{n}$, but also the topological classes $\zeta $ and $\xi $ of
the $1$- form and $3$- form gauge fields $\mathcal{A}_{1}$ and $\mathcal{A}%
_{3}$, i.e 
\begin{equation}
\zeta =\left[ \frac{\mathcal{F}_{2}}{2\pi }\right] \text{, \qquad\ }\xi =%
\left[ \frac{\mathcal{F}_{4}}{2\pi }\right] .
\end{equation}%
Here $\mathcal{F}_{2}=d\mathcal{A}_{1}$ and $\mathcal{F}_{4}=d\mathcal{A}%
_{3} $ are the field strengths of the gauge fields $\mathcal{A}_{1}$ and $%
\mathcal{A}_{3}$ \cite{GVW}. Topologically, these $\mathcal{A}$- fields are
classified by the characteristic classes $\zeta \in H^{2}(X_{n};Z)$ and $\xi
\in H^{4}(X_{n};Z)$. In the case of type IIA superstring on \ the K3
surface, the characteristic classes $\zeta $ and $\xi $ are non trivial.
They are respectively given by the \emph{twenty} two 2- cocycles $\left\{
\zeta ^{a},\text{ }a=1,...,22\right\} $ and the volume form $\Omega ^{\left(
2,2\right) }$ of K3. According to topology, one should then distinguishes
the two following situations:\newline
\textbf{(1)} Compactification with $\mathcal{F}_{2}=\mathcal{F}_{4}=0$. 
\newline
\textbf{(2)} Compactification with $\mathcal{F}_{2}\neq 0,$ $\mathcal{F}%
_{4}\neq 0$. \newline
In the first case where the fluxes are turned off, $\mathcal{F}_{2}=\mathcal{%
F}_{4}=0$, the variations of the K3 metric\ is generated by the two
following: \newline
\textbf{(a)} \emph{Complex deformations}:\qquad\ \newline
the variations of the complex structure of complex dimension Calabi-Yau
manifold are described by the complex moduli%
\begin{equation}
z^{I}=\int_{C_{n}^{I}}\Omega ^{\left( n,0\right) },\qquad I=1,...,h^{\left(
n-1,1\right) },\qquad n=2.  \label{an}
\end{equation}%
From the 4D $\mathcal{N}=1$ superfield theory language, the $z^{I}$'s are
just the leading scalar component fields of 4D $\mathcal{N}=1$ chiral
superfields $Z^{I}\left( \mathrm{y},\theta \right) $ with $\theta $-
expansion is given by%
\begin{equation}
Z^{I}\left( \mathrm{y},\theta \right) =z^{I}\left( \mathrm{y}\right) +\theta
\psi ^{I}\left( \mathrm{y}\right) +\theta ^{2}F^{I}\left( \mathrm{y}\right)
,\qquad I=1,...,h^{\left( n-1,1\right) }.
\end{equation}%
Here $\mathrm{y=x+i\theta \sigma }\overline{\mathrm{\theta }}$ with $\mathrm{%
x}$ being the usual 4D space time coordinates. We also have the antichiral
superfields%
\begin{equation}
\overline{Z}^{I}\left( \overline{\mathrm{y}},\overline{\theta }\right) =%
\overline{z}^{I}\left( \overline{\mathrm{y}}\right) +\overline{\theta }%
\overline{\psi }^{I}\left( \overline{\mathrm{y}}\right) +\overline{\theta }%
^{2}\overline{F}^{I}\left( \overline{\mathrm{y}}\right) ,\qquad
I=1,...,h^{\left( n-1,1\right) }.
\end{equation}%
Notice that in the particular case of K3 surface ($n=2$), these superfields
which for later use we rewrite as a $2\times 2$ matrix as follows, 
\begin{equation}
\left( 
\begin{array}{cc}
0 & Z^{I}\left( \mathrm{y},\theta \right) \\ 
\overline{Z}^{I}\left( \overline{\mathrm{y}},\overline{\theta }\right) & 0%
\end{array}%
\right) ,  \label{ma}
\end{equation}%
are not yet all what we need. These superfields $Z^{I}$ capture in fact just
half of the degrees of freedom of the 6D $\mathcal{N}=1$ hypermultiplet $%
H^{I}$ which is same as 4D $\mathcal{N}=2$ hypermultiplet; that is two
different types of 4D $\mathcal{N}=1$ chiral superfields together with their
complex conjugates. Notice also that in eq(\ref{an}), we have added the
complex dimension $n$ just to make contact with higher dimensional
Calabi-Yau manifolds and also to exhibit the specific feature for $n=2$. In
this particular case, complex $\Omega ^{\left( 2,0\right) }$, its complex
conjugate $\Omega ^{\left( 0,2\right) }$ and Kahler $\Omega ^{\left(
1,1\right) }$ structures combine to give a hyperKahler structure on the
moduli space of 10D type IIA superstring on the K3 surface. This is exactly
what we need to reproduce the hypermultiplets as described below.\newline
\textbf{(b) }\emph{Kahler deformations}:\qquad\ \newline
the\textbf{\ }variation of the (complexified) Kahler structure is described
by the complex moduli,%
\begin{equation}
t^{I}=\int_{C_{2}^{I}}\left( B^{NS}+i\Omega ^{\left( 1,1\right) }\right)
,\qquad I=1,...,h^{\left( 1,1\right) }.
\end{equation}%
From the language of 4D $\mathcal{N}=1$ superfield theory, the parameters $%
t^{I}$ get promoted to 4D $\mathcal{N}=1$ chiral superfields as shown below 
\begin{equation}
T^{I}\left( \widetilde{\mathrm{y}},\widetilde{\theta }\right) =t^{I}+%
\widetilde{\theta }\widetilde{\psi }^{I}+\widetilde{\theta }^{2}\widetilde{F}%
^{I}.
\end{equation}%
A similar expression is valid for the antichiral superfield $\overline{T}%
^{I}=\overline{T}^{I}\left( \overline{\widetilde{\mathrm{y}}},\overline{%
\widetilde{\theta }}\right) $\textrm{. }\newline
The 6D $\mathcal{N}=1$ supersymmetric hypermultiplets, which as noted before
is equivalent to two 4D $\mathcal{N}=1$ chiral superfields, have then the
superspace structure 
\begin{equation}
H^{I}=H^{I}\left( \mathrm{y},\theta ;\widetilde{\mathrm{y}},\widetilde{%
\theta };\overline{\mathrm{y}},\overline{\theta };\overline{\widetilde{%
\mathrm{y}}},\overline{\widetilde{\theta }}\right) .
\end{equation}%
They are made of the superfields Z$^{I}$ and T$^{I}$. Using the
representation (\ref{ma}), the $H^{I}$ superfields can be represented by the
following 
\begin{equation}
H^{I}=\left( 
\begin{array}{cc}
T^{I} & Z^{I} \\ 
\overline{Z}^{I} & \overline{T^{I}}%
\end{array}%
\right) ,
\end{equation}%
where the lowest component fields are precisely the quaternionic moduli $%
w^{+I}$.\newline
\textbf{(c) }\emph{Deformations in the presence of RR- fluxes}:\qquad\ 
\newline
First note that in the case where $\mathcal{F}_{2}=\mathcal{F}_{4}=0$, the
expectation values of the $T^{I}$ and $Z^{I}$ are arbitrary in the
supergravity approximation to 10D type superstring IIA on the K3 surface.
However, in the case where the fluxes are turned on, i.e 
\begin{equation}
\mathcal{F}_{2}\neq 0,\qquad \mathcal{F}_{4}\neq 0,
\end{equation}%
the situation become more subtle since we must adjust the complex and Kahler
structures of the K3 surface so that to stabilize the fluxes. This
adjustment is achieved by adding a potential $\mathcal{G}_{BH}$ whose
variation with respect to complex and Kahler moduli fixes the fluxes to
zero. This will be discussed in the next paragraph.

\paragraph{GVW \textbf{superpotential }$\mathcal{G}_{BH}$\textbf{\ }\ 
\newline
}

\qquad We first consider the adjustment of the complex structure and next
the adjustment of the Kahler one. Obviously in the present case dealing with
10D type IIA superstring on the K3 surface, these two adjustments can be
done altogether in $SU\left( 2\right) $ covariant way. We will turn to this
property later on.

To adjust the complex structure of 10D type IIA superstring on the K3
surface, we should impose the following constraint relations:%
\begin{equation}
\frac{\delta \mathcal{G}_{BH}}{\delta \Omega ^{\left( 2,0\right) }}=\mathcal{%
F}^{\left( 0,2\right) }=0,\qquad \frac{\delta \mathcal{G}_{BH}}{\delta
\Omega ^{\left( 0,2\right) }}=\mathcal{F}^{\left( 2,0\right) }=0.  \label{gc}
\end{equation}%
The field strengths $\mathcal{F}^{\left( 0,2\right) }$ and $\mathcal{F}%
^{\left( 2,0\right) }$ are the holomorphic and antiholomorphic components
appearing the decomposition of the 2-form $\mathcal{F}_{2}$. We recall that
the real 2-form $\mathcal{F}_{2}$ can be usually decomposed as an isotriplet
as shown below, see also \emph{footnote 1}, 
\begin{equation}
\mathcal{F}_{2}=\mathcal{F}^{\left( 1,1\right) }\oplus \mathcal{F}^{\left(
0,2\right) }\oplus \mathcal{F}^{\left( 2,0\right) }.  \label{ff}
\end{equation}%
To adjust the Kahler structure of 10D type IIA superstring on the K3
surface, one has to impose the following constraint relation 
\begin{equation}
\frac{\delta \mathcal{G}_{BH}}{\delta \Omega ^{\left( 1,1\right) }}=\mathcal{%
F}^{\left( 1,1\right) }=0.  \label{da}
\end{equation}%
Notice that the 2-form $\mathcal{F}^{\left( 1,1\right) }$ appearing above
comes precisely from the field strength $\mathcal{F}_{2}$ decomposition (\ref%
{ff}). From the 6D $\mathcal{N}=1$ supersymmetric field theory, the
relations (\ref{gc}-\ref{da}) can be viewed as superfield equations of
motion following from the variation of the potential, 
\begin{eqnarray}
\boldsymbol{G}_{BH} &=&+\frac{1}{2}\left( \int_{K3}\mathcal{F}^{\left(
2,0\right) }\wedge \Omega ^{\left( 0,2\right) }+\mathcal{F}^{\left(
0,2\right) }\wedge \Omega ^{\left( 2,0\right) }\right)  \notag \\
&&\int_{K3}\mathcal{F}^{\left( 1,1\right) }\wedge \Omega ^{\left( 1,1\right)
}+\int_{K3}\mathcal{F}_{4},  \label{bh}
\end{eqnarray}%
where we have added the integral constant $q=\int_{K3}\mathcal{F}_{4}$.
Notice that the variation of $\boldsymbol{G}_{BH}$ with respect to $\Omega
^{\left( 0,2\right) }$, $\Omega ^{\left( 2,0\right) }$ and $\Omega ^{\left(
1,1\right) }$, one recovers precisely eqs(\ref{gc}-\ref{da}). Notice also
that up on rewriting 
\begin{eqnarray}
\mathcal{F}^{\left( 2,0\right) }\wedge \Omega ^{\left( 0,2\right) }
&=&\left( i\right) \mathcal{F}^{\left( 2,0\right) }\wedge \left( -i\right)
\Omega ^{\left( 0,2\right) },  \notag \\
\Omega ^{\left( 1,1\right) } &=&\frac{1}{2}\boldsymbol{K}_{-}+\frac{1}{2}%
\boldsymbol{K}_{+}, \\
\mathcal{F}^{\left( 0,2\right) }\wedge \Omega ^{\left( 2,0\right) }
&=&\left( i\right) \mathcal{F}^{\left( 0,2\right) }\wedge \left( -i\right)
\Omega ^{\left( 2,0\right) },  \notag
\end{eqnarray}%
where we have used 
\begin{equation}
\boldsymbol{K}_{\pm }=B^{NS}\pm i\Omega ^{\left( 1,1\right) },
\end{equation}%
we can put $\boldsymbol{G}_{BH}$ as,%
\begin{eqnarray}
\boldsymbol{G}_{BH} &=&\frac{i}{2}\left( \int_{K3}\mathcal{F}^{\left(
1,1\right) }\wedge \boldsymbol{K}_{-}\right) -\frac{i}{2}\left( \int_{K3}%
\mathcal{F}^{\left( 1,1\right) }\wedge \boldsymbol{K}_{+}\right)  \notag \\
&&+\frac{1}{2}\left( \int_{K3}\left( i\right) \mathcal{F}^{\left( 2,0\right)
}\wedge \left( -i\right) \Omega ^{\left( 0,2\right) }\right) \\
&&+\frac{1}{2}\left( \int_{K3}\left( i\right) \mathcal{F}^{\left( 0,2\right)
}\wedge \left( -i\right) \Omega ^{\left( 2,0\right) }\right) +\int_{K3}%
\mathcal{F}_{4}.  \notag
\end{eqnarray}%
Then setting 
\begin{equation}
\boldsymbol{F}_{2}=\left( 
\begin{array}{cc}
\mathcal{F}^{\left( 1,1\right) } & \mathcal{F}^{\left( 2,0\right) } \\ 
\mathcal{F}^{\left( 0,2\right) } & -\mathcal{F}^{\left( 1,1\right) }%
\end{array}%
\right) ,
\end{equation}%
which reads also in terms of Pauli matrices as 
\begin{equation}
\boldsymbol{F}_{2}=\sigma ^{0}\mathcal{F}^{\left( 1,1\right) }+\sigma ^{-}%
\mathcal{F}^{\left( 2,0\right) }+\sigma ^{+}\mathcal{F}^{\left( 0,2\right) },
\end{equation}%
we can bring $\boldsymbol{G}_{BH}$ to the following SU$\left( 2\right) $
covariant formula,%
\begin{equation}
\boldsymbol{G}_{BH}=\int_{K3}\mathcal{F}_{4}+\frac{i}{2}\mathrm{Tr}\left(
\int_{K3}\boldsymbol{F}_{2}\wedge \mathcal{J}_{-}\right)
\end{equation}%
where we have used 
\begin{equation}
\mathcal{J}_{-}=\left( 
\begin{array}{cc}
\boldsymbol{K}_{-} & -i\Omega ^{\left( 2,0\right) } \\ 
-i\Omega ^{\left( 0,2\right) } & \boldsymbol{K}_{+}%
\end{array}%
\right) =B-i\mathbf{\sigma .J.}
\end{equation}%
Now, substituting $\mathcal{J}_{-}$ by its expansion $\sum w^{-I}J_{I}$,
doing the same for $\boldsymbol{F}_{2}$ 
\begin{equation}
\boldsymbol{F}_{2}=\sum_{J}\mathbf{\sigma \cdot q}^{J}J_{J},\qquad
\end{equation}%
with $\mathbf{q}^{J}=\left( q^{1J},q^{2J},q^{3J}\right) $ and then
factorizing the $\mathbf{q}^{J}$ integer 3- vectors as follows%
\begin{equation}
q^{1J}=p^{1}p^{J},\qquad q^{2J}=p^{2}p^{J},\qquad q^{3J}=p^{3}p^{J}
\end{equation}%
or equivalently in a condensed form as%
\begin{equation}
\mathbf{q}^{J}\mathbf{=p}p^{J},
\end{equation}%
we get 
\begin{equation}
\boldsymbol{G}_{BH}=q+\mathrm{Tr}\left[ \mathbf{\sigma .p}\left(
i\sum_{I,J=1}^{20}w^{-I}d_{IJ}p^{J}\right) \right] .
\end{equation}%
This relation reads also as%
\begin{equation}
\boldsymbol{G}_{BH}=q+\mathrm{Tr}\left[ \mathbf{\sigma .p}\left(
i\sum_{I,J=1}^{20}w^{-I}p_{I}\right) \right] ,
\end{equation}%
where we have set 
\begin{equation}
p_{I}=\sum_{I}d_{IJ}p^{J}.
\end{equation}%
Expanding $w^{-I}=y^{0I}-i\mathbf{x}^{I}.\mathbf{\sigma }$ and computing
trace, we obtain the general form of the GVW potential for type IIA
compactification on K3, 
\begin{equation}
\mathcal{G}_{BH}=q+\mathbf{p.}\left( \sum_{I}\mathbf{x}^{I}p_{I}\right) .
\end{equation}%
Using eq(\ref{ch}), one can re-express the above relation, in terms of the
isotriplet central charge $\mathbf{Z=}\sum_{I}\mathbf{x}^{I}p_{I}$, as
follows: 
\begin{equation}
\mathcal{G}_{BH}=q+\mathbf{p.Z,}  \label{b1}
\end{equation}%
together with 
\begin{equation}
\mathbf{p.Z}=\sum_{i=1}^{3}p^{m}Z^{m}.  \label{b10}
\end{equation}%
Notice finally that in the above relations, $q$ is the number of D4-branes
wrapping the K3 surface, and the $p_{I}$'s give the number of D2-branes
wrapping 2-cycles of the K3 surface.

\subsubsection{6D Black 2-Brane}

\qquad Similarly as for 6D $\mathcal{N}=2$ black hole, the 6D $\mathcal{N}=2$
black 2-brane is given by the $6D$ $\mathcal{N}=2$ supergravity describing
the large distance limit of 10D type IIA superstring on the K3 surface with
near horizon geometry $AdS_{4}\times \ S^{2}$.\ We can \ produce these 6D $%
\mathcal{N}=2$ black 2-brane by considering a D-brane system consisting of:%
\newline
\textbf{(1)} a D4- brane wrapping the C$_{2}^{I}$ 2- cycles in K3 and the
extra directions filling two dimensions $\left( t,x\right) $ in space time,
say the first $x^{0}=t$ and second $x^{1}=x$ dimensions. \newline
\textbf{(2)} a D6-brane wrapping K3 and the remaining two others in the
first and second of space time. \newline
The potential induced by RR fluxes on K3 for 6D\ black 2-brane potential has
a similar structure as for 6D black hole except that now that the black
object is electrically charged. Similar computations lead to 
\begin{equation}
\mathcal{G}_{{\small B2}\text{{\small -}}{\small brane}}=p+\mathbf{q.Z,}
\label{b2}
\end{equation}%
with 
\begin{equation}
\mathbf{Z=}\sum_{I}q_{I}\mathbf{x}^{I},  \label{b3}
\end{equation}%
and 
\begin{equation}
\mathbf{q.Z}=\sum_{i=1}^{3}q^{m}Z^{m}.  \label{b11}
\end{equation}%
In eq(\ref{b2}), the integer $p$ gives the number of wrapped of D6- branes
on K3 and the $q_{I}$'s give the number of D4 wrapping the \textit{I-th} 2-
cycles of the K3 surface.

\subsection{Black Objects in 7D}

\qquad The simplest way to describe 7D $\mathcal{N}=2$ supergravity is to
think about it as the large distance limit of 11D M-theory compactified on
the K3 surface. The 7D $\mathcal{N}=2$ component field action $\mathcal{S}%
_{7D}$ can be obtained by starting from the 11D $\mathcal{N}=1$ supergravity
action and performing the compactification on the K3 surface. The bosonic
part action of the 11D supergravity is given by 
\begin{equation}
\mathcal{S}_{11D}=\frac{-1}{2\kappa _{11}^{2}}\int \left( \mathcal{R}\text{ }%
\left( ^{\ast }1_{11}\right) +\frac{1}{2}\mathcal{F}_{4}\wedge \text{ }%
\left( ^{\ast }\mathcal{F}\right) _{7}+\frac{1}{3!}\mathcal{F}_{4}\wedge 
\mathcal{F}_{4}\wedge \mathcal{C}_{3}\right) ,
\end{equation}%
where $\mathcal{R}$ is the usual scalar curvature in 11D, the 4-form field
strength is $\mathcal{F}_{4}=d\mathcal{C}_{3}$ and the 7-form $\left( ^{\ast
}\mathcal{F}\right) _{7}$ is the Hodge dual of $\mathcal{F}_{4}$. The
coupling constant $\kappa _{D}^{2}$ is related to Newton's constant as $%
\kappa _{D}^{2}=8\pi G_{D}$. By compactifying on the K3 surface preserving
half of the 32 supersymmetries, we can get the explicit expression of $%
\mathcal{S}_{7D}$. As there is no 1-cycle nor 3-cycle within the K3 surface,
the moduli space of the 7D $\mathcal{N}=2$ theory is just the geometric one
given by, 
\begin{equation}
\boldsymbol{M}_{7D}\times SO\left( 1,1\right) ,\qquad \boldsymbol{M}_{7D}=%
\frac{SO\left( 3,19\right) }{SO\left( 3\right) \times SO\left( 19\right) },
\label{mo}
\end{equation}%
where $SO\left( 1,1\right) $ stands for the dilaton. The other \emph{57}
moduli parameterizing the homogeneous space $\boldsymbol{M}_{7D}$ are
associated with scalars within the $7D$ $\mathcal{N}=2$ multiplets. Let us
comment below these $7D$ $\mathcal{N}=2$ multiplets; we have: \newline
\textbf{(1) }One $7D$ $\mathcal{N}=2$ gravity multiplet $\mathcal{G}_{7D}^{%
\mathcal{N}=2}$ consisting of $\emph{40}+\emph{40}$ on shell degrees of
freedom. The component fields of its bosonic sector are given by 
\begin{equation}
\mathcal{G}_{7D}^{\mathcal{N}=2}:\left( \phi ,\text{ }g_{\mu \nu },\mathcal{C%
}_{\mu \nu \rho },\mathcal{A}_{\mu }^{0},\mathcal{A}_{\mu }^{\pm }\right) .
\end{equation}%
They describe, in addition to the dilaton, a spin 2 field, one 3-form gauge
field and three 1-form gauge fields. It involves then one scalar only.%
\newline
(\textbf{2}) Fifty seven (\emph{57}) $7D$ $\mathcal{N}=2$ vector multiplet 
\begin{equation}
\left( \mathcal{V}_{7D}^{\mathcal{N}=2}\right) _{I},\qquad I=1,...,19.
\end{equation}%
Each vector multiplet $\mathcal{V}_{7D}^{\mathcal{N}=2}$ has $\emph{8}+\emph{%
8}$ on shell degrees of freedom. The bosonic sector of $\mathcal{V}_{7D}^{%
\mathcal{N}=2}$ contains a seven dimensional gauge field $\mathcal{C}_{\mu }$
and three scalars 
\begin{equation}
\mathbf{x}=\left( x^{1},x^{2},x^{3}\right) .
\end{equation}%
The bosonic sector of the \emph{57} supermultiplets of the seven dimensional 
$\mathcal{N}=2$ vector superfields $\left( \mathcal{V}_{7D}^{\mathcal{N}%
=2}\right) _{I}$ read then as 
\begin{equation}
\left( \mathcal{V}_{7D}^{\mathcal{N}=2}\right) _{I}:\left( \mathcal{C}_{\mu
}^{I},\mathbf{x}^{I}\right) ,\qquad I=1,...,19
\end{equation}%
where the $\mathbf{x}^{I}$'s ($\left( x^{1I},x^{2I},x^{3I}\right) $) refer
to the \emph{57} isotriplets of 7D scalars. They describe the metric
deformations of the K3 surface.$\ $Notice that the gauge fields $\mathcal{C}%
_{\mu }^{I}$ belong to the representation $\left( 1,19\right) $ of the group 
$SO\left( 3\right) \times SO\left( 19\right) $ and the scalars $\mathbf{x}%
^{I}$ are in the $\left( 3,19\right) $ one.%
\begin{eqnarray}
\mathcal{C}_{\mu }^{I}\qquad &\sim &\qquad \left( 1,19\right) \text{ \ }\in 
\text{ \ }SO\left( 3\right) \times SO\left( 19\right)  \notag \\
\mathbf{x}^{I}\qquad &\sim &\qquad \left( 3,19\right) \ \ \in \text{ \ }%
SO\left( 3\right) \times SO\left( 19\right)
\end{eqnarray}%
Now, we move to discuss $\mathcal{N}=2$ black objects in seven dimensions
from the view of 10D type IIA superstring on the K3 surface and their
uplifting to 7D. This is achieved by starting from the 6D results, in
particular the expression of the central charges in 6D, eqs(\ref{b10}) and (%
\ref{b11}), 
\begin{equation}
Z^{m}=\sum_{I=1}^{20}Q_{I}x^{mI},
\end{equation}%
and minimizing $Z^{m}=Z^{m}\left( x\right) $ by taking into account the
constraint eqs(\ref{c1}-\ref{c3}) fixing the volume of the K3 surface to a
constant ($\mathcal{V}_{K3}=$constant), i.e 
\begin{equation}
\sum_{n=1}^{3}\left( \sum_{I,J=1}^{20}x^{nI}d_{IJ}x^{nI}\right) =\text{ 
{\small constant}.}
\end{equation}%
The electric/magnetic duality equation requires that the space dimensions $%
D_{1}$ and $D_{2}$ of two dual objects in 7D should be as: 
\begin{equation}
D_{1}+D_{2}=3.
\end{equation}%
According to 10D type II superstrings view, there are \emph{four} kinds of
7D black objects: \emph{Two} of them can be described in type 10D IIA
superstring and the \emph{two} others in 10D type IIB superstring. The near
horizon of these geometries are classified as follows: \newline
\textbf{(i)} $AdS_{2}\times \ S^{5}$ describing \ the near horizon geometry
of 7D black holes.\newline
\textbf{(ii)} $AdS_{4}\times S^{3}$ describing 7D black 2-branes. \newline
These black objects are quite similar to the 6D case discussed perviously.
They are described in the context of uplifting of 10D type IIA superstring
on the K3 surface. The two others should be described in the context of
uplifting of 10D type IIB superstring on the K3 surface. Their near horizon
geometries are as follows:\newline
\textbf{(iii)} $AdS_{3}\times S^{4}$ associated with 7D black strings. 
\newline
\textbf{(iv)} $AdS_{5}\times \ S^{2}$ associated with 7D black 3-branes. 
\newline
The electric/magnetic duality in seven dimensions can be understood as
describing the following interchange 
\begin{eqnarray}
AdS_{2}\times \ S^{5}\ \qquad &\leftrightarrow &\qquad \ AdS_{5}\times \
S^{2}  \notag \\
AdS_{3}\times \ S^{4}\qquad &\leftrightarrow &\qquad AdS_{4}\times \ S^{3}.
\end{eqnarray}%
It interchanges the A and B-models and maps 7D black holes to 7D black
3-branes and 7D black strings to 7D black 2-branes. This interchange may
have a T-duality interpretation connecting type IIA and type IIB in odd
dimensional space-time; in particular in 7D. As previously mentioned $%
AdS_{3}\times \ S^{3}$ is a dyonic black string in 6D, and its uplift to 7D
describes a 7D black string with the horizon geometry $AdS_{3}\times \ S^{4}$
and a 7D black 2-brane with the horizon geometry $AdS_{4}\times \ S^{3}$.

On the other hand, $AdS_{2}\times \ S^{4}$ and $AdS_{4}\times \ S^{2}$ is
uplifted respectively to $AdS_{3}\times \ S^{4}$ and $AdS_{4}\times \ S^{3}$
describing type IIA black objects in 7D. From M-theory compactification
view, these 7D black brane configurations can be reproduced by wrapping M2
and M5-branes on appropriate cycles in the K3 surface. Note that in M-theory
background there is no RR 2- form flux since we have no 1-form. We have
rather a 4-form flux where quite similar computations can be done.

\section{Effective potential and Attractor mechanism in 6D and 7D}

\qquad In this section, we study the effective potential and attractor
mechanism of the 6D (7D) supersymmetric black objects. This study requires
considering all the scalar field moduli of the 6D (7D) non chiral
supergravity theory including the dilaton field that we have freezed before.
In 6D\textrm{\ }(7D) space time, there are 81 (58) scalars distributed as
follows: \newline
(i) the dilaton $\sigma $ belonging to the 6D (7D) $\mathcal{N}=2$ gravity
multiplet. It has been neglected before; but below it will be taken as a
dynamical variable. For 7D, the field $\sigma $ has a geometric
interpretation in term of the volume of K3.\newline
(ii) the eighty (fifty seven) other moduli $\omega _{aI}$ ($\rho _{aI}$)
belonging to the 6D (7D) $\mathcal{N}=2$ Maxwell multiplets. \newline
To fix the ideas, we consider in the next two subsections 7.1 and 7.2 the 6D 
$\mathcal{N}=2$ supergravity and study the effective scalar potential $%
\mathcal{V}_{eff}=\mathcal{V}_{eff}\left( \sigma ,\omega _{aI}\right) $ of
the 6D space time black objects. In subsection 7.3, we give the results for
7D.\newline
The effective scalar potential $\mathcal{V}_{eff}$ of the 6D black objects
is given by the Weinhold potential expressed in terms of the \emph{dressed}
charges,%
\begin{equation}
\left( Z_{+},Z_{-},Z_{a},Z_{I}\right) ,\qquad a=1,...,4,\qquad I=1,...,20.
\label{cec}
\end{equation}%
These central charges appear in the supersymmetric transformations of the
fields of the 6D supergravity theory; in particular in \textrm{\cite{6}}:%
\newline
(i) the supersymmetric transformations of the \emph{two} 6D gravitinos and
the \emph{four} gravi-photinos/dilatinos of the supergravity multiplet.%
\newline
(ii) the supersymmetric transformations of the \emph{twenty} photinos of the 
$U^{20}\left( 1\right) $ gauge supermultilets that follow from the
compactification of 10D type IIA superstring on K3.\newline
At the event horizon of the 6D black objects, the potential $\left( \mathcal{%
V}_{eff}\right) _{black}$ attains the minimum. The real $\left( \sigma
,\omega _{aI}\right) $ moduli parameterizing $\widehat{G}=SO\left(
1,1\right) \times G$ with $G=SO\left( 4,20\right) /SO\left( 4\right) \times
SO\left( 20\right) $ are generally fixed by the charges 
\begin{equation}
g^{+},\text{ \ }g^{-},\text{ \ }g^{a},\text{ \ }h^{I},\text{ \ }q_{a},\text{
\ }p_{I},  \label{cah}
\end{equation}%
of the $\mathcal{N}=2$ 6D supergravity gauge field strengths%
\begin{equation}
H_{3}^{+}\text{ \ },\text{ \ }H_{3}^{-}\text{ \ },\text{ \ }F_{2}^{a}\text{
\ },\text{ \ }F_{2}^{I}\text{ \ },\text{ \ }F_{4a}\text{ \ },\text{ \ }%
F_{4I}.
\end{equation}%
The attractor equations of the 6D $\mathcal{N}=2$ black objects will be
obtained from the minimization of the $\left( \mathcal{V}_{eff}\right) _{%
\text{black}}$. Once we write down these attractor eqs, we pass to examine
their solutions at the horizon of the black attractors. These solutions fix
the moduli $\left( \sigma ,\omega _{aI}\right) $ in terms of the charges (%
\ref{cah}),%
\begin{equation}
\left( \sigma \right) _{\text{horizon}}=\sigma \left( g,e\right) ,\qquad
\left( \omega \right) _{\text{horizon}}=\omega \left( q_{a},p_{I}\right) ,
\end{equation}%
and give the expression of the black objects entropies in terms of these
charges.\newline
Notice by the way that there are different ways to derive the gauge
invariant effective scalar potential $\mathcal{V}_{eff}$ of the 6D $\mathcal{%
N}=2$ supersymmetric BPS black objects \textrm{\cite{6,61,62,63}}. Here, we
shall follow the approach used in the works\textrm{\ \cite{F0,F1,F2,F3}}.
The effective scalar potential $\mathcal{V}_{eff}$ of the 6D black object is
expressed as a quadratic form of the central charges (\ref{cec}). \newline
Notice moreover that from the field spectrum of the 6D $\mathcal{N}=2$ non
chiral supergravity, one learns that two basic situations should be
distinguished:\newline
(\textbf{1}) 6D black F- string (BFS) with near horizon geometry $%
AdS_{3}\times S^{3}$. This is a 6D dyonic black F-string solution. The
electric/magnetic charges involved here are those of the gauge invariant 3-
form field strengths%
\begin{equation}
H_{3}^{+}=\frac{1}{2}\left( H_{3}+\text{ }^{\star }H_{3}\right) \quad ,\quad
H_{3}^{-}=\frac{1}{2}\left( H_{3}-\text{ }^{\star }H_{3}\right) ,
\end{equation}%
associated with the usual 6D 2- form antisymmetric $B_{\mu \nu }^{\pm }$
fields. The $\star $ conjugation stands for the usual Poincar\'{e} duality
interchanging $n$- forms with $\left( 6-n\right) $ ones. \newline
(\textbf{2}) 6D black hole (BH) and its black 2- brane (B2B) dual. Their
near horizon geometries were discussed in previous sections. The field
strengths involved in these objects are related by the Poincar\'{e} duality
in 6D space time which interchanges the 2- and 4- form field strengths.%
\newline
Below, we study separately these two configurations.

\subsection{Black F-string in 6D}

\qquad The black BPS object of the 6D $\mathcal{N}=2$ non chiral theory is a
dyonic string charged under both the self dual $H_{3}^{+}$ and antiself dual 
$H_{3}^{-}$ field strengths of the NS-NS B$^{\pm }$-fields. Using the
following bare magnetic/electric charges,%
\begin{equation}
g^{\pm }=\int_{S^{3}}H_{3}^{\pm },\qquad g^{\pm }=\frac{1}{2}\left( g\pm
e\right) ,
\end{equation}%
where $g=\int_{S^{3}}H_{3}$ and $e=\int_{S^{3}}$ $^{\star }H_{3}$, one can
write down the physical charges in terms of the dressed charges.

\subsubsection{Dressed charges}

The dressed charges play an important role in the study of supergravity
theories. They appear in the supersymmetric transformations of the Fermi
fields (here gravitinos), and generally read like%
\begin{eqnarray}
Z^{+} &=&X_{+}^{+}g^{+}+X_{-}^{+}g^{-}  \notag \\
Z^{-} &=&X_{+}^{-}g^{+}+X_{-}^{-}g^{-},  \label{cz}
\end{eqnarray}%
where the real $2\times 2$ matrix 
\begin{equation}
X=\left( 
\begin{array}{cc}
X_{+}^{+} & X_{-}^{+} \\ 
X_{-}^{+} & X_{+}^{+}%
\end{array}%
\right) ,
\end{equation}%
parameterizes the $SO\left( 1,1\right) $ factor of the moduli space $%
\widehat{G}$. \newline
Taking the $\eta _{rs}$ flat metric as $\eta =diag\left( 1,-1\right) $, we
can express all the four real parameters $X_{-}^{\pm }$ and $X_{+}^{\pm }$
in terms of the dilaton $\sigma =\sigma \left( x\right) $ by solving the
constraint eqs $X^{t}\eta X=\eta $ which split into four constraint
relations like%
\begin{eqnarray}
X_{+}^{+}X_{+}^{+}-X_{+}^{-}X_{+}^{-} &=&1  \notag \\
X_{+}^{+}X_{-}^{+}-X_{+}^{-}X_{-}^{-} &=&0  \notag \\
X_{-}^{+}X_{+}^{+}-X_{-}^{-}X_{+}^{-} &=&0  \label{xeq} \\
X_{-}^{-}X_{-}^{-}-X_{-}^{+}X_{-}^{+} &=&1.  \notag
\end{eqnarray}%
These eqs can be solved by, 
\begin{equation}
X_{+}^{+}=X_{-}^{-}=\cosh \left( 2\sigma \right) ,\qquad
X_{+}^{-}=X_{-}^{+}=\sinh \left( 2\sigma \right) .  \label{xx}
\end{equation}%
Putting these solutions back into the expressions of the central charges $%
Z^{+}$ and $Z^{-}$ (\ref{cz}), we get the following dilaton dependent
quantities%
\begin{eqnarray}
Z^{+} &=&\frac{1}{2}\left[ g\exp \left( -2\sigma \right) +e\exp \left(
+2\sigma \right) \right]  \notag \\
Z^{-} &=&\frac{1}{2}\left[ g\exp \left( -2\sigma \right) -e\exp \left(
2\sigma \right) \right] .  \label{zz}
\end{eqnarray}%
Notice that these dressed charges have no dependence on the $\omega _{aI}$
field moduli of the coset $SO\left( 4,20\right) /SO\left( 4\right) \times
SO\left( 20\right) $. This is because the NS-NS B- fields is not charged
under the isotropy group of the above coset manifold.

\subsubsection{BFS potential}

With the dressed charges $Z^{+}$ and $Z^{-}$, we can write down the gauge
invariant effective scalar potential $\mathcal{V}_{BFS}$. It is given by the
so called Weinhold potential,%
\begin{equation}
\mathcal{V}_{BFS}=\left( Z^{+}\right) ^{2}+\left( Z^{-}\right) ^{2}.
\label{ana}
\end{equation}%
Notice that, as far symmetries are concerned, one also have the other \emph{%
"orthogonal"} combination namely $\left( Z^{+}\right) ^{2}-\left(
Z^{-}\right) ^{2}$. This quantity cannot, however, be interpreted as a
supersymmetric BPS potential. First it is not positive definite and second
it has an interpretation in terms of eqs(\ref{xeq}). We will show later that
this gauge invariant combination corresponds just to the electric/magnetic
charge quantization condition.\newline
By substituting eq(\ref{cz}) into the relation (\ref{ana}), we get the
following form of the potential,%
\begin{equation}
\mathcal{V}_{BFS}=\left( g^{+},g^{-}\right) \mathcal{M}\left( 
\begin{array}{c}
g^{+} \\ 
g^{-}%
\end{array}%
\right) ,
\end{equation}%
with%
\begin{equation}
\mathcal{M}=\left( 
\begin{array}{cc}
\left( X_{+}^{+}\right) ^{2}+\left( X_{+}^{-}\right) ^{2} & 
2X_{+}^{+}X_{-}^{+} \\ 
2X_{+}^{-}X_{-}^{-} & \left( X_{-}^{-}\right) ^{2}+\left( X_{-}^{+}\right)
^{2}%
\end{array}%
\right) .
\end{equation}%
From this matrix and using the transformations given in \textrm{\cite{F1}},
we can read the gauge field coupling metric $\mathcal{N}_{+-}$ and $\mathcal{%
N}_{-+}$ that appear in the 6D $\mathcal{N}=2$ supergravity component field
Lagrangian density%
\begin{equation}
\frac{\mathcal{L}_{6D}^{N=2\text{ sugra}}}{\sqrt{-g}}=\mathcal{R}%
_{6}1+\left( \frac{1}{2}\mathcal{N}_{+-}H^{+}\wedge H^{-}+\frac{1}{2}%
\mathcal{N}_{-+}H^{-}\wedge H^{+}\right) +\cdots
\end{equation}%
In this eq, $\mathcal{R}_{6}$ is the usual 6D scalar curvature and $g=\det
\left( g_{\mu \nu }\right) $. By further using (\ref{zz}), we can put the
potential $\mathcal{V}_{BFS}$ into the following form%
\begin{equation}
\mathcal{V}_{BFS}\left( \sigma \right) =\frac{g^{2}}{2}\exp \left( -4\sigma
\right) +\frac{e^{2}}{2}\exp \left( 4\sigma \right) .  \label{bfs}
\end{equation}%
Notice that the self and anti- self duality properties of the field
strengths H$_{3}^{+}$ and H$_{3}^{-}$ imply that the corresponding
magnetic/electric charges are related as follows%
\begin{equation}
g^{+}=e^{+},\qquad g^{-}=-e^{-}.
\end{equation}%
Using the quantization condition for the dyonic 6D black F- string namely,%
\begin{equation}
\left( e^{+}g^{+}+g^{-}e^{-}\right) =2\pi k,\qquad k\text{ integer,}
\end{equation}%
one gets,%
\begin{equation}
\left( g^{+}g^{+}-g^{-}g^{-}\right) =eg=2\pi k.  \label{qn}
\end{equation}%
Then the the quantity $\left( Z^{+}\right) ^{2}-\left( Z^{-}\right) ^{2}$
becomes 
\begin{equation}
\left( Z^{+}\right) ^{2}-\left( Z^{-}\right) ^{2}=2eg,
\end{equation}%
being just the quantization condition of the electric/magentic charges of
the F- string in 6D space time.

\subsubsection{BFS attractor equation}

\qquad The attractor condition on the 6D field $\sigma $ modulus\ at the
horizon geometry of the 6D black F- string is obtained by minimizing the
potential $\mathcal{V}_{BFS}$. The corresponding attractor eq reads then as
follows:%
\begin{equation}
\left( \frac{d\mathcal{V}_{BFS}}{d\sigma }\right) =0,\qquad \left( \frac{%
d^{2}\mathcal{V}_{BFS}}{d\sigma ^{2}}\right) >0.
\end{equation}%
By help of eq(\ref{bfs}), we then have the following condition on the field
modulus $\sigma $ at the horizon geometry of the BFS:%
\begin{equation}
\frac{g^{2}}{2}\left[ \exp \left( -4\sigma \right) \right] _{\text{horizon}}=%
\frac{e^{2}}{2}\left[ \exp \left( 4\sigma \right) \right] _{\text{horizon}}.
\end{equation}%
Its solution is given by%
\begin{equation}
\left[ \exp \left( 4\sigma \right) \right] _{\text{horizon}}=\frac{g}{e}\geq
0.  \label{sol}
\end{equation}%
Notice that in 6D non chiral $\mathcal{N}=2$ supergravity the value of the
dilaton is no longer infinite as it is the case in $\mathcal{N}=\left(
1,0\right) $ and $\mathcal{N}=\left( 2,0\right) $ chiral supergravities with
dilaton field belonging to the tensor multiplet. Indeed switching off the
magnetic charge $g$ (or equivalently the electric charge $e$); i.e $g=0$ ($%
e=0$), we recover the standard picture%
\begin{equation}
\left( \left[ e^{4\sigma }\right] _{\text{horizon}}\right) _{\text{{\small 6D%
} chiral models}}=0,\qquad g=0,
\end{equation}%
which leads to the quite well known result\footnote{%
We thank S. Ferrara and A. Marrani for drawing our attention to this point.
\par
{}} 
\begin{equation}
\left( \left[ \sigma \right] _{\text{horizon}}\right) _{\text{{\small 6D}
chiral models}}=-\infty .
\end{equation}%
Now putting the solution eq(\ref{sol}) back into the central charge
relations (\ref{zz}), we obtain:%
\begin{eqnarray}
\left( Z^{+}\right) _{horizon} &=&\frac{1}{2}\left[ g\sqrt{\frac{e}{g}}+e%
\sqrt{\frac{g}{e}}\right] =\sqrt{eg},  \notag \\
\left( Z^{-}\right) _{horizon} &=&\frac{1}{2}\left[ g\sqrt{\frac{e}{g}}-e%
\sqrt{\frac{g}{e}}\right] =0.
\end{eqnarray}%
Therefore the value of the potential at the horizon is, 
\begin{equation}
\left( \mathcal{V}_{BFS}\right) _{\text{horizon}}=\frac{g^{2}}{2}\frac{e}{g}+%
\frac{e^{2}}{2}\frac{g}{e}=eg
\end{equation}%
which, up on using the quantization condition, can be expressed in terms of
the positive definite integer $k$ of eq(\ref{qn}). The analogous of the
Bekenstein-Hawking entropy $S_{BFS}^{\text{entropy}}$ of the 6D BFS is then
proportional to $eg$,%
\begin{equation}
S_{BFS}^{\text{entropy}}\sim G_{N}^{-\frac{3}{4}}\times \mathcal{A}_{\text{%
area}}\sim eg,
\end{equation}%
where $\mathcal{A}_{\text{area}}$ is the 3d- horizon area and $G_{N}$ is the
Newton constant in 6D. $S_{BFS}^{\text{entropy}}$ vanishes for $e=0$ or $g=0$
as predicted by chiral supergravity theories in 6D.

\subsection{6D Black Hole}

\qquad Contrary to the dyonic BFS, the 6D black hole is magnetically charged
under the $U^{4}\left( 1\right) \times U^{20}\left( 1\right) $ gauge group
symmetry generated by the gauge transformations of the $\left( 4+20\right) $
gauge fields of the 6D $\mathcal{N}=2$ gravity fields spectrum. \newline
Recall that in 6D, the electric charges are given, in terms of the field
strenght $F_{4a}$ and $F_{4I}$, by,%
\begin{eqnarray}
q_{a} &=&\int_{S^{4}}F_{4a},\qquad a=1,...,4,  \notag \\
p_{I} &=&\int_{S^{4}}F_{4I},\qquad a=1,...,20.
\end{eqnarray}%
The corresponding magnetic duals, which concern the black 2- brane, involve
the 2- form field strengths $F_{2}^{\Lambda }$ integrated over 2- sphere, 
\begin{eqnarray}
g^{a} &=&\int_{S^{2}}F_{2}^{a},\qquad a=1,...,4,  \notag \\
h^{I} &=&\int_{S^{2}}F_{2}^{I},\qquad a=1,...,20.
\end{eqnarray}%
Like for BFS, the charges $Q_{\Lambda }=\left( q_{a},p_{I}\right) $ are not
the physical ones. The physical charges; to be denoted like%
\begin{equation}
Z_{a},\qquad Z_{I},
\end{equation}%
appear dressed by the 6D scalar fields $\omega ^{aI}$ parameterizing the
moduli space of the 10D type IIA superstring on K3. Recall that the charges $%
Z_{a}$ and $Z_{I}$ appear respectively in the supersymmetric transformations
of the \emph{four} gravi-photinos/dilatinos and the \emph{twenty} photinos
of the U$^{20}\left( 1\right) $ Maxwell multiplet of the gauge-matter sector.

\subsubsection{Dressed charges}

The dressing of the \emph{twenty four} electric charges $\left(
q^{a},p^{I}\right) $ of the gauge fields $\left( \mathcal{A}_{\mu }^{a},%
\mathcal{A}_{\mu }^{I}\right) $ read as follows:%
\begin{eqnarray}
Z_{a} &=&e^{-\sigma }\left( Y_{ab}q^{b}+\omega _{aJ}p^{J}\right) ,  \notag \\
Z_{I} &=&e^{-\sigma }\left( V_{Ib}q^{b}+Y_{IJ}p^{J}\right) .  \label{czz}
\end{eqnarray}%
Using the real $24\times 24$ matrix $M_{\Lambda \Sigma }$, 
\begin{equation}
M_{\Lambda \Sigma }=e^{-\sigma }\times L_{\Lambda \Sigma },\qquad L_{\Lambda
\Sigma }=\left( 
\begin{array}{cc}
Y_{ab} & \omega _{aJ} \\ 
V_{Ia} & Y_{IJ}%
\end{array}%
\right) ,
\end{equation}%
that defines the moduli space $\widehat{G}$, the dressed charges $Z_{\Lambda
}=\left( Z_{a},Z_{I}\right) $ can be put in the condensed form%
\begin{eqnarray}
Z_{a} &=&M_{a\Sigma }Q^{\Sigma }=e^{-\sigma }L_{a\Sigma }Q^{\Sigma },  \notag
\\
Z_{I} &=&M_{I\Sigma }Q^{\Sigma }=e^{-\sigma }L_{I\Sigma }Q^{\Sigma }.
\label{dzz}
\end{eqnarray}%
Obviously not all the parameters carried by $L_{\Lambda \Sigma }$\ are
independent. The extra dependent degrees of freedom will be fixed by
imposing the $SO\left( 4,20\right) $ orthogonality constraint eqs and
requiring gauge invariance under $SO\left( 4\right) \times SO\left(
20\right) $. The factor $e^{-\sigma }$ of eq(\ref{czz}) is then associated
with the non compact abelian factor $SO\left( 1,1\right) $ considered
previously.\newline
Taking the $\eta _{\Lambda \Sigma }$ flat metric of the non compact group $%
SO\left( 4,20\right) $ as $\eta _{\Lambda \Sigma }=diag\left( 4\left(
+\right) ,20\left( -\right) \right) $, 
\begin{equation}
\eta _{\Lambda \Sigma }=\left( 
\begin{array}{cc}
\delta _{ab} & 0 \\ 
0 & -\delta _{IJ}%
\end{array}%
\right) ,  \label{met}
\end{equation}%
we can express all the $24\times 24=576$ real parameters $L_{\Lambda \Sigma
} $ in terms of eighty of them only; say $\omega _{aJ}$; i.e 
\begin{eqnarray}
Y_{cd} &=&\mathrm{f}\left( \omega _{aI}\right) ,\qquad a,b=1,...,4,  \notag
\\
Y_{JK} &=&\mathrm{g}\left( \omega _{aI}\right) ,\qquad I,J=1,...,20, \\
V_{Jb} &=&\mathrm{h}\left( \omega _{aI}\right) \qquad ,  \notag
\end{eqnarray}%
where $\mathrm{f}\left( \omega _{aI}\right) $, $\mathrm{g}\left( \omega
_{aI}\right) $ and $\mathrm{h}\left( \omega _{aI}\right) $ are some (non
linear) functions that can be worked out explicitly by solving the
constraint eqs on $L_{\Lambda \Sigma }$ orthogonal matrix. Indeed by solving
the constraint eqs 
\begin{equation}
L^{t}\eta L=\eta ,
\end{equation}%
for the $SO\left( 4,20\right) $ group elements, we obtain the following
identities,%
\begin{eqnarray}
Y_{ca}Y^{cb}-V_{Ka}V^{Kb} &=&\delta _{a}^{b},  \notag \\
Y_{KI}Y^{KJ}-\omega _{cI}\omega ^{cJ} &=&\delta _{I}^{J},  \label{csd}
\end{eqnarray}%
and 
\begin{equation}
Y_{ca}\omega ^{cI}=V_{Ja}Y^{JI}.  \label{sd}
\end{equation}%
Notice that the last eq gives the relation between $\omega _{cI}$ and $%
Y_{Jb} $. By introducing the inverse matrices $E^{ab}$ and $E_{IK}$ 
\begin{equation}
Y_{ca}E^{ab}=\delta _{c}^{b},\qquad Y^{JI}E_{IK}=\delta _{K}^{J},
\end{equation}%
we have either 
\begin{equation}
V_{Ia}=\omega ^{cJ}\left( Y_{ca}E_{JI}\right) ,
\end{equation}%
or%
\begin{equation}
\omega ^{aI}=V_{Jc}\left( Y^{JI}E^{ca}\right) .
\end{equation}%
The other constraint relations (\ref{csd}) can be used to fix $300$
parameters of the matrix $L_{\Lambda \Sigma }$ leaving then $276$
parameters. \newline
Moreover using the isotropy symmetry $SO\left( 4\right) \times SO\left(
20\right) $ of the moduli space one can reduce further this number to 
\begin{equation}
276-6-190=80
\end{equation}%
This gauge fixing is done by taking $Y_{ab}$ and $Y_{IJ}$ as given by $%
4\times 4$ and $20\times 20$ symmetric matrices respectively: 
\begin{equation}
Y_{ab}=Y_{ba},\qquad Y_{IJ}=Y_{JI}.
\end{equation}%
In this gauge, one can solve eqs(\ref{csd}) as follows,%
\begin{equation}
Y_{a}^{b}=\sqrt{\delta _{a}^{b}+\sum_{K=1}^{20}V_{Ka}V^{Kb}},\qquad
Y_{I}^{J}=\sqrt{\delta _{I}^{J}+\sum_{c=1}^{4}\omega _{cI}\omega ^{cJ}}.
\label{sal}
\end{equation}%
Notice moreover that setting,%
\begin{eqnarray}
Z_{a} &=&e^{-\sigma }R_{a},\qquad R_{a}=\left( L_{a\Sigma }Q^{\Sigma
}\right) ,  \notag \\
Z_{I} &=&e^{-\sigma }R_{I},\qquad R_{I}=\left( L_{I\Sigma }Q^{\Sigma
}\right) ,  \label{lz}
\end{eqnarray}%
as well as%
\begin{equation}
L_{\Sigma }^{\Upsilon }\cdot E_{\digamma }^{\Sigma }=L_{a}^{\Upsilon }\cdot
E_{\digamma }^{a}-L_{I}^{\Upsilon }\cdot E_{\digamma }^{I}=\delta _{\digamma
}^{\Upsilon },
\end{equation}%
one can compute a set of useful relations. In particular we have%
\begin{eqnarray}
dL_{\digamma \Lambda } &=&L_{\Upsilon \Lambda }\cdot \left( dL_{\Sigma
}^{\Upsilon }\right) \cdot P_{\digamma }^{\Sigma },  \notag \\
\nabla Z_{a} &=&\left( D^{H_{1}}Z_{a}+Z_{a}d\sigma \right) ,  \label{dz} \\
\nabla Z_{I} &=&\left( D^{H_{2}}Z_{I}+Z_{I}d\sigma \right) ,  \notag
\end{eqnarray}%
where%
\begin{eqnarray}
D^{H_{1}}Z_{a} &=&\left( dZ_{a}-\Omega _{a}^{b}Z_{b}\right) ,\qquad H_{1}=%
\mathcal{O}\left( 4\right) ,  \notag \\
D^{H_{2}}Z_{I} &=&\left( dZ_{I}-\Omega _{I}^{J}Z_{J}\right) ,\qquad H_{2}=%
\mathcal{O}\left( 4\right) ,  \label{bz}
\end{eqnarray}%
and where $\Omega _{a}^{b}$ and $P_{a}^{I}$\ are given by%
\begin{equation}
\Omega _{a}^{b}=E_{a}^{\Sigma }\cdot \left( dL_{\Sigma }^{b}\right) ,\qquad
P_{a}^{I}=E_{a}^{\Sigma }\cdot \left( dL_{\Sigma }^{I}\right) ,
\end{equation}%
together with similar relation for $\Omega _{I}^{J}$ and $P_{I}^{a}$. 
\newline
Using (\ref{dz}), we can write down the Maurer-Cartan eqs for the dressed
charge. They read as follows,%
\begin{equation}
\nabla Z_{a}=P_{a}^{I}Z_{I},\qquad \nabla Z_{I}=P_{I}^{a}Z_{a}.  \label{cm}
\end{equation}%
Notice in passing that $Z_{I}=0$ is a solution of $\nabla Z_{a}=0$. The same
property is valid for $Z_{a}=0$ which solves $\nabla Z_{I}=0$. These
properties will be used when we study the minimization of the black hole
potential.

\subsubsection{Effective black hole potential}

\qquad Using the dressed charges (\ref{czz}-\ref{dzz}), we can write down
the gauge invariant effective scalar potential $\mathcal{V}_{BH}$. Following 
\textrm{\cite{F2}}, this effective potential is given by the Weinhold
potential,%
\begin{equation}
\mathcal{V}_{BH}\left( \sigma ,L\right) =\left( Z_{a}Z^{a}\right) +\left(
Z_{I}Z^{I}\right) ,  \label{pot}
\end{equation}%
which can be also put in the form%
\begin{equation}
\mathcal{V}_{BH}\left( \sigma ,L\right) =e^{-2\sigma }\left[ \left(
R_{a}R^{a}\right) +\left( R_{I}R^{I}\right) \right] .
\end{equation}%
Clearly $\mathcal{V}_{BH}$, which is positive, is manifestly gauge invariant
under both: \newline
(\textbf{a}) the $U^{4}\left( 1\right) \times U^{20}\left( 1\right) $ gauge
transformations since the vectors $Z_{a}$ and $Z_{I}$ depend on the electric
charges of the field strengths only which, as we know, are gauge invariant.%
\newline
(\textbf{b}) the gauge transformations of the $SO\left( 4\right) \times
SO\left( 20\right) $ isotropy group of the moduli space. $\mathcal{V}_{BH}$
is given by scalar products of the vectors $Z_{a}$ and $Z^{a}$ (resp $Z^{I}$
and $Z_{I}$).\newline
Using eqs(\ref{czz}), we can express the black hole potential as follows:%
\begin{equation}
\mathcal{V}_{BH}=e^{-2\sigma }\left( q^{a}\mathcal{N}_{ab}q^{b}+q^{a}%
\mathcal{N}_{aJ}p^{J}+p^{I}\mathcal{N}_{Ib}q^{b}+p^{I}\mathcal{N}%
_{IJ}p^{J}\right) ,
\end{equation}%
or in a condensed manner like,%
\begin{equation}
\mathcal{V}_{BH}=e^{-2\sigma }Q^{\Lambda }\mathcal{N}_{\Lambda \Sigma
}Q^{\Sigma }
\end{equation}%
with%
\begin{equation}
\mathcal{N}_{\Lambda \Sigma }=\left( 
\begin{array}{cc}
\mathcal{N}_{ab} & \mathcal{N}_{aJ} \\ 
\mathcal{N}_{aJ} & \mathcal{N}_{IJ}%
\end{array}%
\right)
\end{equation}%
and%
\begin{eqnarray}
\mathcal{N}_{ab} &=&Y_{ca}Y_{b}^{c}+V_{Ka}V_{b}^{K}=\mathcal{N}_{ba}  \notag
\\
\mathcal{N}_{aJ} &=&Y_{ca}\omega _{J}^{c}+V_{a}^{J}Y_{JI}=\mathcal{N}_{Ja} 
\notag \\
\mathcal{N}_{Ib} &=&\omega _{I}^{c}Y_{cb}+Y_{IJ}V_{b}^{J}=\mathcal{N}_{bI}
\label{pr} \\
\mathcal{N}_{IJ} &=&Y_{KI}Y^{KJ}+\omega _{cI}\omega ^{cJ}=\mathcal{N}_{JI}, 
\notag
\end{eqnarray}%
together with the constraint relations (\ref{csd}-\ref{sd}). Notice that,
like for BFS, $\mathcal{N}_{\Lambda \Sigma }$ has a 6D filed theoretical
interpretation in terms of the {\LARGE \ }gauge coupling of the gauge field
strengths $\mathcal{F}_{\mu \nu }^{\Lambda }$; i.e a term like $\frac{1}{4}%
\sqrt{-g}\mathcal{N}_{\Lambda \Sigma }\mathcal{F}_{\mu \nu }^{\Lambda }%
\mathcal{F}^{\mu \nu \Sigma }$ appears in the component fields of the 6D $%
\mathcal{N}=2$ supergravity Lagrangian density.

\subsubsection{ 6D black hole attractors}

\qquad The attractor condition on the 6D field moduli at the horizon
geometry of the 6D black hole is obtained by minimizing the potential $%
\mathcal{V}_{BH}\left( \sigma ,L\right) $ with respect to the field moduli $%
\left( \sigma ,L\right) $. The variation of the potential $\mathcal{V}_{BH}$
can be put in the nice form,%
\begin{equation}
\delta \mathcal{V}_{BH}=\left( \frac{\partial \mathcal{V}_{BH}}{\partial
Z_{a}}\right) \nabla Z_{a}+\left( \frac{\partial \mathcal{V}_{BH}}{\partial
Z_{I}}\right) \nabla Z_{I},
\end{equation}%
where we have used eqs(\ref{dz}-\ref{bz}). Moreover, using eq(\ref{pot}), we
also have%
\begin{equation}
\delta \mathcal{V}_{BH}=2Z^{a}\nabla Z_{a}+2Z^{I}\nabla Z_{I},
\end{equation}%
together with the constraint relations%
\begin{equation}
\nabla Z_{a}=P_{a}^{I}Z_{I},\qquad \nabla Z_{I}=P_{I}^{a}Z_{a},
\end{equation}%
following from Maurer-Cartan relations (\ref{cm}), and%
\begin{equation}
Z_{a}Z^{a}-Z_{I}Z^{I}=e^{-2\sigma }Q^{2}.  \label{ZZ}
\end{equation}%
In the above relation, we have set%
\begin{eqnarray}
Q^{2} &=&Q^{\Lambda }\eta _{\Lambda \Sigma }Q^{\Sigma }=q^{2}-p^{2},  \notag
\\
q^{2} &=&\sum_{i=1}^{4}q_{a}q^{a},\qquad p^{2}=\sum_{I=1}^{20}p_{I}p^{I}.
\end{eqnarray}%
Eq(\ref{ZZ}) follows from the orthogonality condition $\left( L^{t}\eta
L\right) _{\Lambda \Sigma }=\eta _{\Lambda \Sigma }$. Up on multiplying both
sides of this condition by $Q^{\Lambda }Q^{\Sigma }$, that is,%
\begin{equation}
Q^{\Lambda }\left( L^{t}\eta L\right) _{\Lambda \Sigma }Q^{\Sigma
}=Q^{\Lambda }\eta _{\Lambda \Sigma }Q^{\Sigma },
\end{equation}%
we end with eq(\ref{ZZ}). Notice that eq(\ref{ZZ}) has an indefinite sign
since it can be either positive, null or negative in agreement with the sign
of the number $\left( q^{2}-p^{2}\right) $.\newline
The attractor condition for black hole read therefore 
\begin{eqnarray}
\left[ Z^{a}\left( \nabla Z_{a}\right) \right] _{\text{horizon}} &=&0, 
\notag \\
\left[ Z^{I}\left( \nabla Z_{I}\right) \right] _{\text{horizon}} &=&0,
\end{eqnarray}%
together with the constraint eqs,%
\begin{eqnarray}
\left[ \nabla Z_{a}\right] _{\text{horizon}} &=&\left[ P_{a}^{I}Z_{I}\right]
_{\text{horizon}},  \notag \\
\left[ \nabla Z_{I}\right] _{\text{horizon}} &=&\left[ P_{I}^{a}Z_{a}\right]
_{\text{horizon}},  \label{pz}
\end{eqnarray}%
as well as%
\begin{equation}
\left[ Z_{a}Z^{a}-Z_{I}Z^{I}\right] _{\text{horizon}}=\left( e^{-2\sigma
}\right) _{\text{horizon}}Q^{2}.
\end{equation}%
By substituting $Z_{a}$ and $Z_{I}$ by eqs(\ref{lz}-\ref{dz}-\ref{bz}), we
can also express these conditions in terms of the fields of the moduli space.

\emph{Solutions}\newline
Clearly one distinguishes three main classes of solutions minimizing the
black hole potential. These are given by%
\begin{eqnarray}
\left( 1\right) &:&\qquad Z_{a}=0,\qquad Z_{I}=0,  \notag \\
\left( 2\right) &:&\qquad Z_{a}=0,\qquad \nabla Z_{I}=0, \\
\left( 3\right) &:&\qquad Z_{I}=0,\qquad \nabla Z_{a}=0.  \notag
\end{eqnarray}%
The case $\nabla Z_{a}=0$ and $\nabla Z_{I}=0$ is the same as the case (1)
because of the constraint relations (\ref{pz}). \newline
(\textbf{1}) \textbf{Case} $Z_{a}=Z_{I}=0$\newline
This solution corresponds to the two following possibilities:\newline
(\textbf{a}) the field $\sigma \rightarrow \infty $ whatever the other
moduli fields $L_{\Lambda \Sigma }$ are.\newline
(\textbf{b}) the fields $L_{\Lambda \Sigma }=Q_{\Lambda }Q_{\Sigma }$ with $%
Q^{2}=0$.\newline
In both cases (\textbf{a}) and (\textbf{b}), there is no attractor solution.
The value of the potential at the horizon is 
\begin{equation}
\left( \mathcal{V}_{BH}\right) _{\text{horizon}}=0,
\end{equation}%
and so there is no black hole entropy. \newline
(\textbf{2}) \textbf{Case} $Z_{a}=0,$ $\nabla Z_{I}=0$\newline
In this case, a non zero entropy solution can be given by taking the value
dilaton $\sigma $ arbitrary but finite; and the $L_{\Lambda \Sigma }$ moduli
like, 
\begin{equation}
\left( L_{\Lambda \Sigma }\right) _{\text{horizon}}Q^{\Sigma }=0.
\end{equation}%
A remarkable candidate for the value of the dilaton $\left( \sigma \right) _{%
\text{horizon}}^{\text{BH}}$ consists to take it the same as the value of
the black hole horizon of the BFS%
\begin{equation}
\left( \sigma \right) _{\text{horizon}}^{\text{BH}}=\left[ \exp \left(
-2\sigma \right) \right] _{\text{horizon}}^{\text{Black-F- string}}=\sqrt{%
\frac{e}{g}}.  \label{sig}
\end{equation}%
The remaining $L_{\Lambda \Sigma }$ moduli, which solve $L_{\Lambda \Sigma
}Q^{\Sigma }=0$, are as follows,%
\begin{equation}
\left[ L_{\Lambda \Sigma }\right] _{\text{horizon}}=\left( \frac{Q_{\Lambda
}Q_{\Sigma }}{Q^{2}}-\eta _{\Sigma \Lambda }\right) ,\qquad Q^{2}\neq 0.
\end{equation}%
Moreover, since $Z_{a}=0$, the constraint relation 
\begin{equation}
\left[ Z_{a}Z^{a}-Z_{I}Z^{I}\right] _{\text{horizon}}=Q^{2}\left(
e^{-2\sigma }\right) _{\text{horizon}},
\end{equation}%
reduces to, 
\begin{eqnarray}
\left[ Z_{I}Z^{I}\right] _{\text{horizon}} &=&-Q^{2}\left( e^{-2\sigma
}\right) _{\text{horizon}},\qquad  \notag \\
\left( Z_{I}\right) _{\text{horizon}} &=&\frac{p_{I}\left( e^{-\sigma
}\right) _{\text{horizon}}}{\sqrt{\sum_{J=1}^{20}p_{J}p^{J}}}\sqrt{%
\left\vert Q^{2}\right\vert }.
\end{eqnarray}%
So the $q_{a}$ and $p_{I}$ charges should be constrained like%
\begin{equation}
Q^{2}=\left( q^{2}-p^{2}\right) <0.
\end{equation}%
Therefore the value of the potential at the BH horizon is 
\begin{equation}
\left( \mathcal{V}_{BH}\right) _{\text{horizon}}=\left\vert Q^{2}\right\vert
\left( e^{-2\sigma }\right) _{\text{horizon}}=\left( p^{2}-q^{2}\right)
\left( e^{-2\sigma }\right) _{\text{horizon}}>0.
\end{equation}%
It is proportional to the norm $\left\vert Q^{2}\right\vert $ of the central
charge vector $Q=\left( q,p\right) $. Using eq(\ref{sig}), we get 
\begin{equation}
\left( \mathcal{V}_{BH}\right) _{\text{horizon}}=\left\vert \left(
\sum_{I=1}^{20}p_{I}^{2}-\sum_{a=1}^{4}q_{a}^{2}\right) \right\vert \sqrt{%
\left\vert \frac{e}{g}\right\vert }.
\end{equation}%
(\textbf{3}) Case $Z_{I}=0,$ $\nabla Z_{a}=0$\newline
This situation is quite analogous to the previous one. Non zero entropy
solution corresponds to finite values of the dilaton $\left( \sigma \right)
_{\text{horizon}}$ which can be taken as in the BFS eq(\ref{sig}); $\left(
\sigma \right) _{\text{horizon}}=\sqrt{\left\vert e/g\right\vert }$.
Moreover, because of the constraint relation, 
\begin{equation}
\left[ Z_{a}Z^{a}-Z_{I}Z^{I}\right] _{\text{horizon}}=Q^{2}\left(
e^{-2\sigma }\right) _{\text{horizon}},
\end{equation}%
which reduces to%
\begin{equation}
\left[ Z_{a}Z^{a}\right] _{\text{horizon}}=Q^{2}\left( e^{-2\sigma }\right)
_{\text{horizon}},
\end{equation}%
the $q_{a}$ and $p_{I}$ charges should be like%
\begin{equation}
Q^{2}=\left\vert Q^{2}\right\vert =\left( q^{2}-p^{2}\right) >0.
\end{equation}%
The solution for $Z_{a}$ in terms of the central charges is given by%
\begin{equation}
\left( Z_{a}\right) _{\text{horizon}}=\frac{q_{a}\left( e^{-\sigma }\right)
_{\text{horizon}}}{\sqrt{\sum_{b=1}^{4}q^{b}q_{b}}}\sqrt{\left\vert
Q^{2}\right\vert }.
\end{equation}%
As expected the value of the potential on the black hole horizon is $\left( 
\mathcal{V}_{BH}\right) _{\text{horizon}}=\left\vert Q^{2}\right\vert \left(
e^{-2\sigma }\right) _{\text{horizon}}$.\newline
\emph{Beyond Weinhold potential}\newline
We end this study by making two comments: \newline
(\textbf{1}) the Weinhold potential we have been using above has a natural
extension given by,%
\begin{equation}
\mathcal{V}_{BH}\left( \sigma ,L\right) =\left( Z_{a}Z^{a}\right) +\left(
Z^{I}d_{IJ}Z^{J}\right)  \label{nwh}
\end{equation}%
where the $20\times 20$ matrix $d_{IJ}$ stand for the intersection matrix of
the $h^{1,1}$ cycles of K3. Since $d_{IJ}$ can be also defined as the scalar
of some real 20- dimensional vector basis $\left\{ \alpha _{I}\right\} $
like $d_{IJ}=\alpha _{I}\cdot \alpha _{J}$, the black hole potential $%
\mathcal{V}_{BH}\left( \sigma ,L\right) $ is positive. \newline
The constraint relations that go with eq(\ref{nwh}) can be obtained by using
the relations given for the Weinhold potential and substitute the $\delta
_{IJ}$ metric by $d_{IJ}$.\newline
(\textbf{2}) The results for the 6D black hole derived above are valid as
well for the potential eq(\ref{nwh}); all one has to do is to replace the
metric $\eta _{\Lambda \Sigma }$ (\ref{met}) by the new one,%
\begin{equation}
G_{\Lambda \Sigma }=\left( 
\begin{array}{cc}
\delta _{ab} & 0 \\ 
0 & -d_{IJ}%
\end{array}%
\right) ,
\end{equation}%
and think about $Q^{2}$ as given by $Q^{2}=Q^{\Lambda }G_{\Lambda \Sigma
}Q^{\Sigma }$ $=q^{a}\delta _{ab}q^{b}-p^{I}d_{IJ}p^{J}$. The generic
formula for 6D black hole entropy reads as%
\begin{equation}
\mathcal{S}_{BH}^{\text{entropy}}\sim \left\vert \left(
\sum_{I=1}^{20}p^{I}d_{IJ}p^{J}-\sum_{a=1}^{4}q_{a}^{2}\right) \right\vert 
\sqrt{\left\vert \frac{e}{g}\right\vert },
\end{equation}%
where $e$ and $g$ are as before.

\subsection{7D black attractors}

\qquad Here we discuss briefly the effective scalar potential and attractor
mechanism of the black objects in 7D.\ This study is quite similar to the
previous 6D analysis. \newline
Recall that the moduli space of this theory is given by%
\begin{equation}
\frac{SO\left( 3,19\right) }{SO\left( 3\right) \times SO\left( 19\right) }%
\times SO\left( 1,1\right) .  \label{o11}
\end{equation}%
Here an interesting property emerges, the dilaton $\sigma $ parameterizing $%
SO\left( 1,1\right) $ has an interpretation in term of the volume of the K3
surface. Recall also that in 7D space time, the bosonic fields content of
the $\mathcal{N}=2$ supergravity multiplet is given by%
\begin{equation}
\left( g_{\mu \nu },\text{ \ }B_{\left[ \mu \nu \right] },\text{ \ }\mathcal{%
A}_{\mu }^{a},\text{ \ }\sigma \right) ,\qquad a=1,2,3,\qquad \mu ,\nu ,\rho
=0,...,6,
\end{equation}%
where $B_{\left[ \mu \nu \right] }$ is dual to a 3- form gauge field\ $C_{%
\left[ \mu \nu \sigma \right] }$. There is also nineteen U$\left( 1\right) $
Maxwell with the following 6D bosons: 
\begin{equation}
\left( \mathcal{A}_{\mu }^{I}\text{ \ },\text{ \ }\rho ^{aI}\right) ,\qquad
a=1,2,3,\qquad I=1,...,19,\text{\ }
\end{equation}%
where $\rho ^{aI}$ capture $3\times 19$ degrees of freedom. The gauge
invariant $\left( p+2\right) $- forms of the 7D $\mathcal{N}=2$ supergravity
are given by%
\begin{equation}
H_{3}\sim dB_{2},\qquad \mathcal{F}_{2}^{a}\sim d\mathcal{A}^{a},\qquad 
\mathcal{F}_{2}^{I}\sim d\mathcal{A}^{I}.
\end{equation}%
Extending the above 6D study to the 7D case, one distinguishes:\newline
(\textbf{i}) 7D black 2- brane (black membrane BM)\newline
The effective scalar potential of the BM is 
\begin{equation}
\mathcal{V}_{BM}^{7D}\left( \sigma \right) \sim Z^{2}=e^{-4\sigma }g^{2},
\end{equation}%
with%
\begin{equation}
g=\int_{S^{3}}H_{3}.
\end{equation}%
The extremum of this potential is given by $\sigma =\infty $. The value of
the potential at the minimum is 
\begin{equation}
\left[ \mathcal{V}_{BM}^{7D}\left( \infty \right) \right] _{\min }=0,
\end{equation}%
and so the entropy vanishes identically.\newline
(\textbf{ii}) 7D black hole: \newline
The Weinhold potential of this black hole is given by%
\begin{equation}
\mathcal{V}_{BH}^{7D}\left( \sigma ,L\right)
=\sum_{a=1}^{3}Z_{a}Z^{a}+\sum_{I=1}^{19}Z_{I}Z^{I},  \label{ral}
\end{equation}%
where%
\begin{eqnarray}
Z_{a} &=&e^{-\sigma }L_{a\Lambda }g^{\Lambda },  \notag \\
Z_{I} &=&e^{-\sigma }L_{a\Lambda }g^{\Lambda },  \label{rel}
\end{eqnarray}%
satisfying the constraint relation,%
\begin{equation}
\sum_{a=1}^{3}Z_{a}Z^{a}-\sum_{I=1}^{19}Z_{I}Z^{I}=Q^{2},\qquad \left(
\sum_{a=1}^{3}q_{a}q^{a}-\sum_{I=1}^{19}p_{I}p^{I}\right) =Q^{2}
\end{equation}%
and $Q^{\Lambda }=\left( q^{a},p^{I}\right) $ with 
\begin{eqnarray}
q^{a} &=&\int_{S^{2}}\mathcal{F}_{2}^{a},\qquad a=1,2,3,  \notag \\
p^{I} &=&\int_{S^{2}}\mathcal{F}_{2}^{I},\qquad I=1,...,19.
\end{eqnarray}%
The real $22\times 22$ matrix 
\begin{equation}
L_{a\Lambda }=\left( 
\begin{array}{cc}
L_{ab} & \rho _{aI} \\ 
V_{Ia} & L_{IJ}%
\end{array}%
\right) ,
\end{equation}%
is associated with the group manifold $SO\left( 3,19\right) /SO\left(
3\right) \times SO\left( 19\right) $. It is an orthogonal matrix satisfying%
\begin{equation}
L^{t}\eta L=\eta ,\qquad \eta =diag\left[ 3\left( +\right) ,19\left(
-\right) \right] .
\end{equation}%
The $SO\left( 3\right) \times SO\left( 19\right) $ symmetry can be used to
choose $L_{ab}$ and $L_{IJ}$\ matrices as follows:%
\begin{equation}
L_{ab}-L_{ba}=0,\qquad L_{IJ}-L_{JI}=0.
\end{equation}%
Putting the relations (\ref{rel}) back into (\ref{ral}), we get%
\begin{equation}
\mathcal{V}_{BH}^{7D}\left( \sigma ,L\right) =e^{-2\sigma }Q^{\Lambda }%
\mathcal{N}_{\Lambda \Sigma }Q^{\Sigma },\qquad
\end{equation}%
where%
\begin{equation}
\mathcal{N}_{\Lambda \Sigma }=\left( L_{a\Lambda }L_{\Sigma
}^{a}+L_{I\Lambda }L_{\Sigma }^{I}\right) .
\end{equation}%
The attractor equations following from the extremum of the $\mathcal{V}%
_{BH}^{7D}\left( \sigma ,L\right) $, 
\begin{equation}
\delta \mathcal{V}_{BH}^{7D}=2\sum_{a=1}^{3}Z_{a}\nabla
Z^{a}+2\sum_{I=1}^{19}Z_{I}\nabla Z^{I}=0,
\end{equation}%
have the following solutions:\newline
(\textbf{i}) Case 1:$\qquad $\newline
In this case we have $\sigma =\infty $ whatever $L_{\Sigma \Lambda }$ is or $%
\sigma =\sigma _{0}=finite$ and $L_{\Sigma \Lambda }=Q_{\Lambda }Q_{\Sigma }$
with $Q^{2}=0$. In both cases, we have%
\begin{equation}
\left( \mathcal{V}_{BH}^{7D}\left( \sigma =\infty ,L\right) \right) _{\min
}=0,  \label{ssl}
\end{equation}%
and%
\begin{equation}
\left( \mathcal{V}_{BH}^{7D}\left( \sigma _{0},L_{\Sigma \Lambda
}=Q_{\Lambda }Q_{\Sigma }\right) \right) _{\min }=0
\end{equation}%
Since $\sigma $ is just the volume of K3, the solution (\ref{ssl})
corresponds to a large volume limit of K3 and the physically un-interesting. 
\newline
(\textbf{ii}) Case 2:$\qquad \sigma =\sigma _{0}=finite$ and%
\begin{equation}
Z^{a}=0,\qquad \nabla Z^{I}=0,\qquad Z^{I}\neq 0:\qquad L_{\Sigma
}^{a}Q^{\Lambda }=0.
\end{equation}%
(\textbf{iii}) Case 3:$\qquad \sigma =\sigma _{0}=finite$ and%
\begin{equation}
Z^{a}\neq 0,\qquad \nabla Z^{a}=0,\qquad Z^{I}=0:\qquad L_{\Sigma
}^{I}Q^{\Lambda }=0
\end{equation}%
Like in the case of 6D, we have%
\begin{eqnarray}
\text{ {\small Case 2}:\qquad } &&\left( \mathcal{V}_{BH}^{7D}\right) _{\min
}=-Z_{I}Z^{I}=e^{-2\sigma _{0}}\left\vert Q\right\vert ^{2},  \notag \\
\text{ {\small Case 3}:\qquad } &&\left( \mathcal{V}_{BH}^{7D}\right) _{\min
}=Z_{a}Z^{a}=e^{-2\sigma _{0}}\left\vert Q\right\vert ^{2},
\end{eqnarray}%
which depend on $\left\vert Q\right\vert ^{2}=\left\vert
q_{a}q^{a}-p_{I}p^{I}\right\vert $; but also on the inverse of the volume of
K3.

\section{Conclusion and Discussion}

\qquad In this paper we have studied six dimensional $\mathcal{N}=2$
supersymmetric black attractors (black hole, black F-string, black 2- brane)
and their uplifting to seven dimensions. These backgrounds arise as large
distance limits of 10D type IIA superstring (11D M-theory) on K3 and may
encode informations on $\mathcal{N}=4$ black hole attractors in four
dimensions considered recently in \cite{Sen1,Sen2}. After revisiting some
general results on 10D type II superstrings compactificatied on the K3
surface and $\mathcal{N}=2$ supersymmetry in six dimensional space-time, we
have developed a matrix method to exhibit manifestly the special
quaternionic structure of 10D type IIA superstring on the K3 surface. This
matrix formulation relies on the homomorphism between the 4-vector
representation of $SO\left( 4\right) $ and the $\left( \frac{1}{2},\frac{1}{2%
}\right) $ representation of $SU\left( 2\right) \times SU^{\prime }\left(
2\right) $. By identification of the two $SU\left( 2\right) $ factors, the $%
SO\left( 4\right) $ becomes $SU^{2}\left( 2\right) $, the 4- vector
representation, which reads now as $2\otimes 2$, split as $1\oplus 3$ and
the $SO\left( 4\right) $ invariance of the moduli space of type IIA
superstring on the K3 surface is then completely captured by representations
of the basic $SU\left( 2\right) $ symmetry.

Our 2$\times $2 matrix formulation has been shown to be an adequate method
to deal with the underlying special quaternionic geometry of the moduli
space of the 10D type IIA superstring on the K3 surface. The non abelian $%
2\times 2$ matrix formalism with typical moduli 
\begin{equation}
w=y+i\sum_{m=0,\pm 1}\sigma ^{m}x^{m},
\end{equation}%
where $\sigma ^{m}$\ are the usual Pauli 2$\times $2 matrices and 
\begin{equation}
y=\int_{C_{2}}B^{NS},\qquad x^{m}=\int_{C_{2}}\Omega ^{\left( 1+m,1-m\right)
},
\end{equation}%
have a formal similarity with the usual complex formalism 
\begin{equation}
w\qquad \leftrightarrow \qquad z,
\end{equation}%
with 
\begin{equation}
z=y+ix^{0},\qquad y=\int_{C_{2}}B^{NS},\qquad x^{0}=\int_{C_{2}}\Omega
^{\left( 1,1\right) },
\end{equation}%
being the usual complex Kahler moduli of the special Kahler geometry of 10D
type IIA superstring on Calabi-Yau threefolds. The matrix method developed
in present study has allowed us to compute explicitly the hyperKahler
potential 
\begin{equation}
\mathcal{H}=\mathrm{Tr}\left[ \ln \left( \boldsymbol{V}_{0}-\boldsymbol{S}%
\right) \right]  \label{tra}
\end{equation}%
in terms of the \emph{"volume"} $\boldsymbol{V}_{0}$ of the K3 surface and
an isoquintet described by a $3\times 3$ traceless symmetric matrix $%
\boldsymbol{S,}$ see also the discussion around eq(\ref{hyp}). The hermitian
hyperKahler potential $\mathcal{H}$ has remarkable properties which deserves
more analysis. Let us comment briefly some of these specific features
herebelow: \newline
(1) $\mathcal{H}$ is given by the trace of $3\times 3$ matrix $\left( 
\boldsymbol{V}_{0}\mathrm{I}_{3\times 3}-\boldsymbol{S}\right) $. The real
number $\boldsymbol{V}_{0}$, which is proportional to the volume of K3 ($%
\boldsymbol{V}_{K3}=6\boldsymbol{V}_{0}$), has the following realization in
term of the geometric moduli $x^{mI}$, 
\begin{equation}
\boldsymbol{V}_{0}=\frac{1}{6}\sum_{m=1}^{3}\left( \sum_{I,J=0}^{h_{%
{\scriptsize K3}}^{1,1}}x^{mI}d_{IJ}x^{mJ}\right) ,
\end{equation}%
where $d_{IJ}$ is the intersection matrix of 2-cycles in the K3 surface. The 
$3\times 3$ matrix $\boldsymbol{S}$ is an isoquintet captured by the
symmetrization of the tensor product $\sigma ^{m}\sigma ^{n}$ of the Pauli
matrices and has the following realization in terms of the geometric moduli: 
\begin{equation}
\boldsymbol{S=}\frac{1}{6}\left( \sum_{I,J=0}^{h_{{\scriptsize K3}%
}^{1,1}}\sum_{k=1}^{3}x^{kI}d_{IJ}x^{kJ}\right) \delta ^{mn}-\frac{1}{2}%
\sum_{m,n=1}^{3}\sigma ^{m}\sigma ^{n}\left( \sum_{I,J=0}^{h_{{\scriptsize K3%
}}^{1,1}}x^{mI}d_{IJ}x^{nJ}\right) .
\end{equation}%
Notice also that the trace in eq(\ref{tra}) is required by gauge invariance
under the $SO\left( 4\right) =SU^{2}\left( 2\right) $ isometry of the moduli
space $SO\left( 4,20\right) /SO\left( 4\right) \times SO\left( 20\right) $.%
\newline
(2) The quantities $\boldsymbol{V}_{0}$ and $\boldsymbol{S}$ are also
invariant under the $SO\left( 20\right) $ gauge symmetry which rotates the
twenty Kahler moduli of K3. \newline
(3) The form (\ref{tra}) of the hyperKahler potential $\mathcal{H}$ of the
special hyperKahler geometry of type IIA superstring on the K3 surface can
be expanded into a power series as follows 
\begin{equation}
\mathcal{H}=\mathrm{Tr}\left[ \ln \boldsymbol{V}_{0}\right] -\sum_{n\geq 1}%
\frac{g^{n}}{n}\mathrm{Tr}\left( \boldsymbol{S}^{n}\right) ,  \label{exp}
\end{equation}%
where the coupling parameter $g$ is given by the inverse of the volume of
the K3 surface: 
\begin{equation}
g=\frac{1}{\boldsymbol{V}_{0}},
\end{equation}%
Notice that the leading term $\mathrm{Tr}\left[ \ln \boldsymbol{V}_{0}\right]
$ appears as just the Kahler component and next leading vanishes identically
since $\mathrm{Tr}\left( \boldsymbol{S}\right) =0$. This property could be
interpreted as capturing the Ricci flat condition. Higher terms might be
interpreted as the geometric correction one has to have in order to bring a
Kahler manifold to a hyperKahler one. \newline
(4) The expansion can be also viewed as a perturbative definition of the
potential of HyperKahler geometry in terms the potential Kahler geometry $%
\mathrm{Tr}\left[ \ln \boldsymbol{V}_{0}\right] $ plus extra contribution
captured by the trace of powers of isoquintet representation $\boldsymbol{S}$%
.

We have also computed the corresponding \textit{"holomorphic"} matrix
prepotential $\mathcal{G}\left( w\right) $ which is found to be quadratic in
the quaternionic matrix moduli 
\begin{equation}
\mathcal{G}\left( w\right) =\sum_{I=1}^{20}w^{I}d_{IJ}w^{J},
\end{equation}%
in agreement with the structure of the intersection matrix of the 2- cycles
within the K3 surface. The potential $\mathcal{G}\left( w\right) $ is
invariant under $SO\left( 20\right) $ gauge invariance; but transforms in
the adjoint of $SU\left( 2\right) $. We have studied as well the vacuum
background of 10D type IIA superstring on K3 by considering also extra
contributions coming from the non zero topological classes $\left[ \frac{%
\mathcal{F}_{2}}{2\pi }\right] $ and $\left[ \frac{\mathcal{F}_{4}}{2\pi }%
\right] $ associated with the fields strengths $\mathcal{F}_{2}=d\mathcal{A}%
_{1}$ and $\mathcal{F}_{4}=d\mathcal{A}_{3}$ of the $1$- form and $3$- form
gauge fields $\mathcal{A}_{1}$ and $\mathcal{A}_{3}$. These non zero fluxes
on the K3 surface have allowed us to determine the potential of the 6D
supersymmetric black attractors and to get the explicit moduli expression of
the isotriplet $Z^{m}$ of central charges of the 6D $\mathcal{N}=2$
supersymmetric algebra given by eqs(\ref{b1}-\ref{b3}).

We end this study by making four more comments: \newline
The first comment concerns the attractor mechanism of the 6Dblack objects.
The entropy relations following from solving the attractor eqs of the black
f-string and black hole are respectively given by 
\begin{eqnarray}
\mathcal{S}_{\text{black f-string}}^{\text{entropy}} &=&\frac{1}{4}%
\left\vert eg\right\vert , \\
\mathcal{S}_{\text{black hole}}^{\text{entropy}} &=&\frac{1}{4}\left\vert
Q^{2}\right\vert e^{-\sigma _{0}},
\end{eqnarray}%
where $e$ and $g$ are the electric charge and magnetic charge of the 6D
dyonic F-string and $Q^{2}=\left(
\sum_{a=1}^{4}q_{a}^{2}-\sum_{I=1}^{20}p_{I}^{2}\right) $. The charge vector 
$\left( q_{a},p_{I}\right) $ defines the magnetic charge of the 6D black
hole. Notice that $e^{-\sigma _{0}}$\ may be a finite number and can be
taken as $e^{-\sigma _{0}}=\sqrt{\left\vert e/g\right\vert }$. Notice also
that value $\mathcal{S}_{\text{black hole}}^{\text{entropy}}$ can be zero
for $Q^{2}=0$ and or $\sigma _{0}=\infty $. \newline
The second comment concerns the link with the analysis on c-map developed in 
\textrm{\cite{RO1,RO2}}. From the relation between Higgs branch of 10D type
II superstrings on CY3$\times S^{1}$ and 6D\ $N=\left( 1,0\right) $
hypermultiplets, we suspect that the matrix formulation developed in the
present paper could be also used to approach the quaternion-Kahler geometry
underlying the hypermultiplet moduli space of type II superstrings on $%
CY3\times S^{1}$ and the c-map considered in the above mentioned references.%
\newline
The third comment deals with the harmonic space formulation of the special
hyperKahler structure of the moduli space of 10D type IIA\ superstring on
the K3 surface. There, one uses the identification $SU\left( 2\right) \sim
S^{3}$ to \emph{"geometrize"} the $SU\left( 2\right) $ R-symmetry. The 6D
Minkowski space-time $M_{6D}$ with local coordinates $\left( x^{\mu }\right) 
$ and $SO\left( 1,5\right) $ symmetry gets promoted to 
\begin{equation}
M_{6D}\times S^{3},\qquad S^{3}\sim S^{2}\times S^{1},
\end{equation}%
with local coordinates $\left( x^{\mu },u_{i}^{\pm }\right) $. Here the
complex isodoublets $u_{i}^{\pm }$ are harmonic variables parameterising $%
S^{2}\times S^{1}$ whose defining equation as a real 3-dimensional
hypersurface in the complex plane $C^{2}$ is 
\begin{equation}
\sum_{i,j=1}^{2}\frac{1}{2}\epsilon _{ij}\left(
u^{+i}u^{-j}-u^{+j}u^{-i}\right) =1.
\end{equation}%
In terms of these variables, the quaternionified 2-form $\mathcal{J}$ is
given by a special a function on $S^{2}$ as shown below: 
\begin{equation}
\mathcal{J}\left( u\right) =u_{i}^{+}u_{j}^{-}\mathcal{J}%
^{ij}=u_{[i}^{+}u_{j]}^{-}\mathcal{J}^{\left[ ij\right]
}+u_{(i}^{+}u_{j)}^{-}\mathcal{J}^{\left( ij\right) },
\end{equation}%
where one recognizes the antisymmetric part associated with NS-NS B-field
contribution and the geometric one associated with the symmetric part. With
this harmonic 2-form together with other tools that can be found in \cite%
{BS2}, one can go ahead and study the special hyperKahler geometry of the
moduli space of 10D type IIA superstring on the K3 surface.\newline
The fourth comment concerns the study of 6D\ supersymmetric black string. As
noted before, this should be described in the framework of 10D type IIB
superstring on the K3 surface with the field theoretical limit given by 6D $%
\mathcal{N}=\left( 2,0\right) $ chiral supersymmetric gauge theory and the
moduli space given by eq(\ref{2b}). However a quick inspection of the graded
commutation relations of the 6D $\mathcal{N}=\left( 2,0\right) $
superalgebra reveals that the 6D $\mathcal{N}=\left( 2,0\right) $ chiral
superalgebra is very special. It allows no central charges which are
isosinglets under space-time symmetry; see eqs(\ref{chi}). As such, there is
apparently no flux potential of the kind we have obtained in the context of
6D black holes and 6D black 2- branes of 10D type IIA on K3. The fact that
the 6D $\mathcal{N}=\left( 2,0\right) $ chiral superalgebra is special can
be also viewed on its vector representation. Indeed, recall that the on
shell degrees of freedom of the gauge multiplet in the chiral 6D $\mathcal{N}%
=\left( 2,0\right) $ supersymmetry are given, in terms of the $\left(
j_{1},j_{2}\right) $ representations of the $SO\left( 4\right) $ group
transverse to $SO\left( 1,1\right) $ with $SO\left( 1,5\right) $, as follows 
\begin{equation}
\left( 1,0\right) ,\text{ \quad }\left( \frac{1}{2},0\right) ^{4},\text{
\quad }\left( 0,0\right) ^{5},  \label{sm}
\end{equation}%
where there is no standard 1-form gauge field. The representation $\left(
1,0\right) $, which captures three on shell degree of freedom, is described
by a self dual antisymmetric tensor field $B_{\left[ \mu \nu \right] }^{+}$.
The above multiplet should be compared with the gauge multiplet of the non
chiral 6D $\mathcal{N}=\left( 1,1\right) $ superalgebra 
\begin{equation}
\left( \frac{1}{2},\frac{1}{2}\right) ,\text{ \quad }\left( \frac{1}{2}%
,0\right) ^{2},\text{ \quad }\left( 0,\frac{1}{2}\right) ^{2},\text{ \quad }%
\left( 0,0\right) ^{4},  \label{ga}
\end{equation}%
where $\left( \frac{1}{2},\frac{1}{2}\right) $ stands for the gauge field
and where the four scalar have been interpreted in section 2, eq(\ref{213})
as sgauginos along the central charge directions. \textrm{It is then an
interesting task to study 6D\ black D-string attractor mechanism in the
context of type IIB superstring. Progress in this direction will be reported
elsewhere.}

\section{Appendix: $\mathcal{N}=2$ supersymmetry in 6D}

\qquad Because of the self conjugation property of Weyl spinors in 6D, one
distinguishes two kinds of $\mathcal{N}=2$ supersymmetric algebras in six
dimensional space time:\ \newline
(1) the chiral $\mathcal{N}=\left( 2,0\right) $ supersymmetric algebra.%
\newline
(2) the non chiral $\mathcal{N}=\left( 1,1\right) $ supersymmetric algebra. 
\newline
These superalgebras appear as world volume supersymmetries in NS-NS 5-branes
of type IIA and IIB superstring theories. They are also graded symmetries of
6D supersymmetric gauge theories describing the low energy limit of the
compactification of 10D type IIA and IIB superstrings on K3.

6D $\mathcal{N}=\left( 1,1\right) $ and $\mathcal{N}=\left( 2,0\right) $ can
be obtained by special reductions of the 10D IIA and IIB superalgebras down
to 6D by keeping only half of the 32 original supercharges. They can be also
obtained by taking the tensor product of two 6D $\mathcal{N}=\left(
1,0\right) $ ( $\mathcal{N}=\left( 0,1\right) $) chiral superalgebras.
Recall that 6D $\mathcal{N}=\left( 1,0\right) $ supersymmetry is the basic
superalgebra in six space time dimensions, it is the symmetry of the field
theoretical limit of 10D heterotic and type I superstrings on K3. Below, we
give the main relations of these superalgebras.

\subsection{6D $\mathcal{N}=\left( p,q\right) $ superalgebra}

\qquad To begin, we consider $\mathcal{N}=1$ supersymmetry in ten
dimensional space time and work out the reduction down to six dimensions.
Then, we specify to the 10D $\mathcal{N}=2$ case on K3 where $p$ and $q$
positive integers to $p+q=2$.

\subsubsection{Dimensional reduction}

\qquad In the compactification of 10D space time $\mathcal{M}_{10}$ down to
6D space time $\mathcal{M}_{6}$, ($\mathcal{M}_{10}\sim \mathcal{M}%
_{10}\times \mathcal{K}_{4}$, where $\mathcal{K}_{4}$ is a real four compact
manifold), the $SO\left( 1,9\right) $ Lorentz group gets reduced down to 
\begin{equation}
SO\left( 1,9\right) \rightarrow SO\left( 1,5\right) \times SU\left( 2\right)
\times SU^{\prime }\left( 2\right) ,
\end{equation}%
where we have substituted $SO\left( 4\right) $ by $SU\left( 2\right) \times
SU^{\prime }\left( 2\right) $. The 10D vector and the two 10D Majorana-Weyl
spinors $16$ and $16^{\prime }$ decompose as follows:%
\begin{eqnarray}
10 &=&\left( 6,1,1^{\prime }\right) \oplus \left( 1,2,2^{\prime }\right) , 
\notag \\
16 &=&\left( 4,2,1^{\prime }\right) \oplus \left( 4^{\prime },1,2^{\prime
}\right) , \\
16^{\prime } &=&\left( 4,1,2^{\prime }\right) \oplus \left( 4^{\prime
},2,1^{\prime }\right) ,  \notag
\end{eqnarray}%
where $\left( \ast ,2,1^{\prime }\right) $ and $\left( \ast ,1,2^{\prime
}\right) $ denote the two fundamental isospinors of $SO\left( 4\right) \sim
SU\left( 2\right) \times SU^{\prime }\left( 2\right) $. Notice that each of
the term $\left( 4,2,1^{\prime }\right) $ and $\left( 4^{\prime
},1,2^{\prime }\right) $ has eight real components coming into four \emph{%
real }isodoublets. Up on breaking the $SU\left( 2\right) $ groups down to $%
U\left( 1\right) \times U^{\prime }\left( 1\right) $, the real doublet $%
\left( 4^{\prime },2,1^{\prime }\right) $ gives rise to a $SU\left( 4\right) 
$ complex 4- vector which we denote as $4_{+}$ together with its complex
conjugate $\overline{4}_{-}=\overline{\left( 4_{+}\right) }$. Similarly $%
\left( 4^{\prime },2,1^{\prime }\right) $ gives $4_{-}$ and $\overline{4}%
_{+}=\overline{\left( 4_{-}\right) }$. Notice also that the $4_{+}$ nor $%
4_{-}$ can obey a Majorana condition and so are the smallest spinor objects
one can have in 6D.

\subsubsection{Compactification from 10D down to 4D}

\qquad In the toroidal compactification of 10D space time down to 6D, ($%
\mathcal{K}_{4}=T^{4}$), all the original supersymmetries in 10D are
preserved in 6D. Then, 10D $\mathcal{N}=1$ supersymmetry, generated by a
single $16$- component Majorana-Weyl spinor, gives therefore $\mathcal{N}%
=\left( 1,1\right) $ supersymmetry in 6D; and 10D $\mathcal{N}=2$
supersymmetry generated by two superscharges, gives 6D $\mathcal{N}=\left(
2,2\right) $ supersymmetry.

The situation is different for the compactification on the K3 surface where
half of the original 10D supersymmetric charges are broken. There, unbroken
supersymmetries depend on where K3 holonomies lie. If the holonomy lies in
the $SU^{\prime }\left( 2\right) $ factor, then a constant spinor $2$ is
also covariantly constant and gives rise to unbroken supersymmetry in 6D. As
such, we have the following reductions 
\begin{eqnarray}
16\qquad &\rightarrow &\qquad 8=\left( 4,2,1^{\prime }\right) ,  \notag \\
16^{\prime }\qquad &\rightarrow &\qquad 8^{\prime }=\left( 4^{\prime
},2,1^{\prime }\right) .
\end{eqnarray}%
In type 10D IIA superstring compacticication on the K3 surface where one
starts with $16\oplus 16^{\prime }$, we end with $8\oplus 8^{\prime }$ and
so a non chiral 6D $\mathcal{N}=\left( 1,1\right) $ supersymmetric algebra.
For 10D IIB superstring on the K3 surface, we start with $16\oplus 16$ and
end with a 6D $\mathcal{N}=\left( 2,0\right) $ superalgebra. For 10D type I
and heterotic superstrings one starts with a 10D $\mathcal{N}=1$
supersymmetry and ends with a 6D$\ \mathcal{N}=\left( 1,0\right) $ algebra.
Let us give below some useful details on the explicit commutation relations
of these superalgebras.

\subsection{6D $\mathcal{N}=\left( 2,0\right) $ superalgebra}

\qquad The chiral $\mathcal{N}=\left( 2,0\right) $ supersymmetric algebra is
generated by two identical copies of $\mathcal{N}=\left( 1,0\right) $
supersymmetric algebras with supergenerators denoted as $Q_{1\alpha }^{i}$
and $Q_{2\alpha }^{i}$. Each \ one of these supercharges has eight real
components ($\alpha =1,...,4$ ; $i=1,2$) belonging to the $\left( 4,2\right) 
$ representation of the $SO\left( 1,5\right) \times SU\left( 2\right) $
group symmetry. Since 
\begin{equation}
Q_{1\alpha }^{i}\text{ \ \ }\sim \text{ \ \ }\left( 4,2\right) \text{ \ }\in 
\text{ \ }SO\left( 1,5\right) \times SU\left( 2\right)  \label{su}
\end{equation}%
are complex, we have to impose the reality condition 
\begin{equation}
\overline{\left( Q_{1\alpha }^{i}\right) }=\epsilon _{ij}\mathrm{B}_{\alpha
}^{\beta }Q_{1\beta }^{j}.  \label{ccc}
\end{equation}%
In this relation, $\mathrm{B}$ is a $4\times 4$ matrix constrained as $%
\mathrm{B}^{+}\mathrm{B}=-1$. This condition can be checked by computing the
complex conjugate of eq(\ref{ccc}), i.e: 
\begin{equation}
\overline{\left( \epsilon _{ij}\mathrm{B}_{\alpha }^{\beta }Q_{1\beta
}^{j}\right) },
\end{equation}%
and using the identity $\overline{\left( \epsilon _{ij}\right) }=\epsilon
^{ji}$. One can also use complex 4-component notations; But we will not use
it here since it breaks the $SU\left( 2\right) $ symmetry of (\ref{su}) we
have been interested in throughtout this study.

To get the graded commutation relations of the chiral $\mathcal{N}=\left(
2,0\right) $ superalgebra, it is interesting to compute the reduction of the
tensor product of $Q_{1\alpha }^{i}$ and $Q_{2\alpha }^{i}$. From
representation group theoretical language, we have 
\begin{equation}
\left( 4,2\right) \times \left( 4,2\right) =\left( 16,4\right) ,\qquad
16\times 4=64,
\end{equation}%
which decompose into the following irreducible components, 
\begin{equation}
\left( 16,4\right) =\left( 6,1\right) \oplus \left( 6,3\right) \oplus \left(
10,1\right) \oplus \left( 10,3\right) .  \label{tp}
\end{equation}%
This decomposition deserves two comment: \newline
First it involves two SU$\left( 2\right) $ isosinglet components: (i) The $%
\left( 6,1\right) $ representation that captures the energy momentum tensor $%
P_{\left[ \alpha \beta \right] }$ and should be associated with 
\begin{equation}
\left\{ Q_{1\alpha }^{i},Q_{1\beta }^{j}\right\} =\epsilon ^{ij}P_{\left[
\alpha \beta \right] }.
\end{equation}%
(ii) The $\left( 10,1\right) $ representation associated with the commutator 
$\left[ Q_{1\alpha }^{i},Q_{1\beta }^{j}\right] $ which as the tensor
structure $\epsilon ^{ij}D_{\left( \alpha \beta \right) }$. This
representation does not concern directly the defining graded commutation
relations.\newline
Second, the decomposition (\ref{tp}) involves no space time singlet and
priori the chiral $\mathcal{N}=\left( 2,0\right) $ superalgebra allows no
central charge. The graded commutation relations of the $\mathcal{N}=\left(
2,0\right) $ superalgebra read then as follows: 
\begin{eqnarray}
\left\{ Q_{1\alpha }^{i},Q_{1\beta }^{j}\right\} &=&\epsilon ^{ij}P_{\left[
\alpha \beta \right] },  \notag \\
\left\{ Q_{2\alpha }^{i},Q_{2\beta }^{j}\right\} &=&\epsilon ^{ij}P_{\left[
\alpha \beta \right] },  \notag \\
\left\{ Q_{1\alpha }^{i},Q_{2\beta }^{j}\right\} &=&0,  \label{chi} \\
\left[ P_{\left[ \alpha \beta \right] },Q_{1\gamma }^{j}\right] &=&0,  \notag
\\
\left[ P_{\left[ \alpha \beta \right] },Q_{2\gamma }^{j}\right] &=&0.  \notag
\end{eqnarray}%
Notice that one can also define this superalgebra by using complex Weyl
spinor $Q_{\pm \alpha }^{i}=Q_{1\alpha }^{i}\pm iQ_{2\alpha }^{i}$. The
smallest representation of this superalgebra is given by eq(\ref{sm}). The
chiral 6D $\mathcal{N}=\left( 2,0\right) $ supergravity multiplet is given
by 
\begin{equation}
\text{{\small Bosons}}:\left( 1,1\right) \oplus \left( 0,1\right)
^{5},\qquad \text{{\small Fermions}}:\left( \frac{1}{2},1\right) ^{4},
\end{equation}%
with $3\times \left( 3+5\right) =24$ bosonic on shell degrees of freedom and 
$4\times \left( 2\times 3\right) =24$ fermionic ones.

\subsection{6D $\mathcal{N}=\left( 1,1\right) $ superalgebra}

\qquad The non chiral $\mathcal{N}=\left( 1,1\right) $ supersymmetric
algebra consists of two copies of 6D $\mathcal{N}=1$ superalgebra with
opposite chiralties; that is the tensor product of the $\mathcal{N}=\left(
1,0\right) $ and $\mathcal{N}=\left( 0,1\right) $ supersymmetric algebras.
Denoting by $Q_{\alpha }^{i}$ and $S_{a}^{\bar{\alpha}}$ the fermionic
generators of these superalgebras and using eq(\ref{tp}), its complex
conjugates as well as the following decomposition 
\begin{equation}
\left( 4,2\right) \times \left( \overline{4},2^{\prime }\right) =\left(
1,2\times 2^{\prime }\right) \oplus \left( 15,2\times 2^{\prime }\right) ,
\label{sam}
\end{equation}%
one can write down the graded commutation relations. From the above
representation group reduction, one sees that we do have space-time singlet
given by the term $\left( 1,2\times 2^{\prime }\right) $. So non trivial
central charges $Z_{a}^{i}$ are allowed by non chiral 6D $\mathcal{N}=\left(
1,1\right) $ supersymmetric algebra. Therefore, the graded commutation
relations following from eqs(\ref{tp}-\ref{sam}) read as 
\begin{eqnarray}
\left\{ Q_{\alpha }^{i},Q_{\beta }^{j}\right\} &=&\epsilon ^{ij}P_{\left[
\alpha \beta \right] },  \notag \\
\left\{ S_{a}^{\bar{\alpha}},S_{b}^{\bar{\beta}}\right\} &=&\epsilon _{ab}P^{%
\left[ \bar{\alpha}\bar{\beta}\right] },  \notag \\
\left\{ Q_{\alpha }^{i},S_{a}^{\bar{\alpha}}\right\} &=&\delta _{\alpha }^{%
\bar{\alpha}}Z_{a}^{i}  \notag \\
\left[ P_{\left[ \alpha \beta \right] },Q_{\gamma }^{j}\right] &=&\left[ P_{%
\left[ \alpha \beta \right] },S_{a}^{\bar{\alpha}}\right] =0  \label{n2} \\
\left[ P_{\left[ \alpha \beta \right] },Z_{a}^{i}\right] &=&\left[
Z_{a}^{i},Q_{\alpha }^{j}\right] =\left[ Z_{a}^{i},S_{b}^{\bar{\alpha}}%
\right] =0  \notag
\end{eqnarray}%
where $P^{\left[ \bar{\alpha}\bar{\beta}\right] }=\epsilon ^{\alpha \beta
\gamma \delta }P_{\left[ \gamma \delta \right] }$ and where $\epsilon
^{\alpha \beta \gamma \delta }$ is the completely antisymmetric tensor in
real four dimensions. Notice that under the identification of the $SU\left(
2\right) $ automorphism groups of the two $\mathcal{N}=\left( 1,0\right) $
and $\mathcal{N}=\left( 0,1\right) $ sectors, the four central charges split
as 
\begin{equation}
Z^{ij}=Z_{0}\epsilon ^{ij}+Z^{\left( ij\right) },
\end{equation}%
which should be compared with eq(\ref{cc}). The smallest representation of
this algebras is given by eq(\ref{ga}). The 6D $\mathcal{N}=\left(
1,1\right) $ supergravity multiplet is given by 
\begin{eqnarray}
\text{{\small Bosons}} &:&\left( 1,1\right) \oplus \left( 1,0\right) \oplus
\left( 0,1\right) \oplus \left( \frac{1}{2},\frac{1}{2}\right) ^{4}\oplus
\left( 0,0\right) ,  \notag \\
\text{{\small Fermions}} &:&\left( 1,\frac{1}{2}\right) ^{2}\oplus \left( 
\frac{1}{2},1\right) ^{2}\oplus \left( \frac{1}{2},0\right) ^{2}\oplus
\left( 0,\frac{1}{2}\right) ^{2}.
\end{eqnarray}%
The carry $9+3+3+16+1=32$ bosonic on shell degrees of freedom and $6\times
4+4\times 2=32$ fermionic ones.

\begin{acknowledgement}
:\qquad\ \ \newline
This research work is supported by the programme PROTARS\ D12/25/CNRST. AB
and AS are supported by MCYT ( Spain) under grant FPA 2003-02948. EHS thanks
M. Assorey, L. Boya, L. R. Pepe for scientific discussions and Departemento
di Fisica Teorica, Universidad Zaragoza, for kind hospitality. We would like
to thank R. Ahl Laamara and P. Diaz for discussions. AB would like to thank
S. Montanez for hospitality and discussions.
\end{acknowledgement}

\end{document}